\documentclass[a4paper,fleqn,usenatbib]{mnras}
\usepackage[T1]{fontenc}
\usepackage{ae,aecompl}

\usepackage{graphicx}
\usepackage{amsmath}
\usepackage{amssymb}	
\usepackage{widetext}
\usepackage{hyperref}
\usepackage{deluxetable}

\newcommand{\snia}{SN~Ia}
\newcommand{\sneia}{SNe~Ia}

\newcommand{\diag}{\rm{diag}}
\newcommand{\sigmaintC}{\sigma_{\mathrm{int,C}}}
\newcommand{\sigmaintSN}{\sigma_{\mathrm{int,SN}}}
\newcommand{\kmsmpc}{km\,s$^{-1}$\,Mpc$^{-1}$}
\newcommand{\kms}{km\,s$^{-1}$}
\newcommand{\mBcor}{m_B^\dagger}

\title[low-$z$ SNe Ia for $H_0$]{A blinded determination of $H_0$ from low-redshift Type Ia supernovae, calibrated by Cepheid variables}

\author[Zhang et al.]{\parbox{\textwidth}{
Bonnie~R.~Zhang$^{1,2}$\thanks{E-mail:bonnie.zhang@anu.edu.au},
Michael~J.~Childress$^3$,
Tamara~M.~Davis$^{2,4}$,\\
Natallia~V.~Karpenka$^3$,
Chris~Lidman$^{2,5}$,
Brian~P.~Schmidt$^{1,2}$,
Mathew~Smith$^3$}\\\\
\parbox{\textwidth}{
$^{1}$ Research School of Astronomy and Astrophysics, 
Australian National University, 
Canberra, ACT 2611, Australia.\\
$^{2}$ARC Centre of Excellence for All-sky Astrophysics (CAASTRO).\\
$^{3}$School of Physics and Astronomy, University of Southampton, Southampton, SO17 1BJ, UK.\\
$^{4}$School of Mathematics and Physics,
University of Queensland, QLD 4072, Australia.\\
$^{5}$Australian Astronomical Observatory, PO Box 915, North Ryde, NSW 1670, Australia.
}
}

\date{Accepted 2017 June 22. Received 2017 June 22; in original form 2016 November 27}

\pubyear{2017}

\begin{document}
\label{firstpage}
\pagerange{\pageref{firstpage}--\pageref{lastpage}}
\maketitle

\begin{abstract}

  Presently a ${>}3\sigma$ tension exists between values of the Hubble constant $H_0$ derived from analysis of fluctuations in the Cosmic Microwave Background by Planck, and local measurements of the expansion using calibrators of type Ia supernovae (\sneia). We perform a blinded reanalysis of \cite{Riess11} to measure $H_0$ from low-redshift \sneia, calibrated by Cepheid variables and geometric distances including to NGC~4258. This paper is a demonstration of techniques to be applied to the \cite{Riess16} data. Our end-to-end analysis starts from available CfA3 and LOSS photometry, providing an independent validation of \cite{Riess11}. We obscure the value of $H_0$ throughout our analysis and the first stage of the referee process, because calibration of \sneia~requires a series of often subtle choices, and the potential for results to be affected by human bias is significant. Our analysis departs from that of \cite{Riess11} by incorporating the covariance matrix method adopted in SNLS and JLA to quantify \snia~systematics, and by including a simultaneous fit of all \snia~and Cepheid data. We find $H_0 = 72.5 \pm 3.1 (\rm{stat}) \pm 0.77 (\rm{sys})$~\kmsmpc~with a three-galaxy (NGC~4258+LMC+MW) anchor. The relative uncertainties are 4.3\% statistical, 1.1\% systematic, and 4.4\% total, larger than in \cite{Riess11} (3.3\% total) and the \cite{Efstathiou14} reanalysis (3.4\% total). Our error budget for $H_0$ is dominated by statistical errors due to the small size of the supernova sample, whilst the systematic contribution is dominated by variation in the Cepheid fits, and for the \sneia, uncertainties in the host galaxy mass dependence and Malmquist bias. 
 
\end{abstract}





\section{Introduction}
\label{sec:intro}
The Hubble constant $H_0$ has proven difficult to measure since the discovery of the Universe's expansion almost a century ago~\citep{Hubble29}, following the prediction of the latter in Friedmann's equations~\citep{Friedmann22}. As given in the Hubble law $v = H_0 D$~\citep[first derived by][]{Lemaitre27}, $H_0$ sets the cosmic distance scale via the present expansion rate of the local Universe. The quest to make precise measurements of $H_0$ has been a continual challenge in observational cosmology, due to the difficulty of making accurate distance measurements.
\\
\\
Recently, discrepant values obtained from local and global measurements have propelled the Hubble constant back into the spotlight. Observations of cosmic microwave background (CMB) anisotropies with the Planck satellite found $H_0 = 67.3\pm 1.2$~\kmsmpc~\citep{Planck14}, assuming a standard $\Lambda$CDM cosmology. This value is 
${\sim}2.7\sigma$ lower than in \citet[][hereafter R11]{Riess11}, who measure $H_0 = 73.8\pm 2.4$~\kmsmpc~from observations of type Ia supernovae (\sneia) in the more local Universe. While the Planck measurement is dependent on an underlying cosmological model, the \snia-based measurement is model-independent. The precision of these values highlights the importance of the tension between the two modes of measurements, which has increased to over $3\sigma$ significance in the updated analyses in \citet[][hereafter R16]{Riess16} (finding $H_0 = 73.0 \pm 1.8$~\kmsmpc), and \citet{Planck16} (finding $H_0 = 67.8 \pm 0.9$~\kmsmpc).
\\
\\
Numerous reanalyses of the \snia-based measurement have followed, many of which have focussed on the methods for the rejection of Cepheid outliers. \citet[][hereafter E14]{Efstathiou14} 
questions and revises the outlier rejection algorithm in R11, concluding $H_0 = 72.5\pm 2.5$~\kmsmpc~assuming a null metallicity dependence of the Leavitt law. Recently, \cite{Cardona16} uses Bayesian hyper-parameters to down-weight portions of the Cepheid data for both R11 and R16 data sets, finding $H_0 = 73.75\pm 2.11$~\kmsmpc~for the R16 data. Moreover, the dependence of the intrinsic magnitude of \sneia~on host galaxy properties has been explored 
in recent years \citep[e.g.][]{Sullivan10}. \cite{Rigault13, Rigault15} find a relationship between peak brightness and star formation rate, and infer an overestimate of ${\sim}3$~\kmsmpc~in the R11 value of $H_0$ arising from the fact that the calibration set of \sneia~exist in galaxies which necessarily contain Cepheids, hence are likely to be late-type galaxies. However, \cite{Jones15} repeat the same analysis, with an increased sample size and the R11 selection criteria applied, and find no significant difference in the brightness of \sneia~in star-forming and passive environments.
\\
\\
The CMB data in Planck has been reanalysed in \cite{Spergel15}, who find a similar value to \cite{Planck14}, of $H_0 = 68.0\pm 1.1$~\kmsmpc. \cite{Bennett14} provides a CMB-based measurement which is independent of Planck, by combining data from WMAP9, the South Pole Telescope (SPT) and Atacama Cosmology Telescope (ACT), and baryon acoustic oscillation (BAO) measurements from BOSS, finding a value of $H_0=69.3\pm 0.7$~\kmsmpc~(with a slight increase to $H_0 = 69.7\pm 0.7$~\kmsmpc~if \snia~data from R11 are included), which is slightly less discrepant with \snia-based values. Strong lensing provides an independent but model-dependent local measurement of $H_0$: the \citet[][(H0LiCOW)]{Suyu16} program studies time delays between multiple images of quasars in strong gravitational lens systems, and find $H_0 = 71.9^{+2.4}_{-3.0}$~\kmsmpc~\citep{Bonvin17} in flat~$\Lambda$CDM. It is noteworthy that the H0LiCOW analysis was performed blind to derived cosmological parameters \citet[e.g.\ ][section~4.4]{Bonvin17}; we discuss the importance of blinding in our analysis in Section~\ref{sec:blind}. 
\\
\\
One of the biggest open questions in cosmology today is whether the tension in $H_0$ signifies new physics -- where inconsistencies between results from supernovae and the CMB arise from the model-dependence of the measurement, and disappear when the correct model is used -- or is the result of some systematic error in one or both measurements that has yet to be accounted for. Possible theoretical modifications to standard $\Lambda$CDM to reconcile the tension in $H_0$ include an increased neutrino effective number (the existence of dark radiation), and/or a more negative dark energy equation-of-state parameter $w$ at late times. \cite{DiValentino16} explore these scenarios in a higher-dimensional parameter space, with their findings supporting phantom dark energy with $w{\sim}-1.3$, while \cite{Wyman14, Dvorkin14, Leistedt14} focus on the implications of an additional massive sterile neutrino species. Meanwhile, \cite{Bernal16} examine the model dependence of the Universe's distance scale (anchored by $H_0$ and by the scale $r_S$ of the sound horizon at radiation drag, at late and early times respectively) by reconstructing its expansion history with minimal cosmological assumptions.\footnote{This is possible as the combination of \sneia~and BAO as probes constrains the product $r_SH_0$ in a model-independent way.} They conclude that the tension in $H_0$ translates to a mismatch in the normalisations provided by $H_0$ and $r_s$ at two opposite ends of the distance ladder.
\\
\\
A genuine inconsistency in the value of the Hubble constant at low and high redshifts would have profound consequences. Therefore it is imperative to fully understand uncertainties in the measured values of $H_0$, and to preclude possible human biases on the result. The most effective way of achieving the latter is to blind the value of $H_0$ throughout the analysis.
\\\\
The use of data from R11 is for 
a proof of concept, necessary for our blind analysis technique, and to be followed shortly with the same analysis applied to R16 data. 
Numerous improvements over R11 have been made in R16, in the analysis as well as the size and quality of data. Changes to the outlier rejection and the Cepheid metallicity calculations have addressed some of the concerns raised in E14. However, our analysis involves both a simultaneous fit to all data sets, and the accepted methodology of recent supernova cosmology analyses \citep{Conley11, Betoule14} for considering \snia~systematics. Both of these points carry significant differences from the R11 and R16 analysis chains, and have yet to be included in a reanalysis. Nor has the supernova data been revisited in its entirety, starting from the available photometry. Thus, we are motivated by the desire to provide such a validation of the supernova data, and by the current relevance and importance of the Hubble constant, to produce in this work an independent, blinded, end-to-end reanalysis of the R11 data to determine $H_0$ and its uncertainty. \\\\
In summary, we combine the framework for calibrating a \snia~Hubble diagram with Cepheid variables, with the best estimates of supernova systematics via covariance matrices. We determine $H_0$ using the magnitude-redshift relation (i.e.\ a Hubble diagram) of low-redshift \sneia, with their zero point set by Cepheid variables in host galaxies of eight nearby \sneia, which are in turn calibrated by very long baseline interferometry (VLBI) observations of megamasers in NGC~4258, and other geometric distances to the Large Magellanic Cloud (LMC) and Cepheids in our Galaxy.
\\\\
This paper is structured as follows. First we present an overview of our methods in \textsection\ref{sec:methods}, followed by the distinct sets of data with the equations relating them in \textsection\ref{sec:data}. In \textsection\ref{sec:cepheids} we focus on the Cepheid variables and perform a fit to the Cepheid data only, comparable to the E14 reanalysis of R11. Next in \textsection\ref{sec:SNe} we discuss type Ia supernovae, including details of fitting \sneia~on a Hubble diagram and results of a preliminary SN-only fit. This fit relies on computations of individual supernova systematic terms in the form of covariance matrices, which are examined in depth in Appendix~\ref{sec:systematics}. Finally, \textsection\ref{sec:results} ties together the Cepheid and \snia~information into a combined and simultaneous fit of all data; we conclude with a discussion of these results in \textsection\ref{sec:conclusion}.


\section{Methods}
\label{sec:methods}
In this section we paint a broad picture of our approach to measuring $H_0$, postponing specific details of and equations relating to the data to Section~\ref{sec:data}. We begin with the theory and mathematics of finding $H_0$ in the cosmology analysis, followed by the astronomy that enables this: distance measurements with Type Ia supernovae and Cepheid variables as standard candles. Equally important are the Bayesian statistics that underpin the analysis, and the method for blinding the result.

\subsection{Theory of extracting $H_0$}
\label{sec:H0}

In its traditional formulation Hubble's law states that the recession velocity of objects is proportional to their distance:
\begin{align}
v(z) = H_0 D(z)\label{eq:HL}
\end{align}
where the constant of proportionality $H_0$ represents the present expansion rate of the Universe, scaled by its size (i.e.\ $H_0 = \frac{\dot{a}}{a}$ where $a$ is the scale factor and overdot indicates differentiation with respect to time, $t$). Methods of determining $H_0$ typically involve taking the ratio of the two sides of Equation~\ref{eq:HL}. We expand on the subtleties of this below. \\
\\
The distance in Hubble's Law is related to the luminosity distance by 
\begin{align}
D(z)=\frac{1}{1+z}D_L(z) \label{eq:DL},
\end{align} 
assuming a flat Universe.\footnote{To include curvature note that the present distance to an object at redshift $z$ is given by $D(z) = R_0\chi$, with $\chi$ being the comoving coordinate and $R_0$ the scale factor at the present day with dimensions of distance (in the equation for $H_0$ above $a(t)\equiv R(t)/R_0$). Then luminosity distance is defined as
\begin{align}
D_L(z)\equiv(1+z)R_0 S_k(\chi),
\end{align}
 with $R_0\equiv c/(H_0 \sqrt{|\Omega_k|})$ and
  \begin{align}S_k(x) &= \begin{cases}\sin x &k=1,\\ x &k=0,\\\sinh x &k=-1,
    \end{cases}\end{align}
    so $D(z) = \frac{1}{1+z} \frac{\chi}{S_k(\chi)} D_L(z)$.
      }

The luminosity distance $D_L(z)$ can be determined observationally (i.e.\ with no knowledge of cosmological parameters) using standard candles. These have known absolute magnitudes $M$, so taking the difference between $M$ and the apparent magnitude $m$ gives the distance modulus $\mu \equiv m - M$ and hence the luminosity distance $D_L \equiv 10^{\frac{\mu -25}{5}}$~Mpc. In practice the process of measuring distances is far from straightforward, and is outlined in Section~\ref{sec:distances}.\\ 
\\
On the left hand side, $v(z)$ is the predicted velocity due to expansion for a galaxy at redshift $z$.\footnote{For simplicity, we do not distinguish here between redshifts in different reference frames, and only use one redshift $z$. We distinguish between different redshifts, particularly in Equation~\ref{eq:DL}, in Appendix~\ref{sec:velcorrection}.} The exact expression for $v(z)$ is given by integrating the Universe's expansion up to redshift $z$:
\begin{align}
  v(z) = c\int_0^z \frac{dz'}{E(z')}, \label{eq:vofz}
\end{align}
where $E(z) \equiv H(z)/H_0$ is a function of cosmological parameters, as defined in \cite{Peebles93},\footnote{In Friedmann-Lema\^{i}tre-Robertson-Walker (FLRW) cosmologies, $E(z)$ is given by~\citep[e.g.][]{CarrollPressTurner92}
\begin{align}
  E(z) = 
  \sqrt{\Omega_M(1+z)^3+\Omega_k(1+z)^2 + \Omega_\Lambda}
\end{align}
where $\Omega_M$ and $\Omega_\Lambda$ are respectively the densities of normal matter and the cosmological constant (relative to the critical density), $k$ is the curvature, and $\Omega_k \equiv 1 - \Omega_M -\Omega_\Lambda$ (zero in a flat Universe).  If dark energy is something other than a cosmological constant, with a generic equation of state $w$, replace $\Omega_\Lambda$ with $\Omega_{\rm de}(1+z)^{3(1+w)}$.}  
and $v(z)$ is independent of $H_0$.\footnote{It is interesting to note that $v(z)$ is independent of $H_0$; it depends only on redshift and cosmological parameters such as $\Omega_M$ and $\Omega_\Lambda$.  That may seem unintuitive, but it is velocity as a function of distance $v(D)$ that is function of $H_0$ (things that are moving faster have gone further).  Velocity as a function of redshift $v(z)$ works differently since redshift is not proportional to distance.  A galaxy's redshift is determined by how much the Universe has expanded since the light was emitted.  That depends on the travel time, which does depend on the densities that cause the Universe to accelerate or decelerate (and thus for the light to take longer or shorter times to reach us), but not on $H_0$.}
\\
\\
At low redshifts the cosmological dependence of $v(z)$ is very weak and it is a good approximation to use a second order Taylor expansion in terms of the deceleration and jerk parameters $q_0$ and $j_0$\footnote{We assume a standard $\Lambda$CDM cosmology with $\Omega_M{\sim}0.3, \Omega_\Lambda{\sim}0.7$, fixing $q_0 = -0.55$ and $j_0 = 1$.\label{foot:1}}. Thus we follow R11 and use, 
\begin{align}
v(z) = \frac{cz}{1+z}\left[1+\frac{1}{2}(1-q_0)z -\frac{1}{6}(1-q_0 - 3q_0^2 + j_0)z^2\right]. \label{eq:vzapprox}
\end{align}
At low redshift Equations~\ref{eq:vofz} and~\ref{eq:vzapprox} both reduce to the familiar $v(z) \approx cz$. At moderate redshifts ($z<0.1$), Equation~\ref{eq:vzapprox} closely approximates most observationally reasonable cosmological models. We explored the uncertainty associated with assuming Equation~\ref{eq:vzapprox} and the cosmology stated in Footnote~\ref{foot:1}, finding the impact to be small: varying either $\Omega_M$ or $w$ by 0.1 changes $H_0$ by 0.015~\kmsmpc~or 0.1~\kmsmpc, respectively, in the sense that an increase in $\Omega_M$ or $w$ causes an increase in $H_0$. The maximal difference in $\mathcal{M}$ induced by varying $q_0, j_0$ within values allowed by 1$\sigma$ contours in \cite{Betoule14} is an order of magnitude smaller than its statistical uncertainty.
\\\\
Rearranging Equations~\ref{eq:HL}, \ref{eq:DL}, and \ref{eq:vzapprox} gives us the equation for $H_0$ as a function of observables, $z$ and $D_L$,\footnote{For non-zero curvature, Equation~\ref{eq:H0obs} becomes
\begin{align}
H_0 = \frac{v(z)(1+z)}{D_L(z)}\frac{S_k(\chi)}{\chi}.
\end{align}
}
\begin{align}
H_0 &= \frac{v(z)(1+z)}{D_L(z)}\notag\\
&= \frac{cz}{D_L(z)}\left[1+\frac{1}{2}(1-q_0)z -\frac{1}{6}(1-q_0 - 3q_0^2 + j_0)z^2\right]\label{eq:H0obs}.
\end{align}
Thus determining $H_0$ amounts to comparing the velocity in Equation~\ref{eq:vzapprox} -- derived from the measured redshift -- to the observed luminosity distance, measured with standard candles. The equations encapsulating this process are detailed in Section~\ref{sec:data}.
\subsection{Measuring distance}
\label{sec:distances}
Astronomical distances can be measured using standard candles: standardisable objects with known absolute magnitude which, combined with an apparent magnitude, give the distance modulus. These distances are often relative rather than absolute. Since each mode of measurement is useful only over a limited range of distances, multiple standard candles are tied together to form a so-called distance ladder. At the bottom of the ladder are absolute distances determined from geometric methods (i.e.\ trigonometric parallax), only accurate at relatively small distances. Then nearby standard candles (i.e.\ Cepheid variables) give distances relative to this geometric scale; similarly, each rung of the ladder is calibrated on the previous.
\\
\\
Standard candles (a distance scale) provide one approach to measuring cosmological parameters including $H_0$. Alternatively, standard rods (a length scale) in the form of baryon acoustic oscillations~\citep[BAO; e.g.][]{Eisenstein13} provide complementary (and for some parameters, orthogonal) constraints, most recently in \cite{Planck16}. \cite{Weinberg13} provides a review of cosmological probes; we refer the interested reader to its section~4 for a review of BAO. 
\\\\
In our determination of $H_0$ we rely on two standard candles: type Ia supernovae and Cepheid variables. These together prescribe a relative distance scale for the low-$z$ \sneia. The absolute calibration is given by the geometric maser distance of NGC~4258 from \cite{Humphreys13}. The Cepheid variables lie in this galaxy and eight other galaxies containing nearby \sneia, calibrating the supernovae. The absolute distances and measured redshifts of the low-$z$ supernovae are combined to determine $H_0$ as described in Section~\ref{sec:H0}, through equations detailed in Section~\ref{sec:data}. Next we briefly describe each standard candle.
\\
\\
Cepheid variables are pulsating supergiants with periods of days to hundreds of days, well-characterised by their luminosity via the empirical Leavitt law~\citep{Leavitt08, LeavittPickering} -- also commonly known as the Period-Luminosity relation. The brightness and regular pulsation of Cepheid variables as well as their ease of discovery and classification make Cepheids reliable distance indicators in the nearby Universe, and the basis of the cosmic distance ladder~\citep{FreedmanMadore01}. Some difficulties and systematics include crowding and confusion (which necessitate outlier rejection), metallicity, and extinction; these are discussed further in Section~\ref{sec:cepheidsystematics}.\\
\\
\sneia~are thought to be thermonuclear explosions of accreting white dwarfs, with two qualities which recommend them as excellent distance indicators up to high redshift: they are intrinsically very bright, and highly standardisable in terms of their apparent peak brightness, lightcurve shape, and colour~\citep{Phillips93, Hamuy96, Riess96} -- see Section~\ref{sec:dataSNIa} for more details. Indeed \sneia~have played a pivotal role in recent observational cosmology, particularly in the discovery of the accelerating Universe \citep{Riess98, Schmidt98, Perlmutter99}. In the past decade \snia~samples have greatly expanded, reducing statistical uncertainty. However, observations of \snia~are still subject to numerous systematics, which can be significant and correlated between SNe. These systematics include: calibration uncertainties, dust, and corrections for peculiar velocities, and host galaxy mass, and will be discussed in Appendix~\ref{sec:systematics}. 
\subsection{Bayesian statistical methods}
\label{sec:bayesian}
We estimate the fit parameters $\Theta$ (given in Section~\ref{sec:data}) in a Bayesian framework, relying on the principle of sampling the likelihood $\mathcal{L}(\Theta)$ over the parameter space to determine the posterior distribution function (PDF). The generalised likelihood is determined from the $\chi^2$ statistic, a function of $\Theta$: 
\begin{align}
  \mathcal{L}(\Theta) &= \exp\left(-\frac{\chi^2(\Theta)}{2}\right) \label{eq:likelihood}\\
    \chi^2(\Theta) &= (\boldsymbol{\hat{m}} -\boldsymbol{m_\mathrm{mod}})\boldsymbol{\cdot}\mathbf{C}^{-1}\boldsymbol{\cdot}(\boldsymbol{\hat{m}} -\boldsymbol{m_\mathrm{mod}})^T.\label{eq:chisqgeneral}
\end{align}
Here $\boldsymbol{\hat{m}}$ and $\boldsymbol{m_{\rm{mod}}}$ are the observed and theoretical magnitude vectors (over all data) respectively,\footnote{We retain this convention where it is necessary to explicitly distinguish the data from the model.} and $\mathbf{C}$ is the covariance matrix of uncertainties in $\boldsymbol{\hat{m}}$. The model $\boldsymbol{m_{\rm{mod}}}$ is implicitly a function of $\Theta$. In each fit outlined in Section~\ref{sec:steps} an expression for $\chi^2$ will be given explicitly, i.e.\ Equations~\ref{eq:chisqcepheids}, \ref{eq:chisqlowz}, and \ref{eq:chisqglobal}. When uncertainties are uncorrelated (i.e.\ $C$ is diagonal) Equation~\ref{eq:chisqgeneral} reduces to the more familiar 
\begin{align}
    \chi^2(\Theta) &= \sum_i \frac{(\hat{m}_i-m_{{\rm{mod}}i})^2}{\sigma_i^2}.
\end{align}

\subsubsection{PDF estimation}
\label{sec:multinest}
In higher dimensional parameter spaces the computational expense of calculating and integrating the likelihood necessitates Monte Carlo techniques to statistically sample the parameter space, the most common being Markov Chain Monte Carlo (MCMC). These techniques are useful for parameter estimation or model selection \citep[see e.g.][]{DavisParkinson16}. Nested sampling~\citep{Skilling04} is another such technique, in which the likelihood is evaluated at sample `live' points drawn from an iteratively replaced distribution until convergence, where the posterior is recovered. The MultiNest algorithm \citep{Feroz08,Feroz09,Feroz13} is a robust nested sampling tool for retrieving posterior samples from distributions which may have multiple peaks or `modes'. We use the implementation \texttt{PyMultiNest} described in \cite{Buchner14} for most fits (details in Section~\ref{sec:steps}). For some lower-dimensional fits (Section~\ref{sec:snfit}) we use \texttt{emcee} \citep{emcee}, a Python implementation of MCMC.\\
\\
Each algorithm takes as inputs the data, a prior distribution within the parameter space (which live points are drawn from), and the likelihood as a function of the data and parameters. In addition we select the sampling efficiency for parameter estimation and the number of live points (walkers). MultiNest outputs include the best fit (maximal likelihood) parameters and the marginalised posterior distribution for each parameter. In our fits the marginalised PDFs appear symmetrical and Gaussian (e.g.\ Figs~\ref{fig:8dim} and \ref{fig:3dim}), so we take our best estimates of values and uncertainties of each parameter from the mean and standard deviation of the marginalised PDF.

\subsection{Blind analysis}
\label{sec:blind}
To perform a blind analysis is to obscure the principal aspects of the result until the analysis is complete. The overarching motivation for blinding is to eliminate the impact of human biases on the result, including confirmation bias. Preconceptions about the `correct' value for a result are irrelevant to the validity of the analysis and can only reduce the value of the findings. Conversely a blind analysis has all the more bearing for having reached its conclusion blind. \cite{CroftDailey11} find evidence of confirmation bias in recent measurements of cosmological parameters and recommend blinding; similarly \cite{MaccounPerlmutter15} argue for its necessity.  In recent years the practice of blind analysis has become standard in particle physics, and is increasingly adopted in cosmology. 
\\\\
Our priority is to hide the value of $H_0$ so as to not influence its result, so we blind the parameter $\mathcal{H}$ which contains equivalent information.\footnote{We fit for the parameter $\mathcal{H}:= 5\log_{10}H_0 - 25$, which is linear in magnitude (Equation~\ref{eq:Hdefn}), instead of $H_0$.} We also blind the \snia~magnitude zero point $M_B$ which has the most interaction with $\mathcal{H}$, and is the best-constrained in the literature, relative to other parameters in $\Theta$. We implement these blinds in the analysis and data respectively. For any likelihood function containing $\mathcal{H}$ (i.e.\ involving the low-$z$ SNe) we make the shift $\mathcal{H} \mapsto \mathcal{H} + \mathfrak{o}_H$ for an offset $\mathfrak{o}_H$. Meanwhile we effectively shift $M_B$ by adding another offset $\mathfrak{o}_M$ to all SN magnitudes $m_B$. Both offsets $\mathfrak{o}_H$ and $\mathfrak{o}_M$ are unknown real numbers, randomly drawn from normal distributions and never printed. These are seeded by distinct known numbers to ensure that the offsets are constant and can be retrieved. Our method allows the recovery of the true unblinded values by simply subtracting the offsets once the blind is lifted.\\
\\
We choose to not blind the other parameters which appear in the preliminary Cepheid- or SN-only fits, primarily because these parameters do not have strong enough priors from the literature to introduce human bias. Moreover, the variation we observe in the preliminary values of the nuisance parameters $\{b_W, Z_W, \alpha, \beta\}$ is useful for informing which preliminary fits to carry forward to the global fits. Knowing the preliminary nuisance parameters will not bias our results because `best' versions of the preliminary fits are not chosen; instead we select a representative sample of these fits and use the scatter to quantify the systematic uncertainties. 


\section{Data and Analysis Techniques}
\label{sec:data}
This section describes our Cepheid and \snia~data, using equations for the apparent magnitude of each data set to demonstrate the relationships between them. These are followed by an outline of the steps of the fit.

\subsection{Data samples}

Our analysis uses three sets of data: 
\begin{enumerate}
\item {\bf Cepheid variables:} 570 spread between nine nearby galaxies (see Table~\ref{tab:nearby}), namely:
\begin{itemize}
\item 165 in the distance anchor NGC~4258, and
\item 405 in eight galaxies that host recent nearby \sneia.
\end{itemize}
\item {\bf Anchor (`nearby') supernovae:} 8 recent \sneia\ in the nearby galaxies (also in Table~\ref{tab:nearby}).
\item {\bf Low-$z$ \sneia:} 280 low-redshift ($z<0.06$) \sneia\ from the CfA3~\citep{Hicken09a} and LOSS~\citep{Ganeshalingam10} samples.   
\end{enumerate}

Together these three data sets allow us to calibrate our distance ladder. The galaxy NGC~4258 hosts the water masers that give us a precise absolute local distant measurement~\citep{Humphreys13}, and allows us to calibrate the Cepheids. As in R11, we also use the LMC and Milky Way (MW) as distance anchors in combination with NGC~4258, relying on independent distances measured from detached eclipsing binaries~\citep{Pietrzynski13} to Cepheids in the LMC, and Hipparcos and HST parallax measurements of Cepheids in our Galaxy~\citep{vanLeeuwen07}. The Cepheids in turn enable us to calibrate the absolute magnitudes of the eight supernovae that occurred in nearby galaxies with quality Cepheid measurements. These then allow us to calibrate the whole supernova sample, which ultimately gives most of the constraining power for our $H_0$ measurement. In practice we perform a global fit to all of these samples together. In the next section we outline the equations needed to relate all of these standard candles and extract a measurement of $H_0$ following the theory in Section~\ref{sec:H0}. \\
\\
Since the purpose of this paper is to provide an independent analysis of the data in R11, we adopt an identical sample in order to make a faithful comparison.  Our aim is to use the same framework to analyse newer data sets including \sneia~in the CfA4 survey~\citep{Hicken12} and Cepheids in R16 at a later stage.

\subsection{Equations for apparent magnitude}
\subsubsection{Cepheids}
\begin{table}
\begin{center}
  \caption{Recent nearby \sneia~and their host galaxies used in R11, 
    along with observations of Cepheids in these galaxies.
\label{tab:nearby}}
\begin{tabular}{lll}
\hline
\hline
Galaxy&\snia&$N_{\rm Cepheids}$\\
\hline
NGC 4536&SN 1981B&69\\
NGC 4639 &SN 1990N&32\\
NGC 3370 &SN 1994ae &79\\
NGC 3982 &SN 1998aq &29\\
NGC 3021&SN 1995al& 26\\
NGC 1309 &SN 2002fk &36\\
NGC 5584 &SN 2007af&95\\
NGC 4038 &SN 2007sr &39\\
NGC 4258 & - &165\\
\hline
Total & & 570\\
\hline
\end{tabular}
\end{center}
\end{table}

Our first data set, the Cepheid variables, allow us to infer distances to the nearby galaxies via the Leavitt law (also commonly known as the period-luminosity relation):
\begin{equation}\label{eq:pl1}
m_W = b_W (\log_{10} P - 1) + Z_W \Delta \log_{10}[O/H]_{ij} + M_W +\mu.\\
\end{equation}
Equation~\ref{eq:pl1} relates the apparent `extinction-free' (Wesenheit) magnitude $m_W$,\footnote{We use the quantity $M_W \equiv V - R_V(V-I)$ constructed in \cite{Madore82} from the Wesenheit function~\citep{vandenBergh75}, from {\em{V}}- and {\em{I}}-band absolute magnitudes. Assuming constant ratio $R_V$ of total to selective absorption, $M_W$ is independent of extinction. We fix $R_V=A_V/E(V-I) = 3.1$ as in R11.} period ($P$; in days), and metallicity of a Cepheid at distance modulus $\mu$. The slopes $b_W$ and $Z_W$ represent the dependence of the magnitude on period and metallicity; the zero point $M_W$ physically represents the Wesenheit magnitude of a Cepheid in our Galaxy (at a distance of 10~pc), with a period of 10 days. We use relative values of the metallicity ($\Delta \log_{10}[O/H]_{ij}:=\log_{10}[O/H] - 8.9$) and period to pivot the fit near the data.

\subsubsection{Type Ia supernovae}
\label{sec:dataSNIa}
Type Ia supernovae comprise our remaining data. A spectroscopically normal \snia~has a lightcurve parametrized by its brightness (hence distance), observed colour and decline rate. These measures are represented by different quantities in various \snia~frameworks; in SALT2~\citep{Guy07} these are the apparent magnitude $m_B$ at time of {\em{B}}-band maximum, `stretch' $X_1$ and colour $C$ (roughly corresponding to $B-V$ at maximum), related by:
\begin{align}\label{eq:snia}
  m_B = M_B - \alpha X_1 + \beta C + \mu
\end{align}
where $M_B$ is the canonical \snia~absolute magnitude, and $\alpha, \beta$ are SALT2 nuisance parameters for the stretch and colour dependences.
\\
\\
\sneia~in more massive galaxies are brighter after these standard corrections for colour and stretch, as discussed in Appendix~\ref{sec:hostcorrection}. To account for this we replace $M_B$ in Equation~\ref{eq:snia} with the corrected absolute magnitude $M_B^*$, which can take two discrete values depending on the host galaxy mass: $M_B$ or $M_B + \Delta M_B$. We will fix $\Delta M_B$ (see Appendix~\ref{sec:hostcorrection}) and fit for the three global parameters $\{\alpha, \beta, M_B\}$.
\\
\\
Our second data set contains the eight `nearby' \sneia~in Table~\ref{tab:nearby}, with apparent magnitudes given by Equation~\ref{eq:snia} (with $M_B^*$ instead of $M_B$). A \snia~and Cepheid in the same galaxy have common distance modulus $\mu$ in Equations~\ref{eq:snia} and Equation~\ref{eq:pl1}; thus, the Cepheids calibrate the nearby SNe, which in turn determine the \snia~magnitude zero point $M_B$. 
\\\\
The much larger sample of 280 \sneia~makes up our third data set. These `low-$z$' supernovae originate from CfA3 and LOSS, with details to follow in Section~\ref{sec:SNobs}. Once we have calibrated their absolute magnitudes using the eight `nearby' supernovae, we can use the theory derived in Section~\ref{sec:H0} to relate their measured magnitudes to the value of $H_0$. Assuming Equation~\ref{eq:H0obs} and writing $f(z) \equiv 1 + \frac{(1-q_0)z}{2} -\frac{(1-q_0 - 3q_0^2 + j_0)z^2}{6}$, we have in place of Equation~\ref{eq:snia}
\begin{equation}\label{eq:sn1}
m_B = M_B^* - \alpha X_1 + \beta C + 5\log_{10}\left(\frac{czf(z)}{H_0}\right) + 25.
\end{equation}

\subsection{Global fit}
\label{sec:globaleqns}
We will fit Equations~\ref{eq:pl1}, \ref{eq:snia}, \ref{eq:sn1} simultaneously for a combined fit to all Cepheid and \snia~data. We rewrite these equations, making explicit the indexing: $i$ varies over the eight nearby galaxies (and the \sneia~they contain), $j$ varies over Cepheids in these galaxies and NGC~4258, $k$ varies over the low-$z$ SNe.

\begin{widetext} 
\begin{align}
m_{Wij} &= b_W (\log_{10} P_{ij} - 1) + Z_W \Delta \log_{10}[O/H]_{ij} + M_W +\mu_{4258} + \Delta\mu_i\label{eq:pl2}\\
m_{Bi} &= M_B^* - \alpha X_{1i} + \beta C_i + \mu_{4258} + \Delta\mu_i\label{eq:nearby2}\\
m_{Bk} &= M_B^* -\alpha X_{1k} + \beta C_k + 5\log_{10}(cz_kf(z_k)) -\mathcal{H}\label{eq:sn2}.\\\notag
\end{align}
\end{widetext}
In Equation~\ref{eq:sn2} we separate the intercept of Equation~\ref{eq:sn1} into parameters $M_B^*$ (also appearing in Equation~\ref{eq:nearby2}) and a constant term $\mathcal{H}$, which contains the same information as $H_0$:
\begin{align}
\label{eq:Hdefn}
\mathcal{H} := 5\log_{10}H_0 - 25.
\end{align}
We fit for all 16 parameters appearing in Equations~\ref{eq:pl2}--\ref{eq:sn2}; explicitly these are $\Theta=\{\alpha, \beta, \mathcal{H}, M_B, b_W, Z_W, M_W, \mu_{4258}, \Delta\mu_i\}$ where $i$ varies over the eight nearby galaxies. Note that we fit for $M_B$ instead of $M_B^*$ as the latter is not a constant. The distance moduli in Equations~\ref{eq:pl2} and \ref{eq:nearby2} are expressed as offsets $\Delta\mu_i\equiv \mu_i - \mu_{4258}$, relative to NGC~4258. \\
\\
Equations~\ref{eq:pl2}--\ref{eq:sn2} assume a distance anchor of NGC~4258. The use of the LMC and MW as alternate or additional anchors is explored, and discussed in Appendix~\ref{sec:anchors}. We impose a strong Gaussian prior $\mu_{4258}= 29.404 \pm 0.066$ on the distance, measured from VLBI observations of megamasers in \cite{Humphreys13}\footnote{This distance is slightly higher than the older value $\mu_{4258}=29.31$ assumed in \cite{R11erratum}; this increase acts to decrease $H_0$ relatively.\label{foot:2}} whenever NGC~4258 is used as an anchor, and similarly $\mu_{\rm{LMC}} = 18.494 \pm 0.049$ if the LMC is included. 

\subsection{Steps in fitting process}
\label{sec:steps}
We break the process of fitting all data to Equations~\ref{eq:pl2}--\ref{eq:sn2} into three steps to streamline the process: the data and parameters are separated into spheres of influence so that results from the Cepheid- and SN-only fits -- in particular their dependences on factors such as rejection, cuts, and distance anchors -- can be isolated, inspected, and selectively carried forward to the global fit.
\\\\
The three steps are as follows. First we fit all Cepheid data simultaneously for parameters $\left\{b_W, Z_W, M_W, \{\mu_i\}\right\}$ to Equation~\ref{eq:pl2}. Separately, we fit only the low-$z$ \sneia~to Equation~\ref{eq:sn2}. The parameters $M_B$ and $\mathcal{H}$ are degenerate when constrained by only the low-$z$ data, so we fit for their difference $\mathcal{M} := M_B - \mathcal{H}$, as well as \snia~parameters $\alpha, \beta$. Finally, a global fit is performed (independent of the first two steps) of both data sets \emph{and} the nearby \sneia~to Equations~\ref{eq:pl2}, \ref{eq:nearby2}, \ref{eq:sn2} simultaneously. This step is similar to the Cepheid-only fit but also includes $\mathcal{H}$ and the SN parameters $\{\alpha,\beta,M_B\}$. Final values for all parameters including $H_0$ are extracted from this global fit. Each preliminary fit is described in detail in Sections~\ref{sec:cepheidfit} and \ref{sec:snfit}, and the global fit in Section~\ref{sec:globalfit}. 
\\\\
The Bayesian methods for parameter estimation (MultiNest for the high-dimensional Cepheid-only and global fits, and MCMC for the SN-only fit -- described in Section~\ref{sec:bayesian}) require priors, which may be uniform, on each parameter in $\Theta$. While some other parameters are predominantly influenced by a subset of the data (namely the nuisance parameters $b_W$ and $Z_W$ which only appear in the Cepheid-only fit, and $\alpha$ and $\beta$ which are predominantly determined by low-$z$ SNe), it would be statistically invalid to place Gaussian priors on these parameters in the global fit based on results of either preliminary fit. However non-uniform priors based on external data are allowed; our priors on $\mu_{4258}$ (and $\mu_{\rm{LMC}}$) are Gaussian if these galaxies are included as calibrators, and we constrain $b_W$ with a Gaussian prior informed by the LMC Cepheids in fits which are not anchored on the LMC (discussed in Appendix~\ref{sec:cepheidpriors}). For the remaining parameters in $\Theta$ we set uniform priors over generous intervals.\\
\\
Our approach differs from the R11 and E14 analyses, which both perform two independent steps: (i) using only the low-$z$ SN data, determine and fix $a_V$ (the intercept of the \snia~$m-z$ relation equivalent to $0.2\mathcal{M}$ in our analysis) and (ii) from the Cepheids only, determine the Leavitt law parameters $b_W, Z_W$, and $zp_{4258}$ (a zero point comparable to our $M_W$). The Cepheid parameters are combined with the nearby \sneia~lightcurves to find the quantity $m^0_{v,4258}$ which signifies the fiducial peak apparent magnitude of a \snia~in NGC~4258; this quantity is then combined with $a_V$ and $\mu_{4258}$ to give $H_0$ (R11, equation~4). We emphasize that, in contrast, our final global fit is truly simultaneous in that it allows each parameter in $\Theta$ to be influenced by all Cepheid and supernova data in the nearby galaxies and low-$z$ sample. Consequently we allow the data sets to interact freely with each other, enabling us to capture covariances between parameters.

\section{Cepheid Leavitt law fit}
\label{sec:cepheids}
Here we describe an initial simultaneous fit of the Cepheids in all nine galaxies to the Leavitt law (Equation~\ref{eq:pl2}). This has two purposes: to estimate the parameters $\left\{b_W, Z_W, M_W, \{\mu_i\}\right\}$ for each fit (which uniquely define a Cepheid data set), and to examine the dependence of these parameters (particularly the period and metallicity coefficients $b_W$ and $Z_W$) on factors explored in R11 and E14 -- namely the rejection algorithm and threshold, distance anchor, and inclusion of longer-period Cepheids, discussed in Appendices~\ref{sec:rejection}-\ref{sec:longperiod}. Some of these fits, with associated Cepheid data sets, are selected to be carried forward to the global fit.
\\\\
We emphasize that the process of choosing these fits is motivated by the desire to capture and quantify variation that arises in results when different (but also valid) choices are made in the fitting process, rather than by the aim of choosing a `best' fit; this will become clear in Fig.~\ref{fig:cepheidall} and its discussion. Thus we do not blind this part of the analysis (the Cepheid-only fit), because the results are not final, and also because they do not directly reveal or affect the value of $H_0$. 

\subsection{Observations}
\label{sec:cepheidobs}
The Cepheids in the nine galaxies in Table~\ref{tab:nearby} were discovered or reobserved in the Supernovae and $H_0$ for the Equation of State (SH0ES) project \citep{Riess09a} on the HST, from Cycle 15. Infrared (F160W) observations of the \snia~host galaxies were made using the Wide Field Camera 3 (WFC3). We refer the reader to R11, section~2 for descriptions of observations and data reduction. Our initial data set consists of 570 Cepheids from R11, table~2, excluding those marked `low~P'; this number is reduced to 488 if we adopt the $P<60$~day cut on Cepheids, following E14. 
\\
\\
We supplement the sample of 157 Cepheids in NGC~4258 with LMC and MW Cepheids, used as alternative anchors (discussed in Appendix~\ref{sec:anchors}). \cite{Persson04} presents near-infrared photometry of 92 Cepheids, of which 53 have optical measurements in \cite{Sebo02}, which we use for determining Wesenheit magnitudes. Two of these 53 Cepheids have period greater than 60 days, which we exclude if we impose the period cut on the Cepheids in the supernova host galaxies. We also make use of 13 Cepheids in the Milky Way from \citet[table~2]{vanLeeuwen07} (excluding Polaris, an overtone pulsator), which have combined parallaxes from Hipparcos and HST data.

\subsection{Cepheid systematics}
\label{sec:cepheidsystematics}
Cepheid variables are powerful distance indicators to nearby galaxies, however they are subject to systematics. We briefly mention those that affect our method, and refer the reader to \citet[section~3.1]{FreedmanMadore01} and references therein for further discussion of Cepheid systematics. In Appendices~\ref{sec:rejection} to ~\ref{sec:longperiod} we test and report the dependence of the Leavitt law parameters on aspects of the Cepheid fit, namely outlier rejection, distance anchor, and cut on Cepheid period.
\\\\
Careful treatment of Cepheids starts with their discovery and identification, where crowding and confusion can lead to misidentification. Light from a Cepheid can be blended with nearby or background sources, and aliasing or sampling problems can cause the wrong period to be inferred. Thus, outliers from the Leavitt law fit must be identified and rejected. Moreover the intrinsic scatter in the Leavitt law must be taken into account in assessing the goodness-of-fit; outliers that are rejected should lie well outside the so-called instability strip.
\\
\\
The secondary dependence of Cepheid luminosities on atmospheric metallicity is an ongoing area of research, and remains contentious. This effect arises from changes in the atmospheres and structure of Cepheids with their chemical composition, which impacts colours and magnitudes. There is evidence of a mild metallicity dependence at optical wavelengths \citep{Kennicutt98, Sakai04, Macri06, Scowcroft09}, which is weaker in the infrared. In the LMC, using spectroscopic [Fe/H] measurements, \cite{FreedmanMadore11} find that $Z_H$ (the metallicity dependence in the {\em{H}}-band) is close to zero. \citet[][section~3.2]{Efstathiou14} argues that these LMC observations, along with theoretical considerations, give cause to applying an external prior on the metallicity dependence centred at $Z_W{\sim}0$. We discuss this prior, which we find is inconsistent with the R11 data, in Appendix~\ref{sec:cepheidpriors}.
 \\\\ 
Historically the zero point of the Leavitt law has proven difficult to measure, due to uncertainties in parallax measurements. To circumvent this, more accurate absolute distances have been pursued, including VLBI measurements of water megamasers in NGC~4258~\citep{Humphreys13}. Multiple distance anchors are also tested and combined to reduce the impact of any single distance anchor. The effects of varying and combining anchors is explored in this analysis in Appendix~\ref{sec:anchors}, following R11 and E14.

\subsection{Cepheid-only fit}
\label{sec:cepheidfit}
Our fit to all Cepheid data is based on E14 with the difference that we do not assume the \snia~zero point (the quantity $a_V$ in R11 and E14) or indeed any SNe data. This is because we intend to fit the Cepheids separately from the SN data, whereas E14 calculates values of $H_0$ from the Cepheid fits, assuming \snia~data from R11. All Cepheid data are fit to the Leavitt law (Equation~\ref{eq:pl2}) with MultiNest (Section~\ref{sec:multinest}). The 12 parameters of fit include the three nuisance parameters $\{b_W, Z_W, M_W\}$, the strongly constrained distance $\mu_{4258}$, and the eight distance modulus offsets $\{\Delta\mu_i\}$. We set an external Gaussian prior on $\mu_{4258}$, and by default place uniform priors for all other parameters over generous intervals. The $\chi^2$ function for the Cepheid fit is a function of $\{b_W, Z_W, M_W, \mu_{4258}\}$, and $\{\Delta\mu_i\}$, and takes the form 
\begin{align}
\label{eq:chisqcepheids}
\chi^2_c= \sum_{ij}\frac{(\hat{m}_{Wij} -m_{Wij,\mathrm{mod}})^2}{\hat{m}_{Wij,\mathrm{err}}^2 + \sigmaintC^2}.
\end{align}
Here $m_{Wij,\mathrm{mod}}(b_W, Z_W, M_W, \mu_{4258},\{\Delta\mu_i\})$ is the model magnitude of the $j$-th Cepheid in galaxy $i$ (given by Equation \ref{eq:pl2}) and $\sigmaintC$ is the intrinsic scatter in Cepheid magnitude, from the width of the instability strip. For clarity, measured quantities are denoted with hats to distinguish them from model quantities. The logarithm of the likelihood $\mathcal{L} = e^{-\chi^2_c/2}$ and the priors on the fit parameters are inputs for MultiNest. We use 1000 live points in MultiNest and confirm that the precision is sufficient.\footnote{For selected fits we repeat the outlier rejection and fitting steps, and find that the scatter in final parameters within ten runs is $<1\%$ of the statistical uncertainty.}

\subsection{Results of Cepheid-only fit}
\label{sec:cepheidresults}

\begin{figure*}
\begin{center}
  \includegraphics[width=\textwidth]{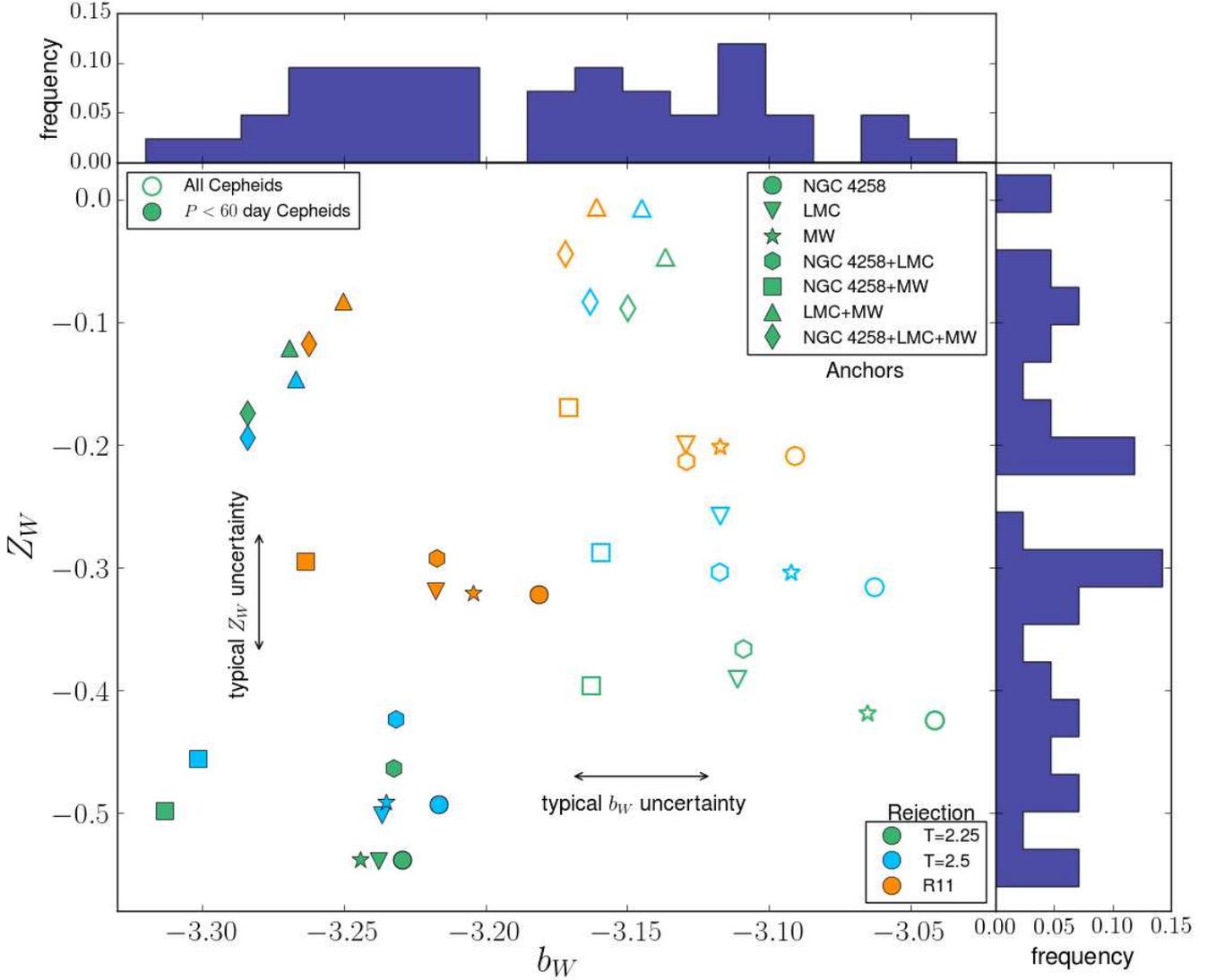}
  \caption{The best-fitting values for $b_W, Z_W$ from all Cepheid-only fits to the Leavitt law (Equation~\ref{eq:pl2}), assuming various distance anchors and rejection algorithms, with and without a cut on the period. The different markers represent these properties as indicated in the legends, with the colour representing the outlier rejection algorithm, shape representing the distance anchor, and solidness reflecting the period cut. We consider all seven combinations of distance anchor galaxies NGC~4258, LMC, and MW (Appendix~\ref{sec:anchors}), and all three rejection algorithms (Appendix~\ref{sec:rejection}). This figure shows: (i) including the longer-period Cepheids increases both $b_W$ and $Z_W$ (empty markers lie up and to the right of solid markers) (ii) Systematic variation in parameters with distance anchor (e.g.\ for each choice of period cut, the NGC~4258 + MW anchor gives the lowest $b_W$ and the NGC~4258-only anchor gives the highest); meanwhile fits with both the LMC and MW as anchors (diamonds and upward triangles, with and without NGC~4258 respectively) are clustered tightly, indicating that these two galaxies together provide a strong constraint on both parameters. (iii) The R11 rejection results in less negative $Z_W$ and to a lesser extent $b_W$ (reflected in orange markers concentrated in the upper-right portion of the figure), while the E14 algorithm with rejection threshold $T=2.5$ (turquoise) results in higher $Z_W$ compared to $T=2.25$ (green) for fits other than those with both the LMC and MW anchors. The typical uncertainties, indicated by the arrows, are ${\sim}0.05$ for $b_W$ and ${\sim}0.1$ for $Z_W$ for most fits, but can be larger for some anchors or rejection algorithms. Evidently the scatter arising from varying the distance anchor, cut on period, and rejection far exceeds the statistical uncertainty. The histograms in the margins display distributions of $b_W$ and $Z_W$ values over all fits. The histogram for $b_W$ shows that values are clustered around $b_W{\sim}-3.25$ for fits with a $P<60$~day cut (reflective of the influence of the LMC Cepheids) and $b_W{\sim}-3.10$ for fits without. The histogram for $Z_W$ shows a spread centred at $Z_W{\sim}-0.3$, dependent on distance anchor; fits with both the LMC and MW anchors lie with $-0.2 < Z_W < 0$.}
\label{fig:cepheidall}
\end{center}
\end{figure*}
The results of all Leavitt law fits, for all combinations of distance anchor, outlier rejection, and upper period limit, are presented in Table~\ref{tab:cepheidall}. The details of these choices are given in Appendices~\ref{sec:rejection}--\ref{sec:longperiod}, along with the effect they have on fit results. The variation in the fits is visualised in Fig.~\ref{fig:cepheidall} in $b_W,Z_W$-space. The choice of these two parameters is obvious as they characterise the Leavitt law and are solely influenced by the Cepheid sample -- all other parameters in $\Theta$ are influenced by the SN data, even the zero point $M_W$. Fig.~\ref{fig:cepheidall} allows us to identify which of the Cepheid fits lie at the edges of the parameter space. The resultant scatter observed in Fig.~\ref{fig:cepheidall} far exceeds the statistical uncertainties reported in Table~\ref{tab:cepheidall}. Therefore it is paramount that the systematic associated with varying the choices made in Appendices~\ref{sec:rejection} through \ref{sec:longperiod} is propagated carefully through the entire analysis process.
\\
\\
The choice of whether or not to apply the upper period limit of $P<60$~days has the most effect on the parameters, especially $b_W$. Fig.~\ref{fig:cepheidall} reveals clearly the impact of including the longer-period Cepheids on parameters $b_W$ and $Z_W$, most notably splitting Fig.~\ref{fig:cepheidall} down the middle vertically, i.e.\ by Leavitt law slope. Both parameters are smaller in magnitude by ${\sim}0.1$ when the longer-period Cepheids are included, indicating a weaker dependence of Cepheid magnitude on both period and metallicity. For the slope $b_W$, this difference dominates the statistical uncertainty and any other variation in $b_W$, whereas for $Z_W$ the resultant change from changing the period cut is comparable in size to the dependence on rejection algorithm, and the statistical uncertainty.
\\
\\
When the longer-period Cepheids are included, each of $b_W$ and $Z_W$ is better-constrained by the distance anchor, and the rejection algorithm, respectively: this is reflected in the vertical lines of empty markers with the same shape, and near-horizontal lines of markers with the same colour. That is, when the $P<60$~day cut is applied, the fit results are more sensitive to the choice of rejection algorithm and distance anchor. However, even without the cut, there remain strong dependences of $Z_W$ on rejection, and of $b_W$ on distance anchor.
\\\\
Within each choice of period cut, the slope $b_W$ varies systematically with distance anchor: the NGC~4258 + MW and NGC~4258 anchors result in the lowest and highest $b_W$ respectively, with results form the other anchor combinations lying in between. The fits with both the LMC and MW in the anchors (upward triangles and diamonds) have the least spread in both parameters. With the exception of these fits, the data suggest a reasonably strong metallicity dependence with $-0.5< Z_W < -0.2$. As noted above, the results are sensitive to rejection algorithm, with the R11 rejection resulting in less negative values for $Z_W$ (and for $b_W$ with the $P<60$~day cut), followed by the E14 rejection with $T=2.5$ to a smaller extent. 
\\
\\
We observe (Table~\ref{tab:cepheidall}) that there is little difference in values for $M_W$ between fits with and without the $P<60$~day cut, with the difference decreasing to zero for fits anchored on both n4258 and the Milky Way. However we defer further comment on $M_W$ (as well as $\{\Delta\mu_i\}$) to the discussion of global fit results. As $M_W$ is a magnitude zero point, and the $\Delta\mu_i$ are affected by the nearby SNe, the values of these parameters have potential to be influenced by the \snia~data, and are expected to change with their inclusion. 
\subsubsection{Comparison to R11 and E14}
\label{sec:cepheidcomparison}
We compare our fits in Table~\ref{tab:cepheidall} to equivalent results in R11 and E14: our fits with R11 rejection and no period cut are compared to bolded fits in R11, table~2, and we compare our fits with the $P<60$~day cut to the results in E14, tables~2--4 without priors on $b_W$ and $Z_W$. Relative to R11, our $b_W$ values with LMC-only or MW-only anchors are slightly lower in magnitude (${\sim}-3.12$ instead of $-3.19$ in R11, a ${\sim}1\sigma$ difference). Moreover our fits with LMC + MW anchors result in a lesser metallicity dependence ($-0.2<Z_W<-0.1$ instead of $Z_W {\sim}-0.3$ in E14) -- however uncertainties in $Z_W$ in these E14 fits exceed 0.1, and our $Z_W$ values (without the period cut) are supported by R11. Aside from these differences our results are in good agreement with R11 and E14, lying well within ranges allowed by statistical uncertainties. We retain 444 Cepheids when adopting the rejection flagged in R11, table~2 (close to the minimum of 448 reported in table~4 of R11) and only 379 with the $P<60$~day restriction. Applying the E14 rejection algorithm, our fits consistently result in lower numbers of remaining Cepheids by $10-20$, and consequently slightly lower $\sigmaintC$. It is worth noting that our methodology differs from E14 (and R11) in that we do not involve any SNe in the fit (omitting the third term in equation~14 of E14), whilst E14 includes the SN fit results by assuming a value of $a_V$ taken from R11. Presumably the complex ways of probing the multi-dimensional parameter space are leading to differences, albeit slight, between this work, R11, and E14, that cannot be easily reconciled.  We believe a solution for the future is for authors to provide code and data sets used for calculations as part of publication, that can be used to better understand differences.

\subsubsection{Selection for global fits}
\label{sec:selection}
\begin{table}
\begin{center}
\caption{Summary of selected Cepheid fits to carry forward to global fit (i.e. rejection, anchors, and period cut used). The positions of the best fit values for $b_W$ and $Z_W$ in the $b_W, Z_W$-plane (represented in Figure~\ref{fig:cepheidall}) are also given, as well as the symbols for these fits in Figure~\ref{fig:cepheidall}. The top half of the table (solid symbols) lists fits with the $P<60$~day cut, whilst the bottom half (empty symbols) contains fits without.
  \label{tab:cepheidsymbols}}
\footnotesize
\begin{tabular}{llll}
\hline
\hline
& Rej ($T$) & Anchor$^a$&  Symbol\\
\hline
\footnotesize{top left} & 2.25 & n4258+LMC+MW  & solid green diamond\\
\footnotesize{top left} & 2.5 & n4258+LMC+MW  & solid turquoise diamond\\
\footnotesize{top left} & R11 & n4258+LMC+MW  & solid orange diamond\\
\footnotesize{top left} & R11 & LMC+MW  & solid orange $\Delta$ \\
\footnotesize{middle} & 2.5 & n4258+LMC & solid turquoise hexagon\\
\footnotesize{middle} & R11 & MW & solid orange star\\
\footnotesize{lower left}& 2.25 & n4258+MW & solid green square\\
\footnotesize{lower} & 2.25 & LMC & solid green $\nabla$\\
\footnotesize{lower} & 2.25 & n4258 & solid green circle \\
\footnotesize{top} & 2.25 & n4258+LMC+MW  & empty green diamond\\
\footnotesize{top} & 2.5 & n4258+LMC+MW & empty turquoise diamond \\
\footnotesize{top} & R11 & n4258+LMC+MW  & empty orange diamond\\
\footnotesize{top} & R11 & LMC+MW & empty orange $\Delta$\\
\footnotesize{top}& 2.25 & LMC+MW &  empty green $\Delta$\\
\footnotesize{middle} & 2.25 & n4258+MW & empty green square\\
\footnotesize{right}& R11 & n4258 &  empty orange circle \\
\footnotesize{right} & 2.5 & n4258 & empty turquoise circle\\
\footnotesize{lower right} & 2.25 & n4258 & empty green circle \\
\hline
\end{tabular}
\end{center}
$^a$ For typographic ease we abbreviate `NGC' to `n'.\\
\end{table}

The choice of Cepheid fits to carry forward to the global fit is informed by their results, i.e.\ Leavitt law slope $b_W$ and metallicity dependence $Z_W$, as these parameters are only influenced by the Cepheid sample and are very minimally affected by the SN data. We are interested in the effect the choice of Cepheid sample (through varying aspects of the fit such as distance anchor, rejection, and upper period limit) has on these parameters in the global fit. In particular it is essential to quantify the systematic uncertainty in $\mathcal{H}$ with varying these choices.
\\
\\
We select 18 fits in total, summarised in Table~\ref{tab:cepheidsymbols}. To span the full range of uncertainty induced by various Cepheid fits, we select fits at extremes of the parameter space (Fig.~\ref{fig:cepheidall}), with a selection of anchors and rejection algorithms. The combination of all three distance anchors has the most constraining power, so we include all of these fits to quantify the uncertainty within them.
\\
\\ 
Each fit has an associated set of best-fitting parameters with uncertainties, as well as (unless using the R11 rejection) the values of the intrinsic scatter and rejection threshold, which together uniquely define a set of Cepheids remaining after outlier rejection. These then make up the Cepheid data and priors for some parameters in $\Theta$, going into the global fit (Section~\ref{sec:results}).


\section{Supernova fit}
\label{sec:SNe}

We now focus on the Type Ia supernovae. First we outline the data set and discuss systematics, then we describe various cuts on the \sneia~and present the preliminary SN-only fit. This section is supplemented by details provided in Appendices~\ref{sec:SALT2}, \ref{sec:malmquist}, \ref{sec:hostcorrection}, \ref{sec:velcorrection} on the lightcurve fitting method, and the corrections applied for Malmquist bias, host galaxy mass and peculiar velocities respectively. Also central to the subject are the computations of SN systematics in covariance matrices, which are also relegated to Appendix~\ref{sec:systematics} for detailed discussion.

\subsection{Observations}
\label{sec:SNobs}
Our supernova data are identical to R11, consisting of eight `nearby' \sneia~in the galaxies hosting Cepheids (Table~\ref{tab:nearby}), and 280 unique `low-$z$' \sneia~from the 185 CfA3~\citep{Hicken09a} and 165 LOSS~\citep{Ganeshalingam10} samples.\footnote{There are 69 SNe in common between the samples; however SN 1998es was discarded because the lightcurve quality was so poor that the SALT2 lightcurve fit failed.} Details of sources of photometry for the nearby supernovae are presented in Table~\ref{tab:nearbySN}. Natural photometry was not available for the oldest two, SN 1981B and SN 1990N. The most recent SNe are already in both CfA3 and LOSS, so we used combined photometry from both sources as described in Appendix~\ref{sec:consistency}. The remaining three (SN 1994e, SN 1995al, SN 1998aq) were observed on the FLWO 1.2~m telescope with a variety of CCDs; we construct SALT2 instruments (including transmissions and zero points) using data from \cite{Jha06}.
\\\\
CfA3 ran from 2001 to 2008 on the 1.2m telescope at FLWO almost entirely with the CfA3 4Shooter2 and Keplercam imagers (in {\em{UBVri}} and {\em{UBVRI}} filters respectively), while LOSS took place on the NICKEL and KAIT telescopes from 1998 to 2008 (in {\em{BVRI}}). Unlike more recent magnitude-limited surveys, CfA3 and LOSS targeted known galaxies and include SNe discovered by other sources, resulting in a more complex selection function and generally resulting in higher host galaxy masses (Appendix~\ref{sec:hostcorrection}). We refer the reader to the above works for further details of observations. Newer low-$z$ \sneia~samples have since been published, notably CfA4~\citep{Hicken12}, Carnegie Supernova Project \citep[CSP-II;][]{CSP}, Pan-STARRS~\citep{Rest14}, Palomar Transient Factory \citep[PTF;][]{PTF}, and La~Silla-QUEST Supernova Survey \citep[LSQ;][]{Walker15}. However we retain the older CfA3-LOSS sample for this analysis to more faithfully compare our results to R11 and E14. 
\\\\
Photometry for the low-$z$ sample is sourced from \cite{Hicken09b} and \cite{Ganeshalingam13} in the natural systems of each filter set, with the exception of the CfA3 4Shooter2 and Keplercam {\em{U}} filters for which reliable measurements do not exist -- we use photometry in the standard Johnson-Cousins {\em{UBVRI}} system as presented in \cite{Bessell90} for these passbands only, as well as the nearby SN 1981B and SN 1990N. We use SALT2~\citep{Guy07} to fit these \snia~lightcurves for the quantities $m_B, X_1, C$, which are used to derive distances via Equation~\ref{eq:snia}. Details of the lightcurve fitting are given in Appendix~\ref{sec:SALT2}.\\
\\
One reason for our choice of SALT2 as a lightcurve fitter is that our framework for assessing \snia~systematic uncertainties with covariance matrices (Section~\ref{sec:SNsyst} and Appendix~\ref{sec:systematics}) follows that in the SNLS-SDSS Joint Lightcurve Analysis \citep[hereafter JLA;][]{Betoule14}, which relies on the SALT2 model. In addition, SALT2 is the most modern fitter and used ubiquitously in cosmology analyses; thus our use allows for easier comparison and greater consistency. While R11 test the effects of fitting lightcurves with both SALT2 and MLCS2k2~\citep{Jha07}\footnote{SALT2 differs from MLCS2k2 substantially in its treatment of extinction: instead of prescribing a reddening parameters $R_V$, all of the colour information (including the SN's intrinsic colour and host extinction) is included in the single colour parameter $C$.} lightcurve fitters, we use SALT2 only. This is justified, as the latest version SALT2.4 \citep[described in][]{Betoule14} was released in parallel with simulations in \cite{Mosher14} which assess and quantify the uncertainty associated with the choice of lightcurve fitter (and the lightcurve model itself) in covariance matrices (Appendix~\ref{sec:covmat}). Hence it is unnecessary to use of multiple fitters to assess the the aforementioned systematic uncertainty.

\begin{table}
\begin{center}
\caption{Observations of nearby \sneia~in Table~\ref{tab:nearby}, including sources of photometry, SALT2 instruments, magnitude systems (including filters) where available. Lightcurves of the two earliest supernovae were given as standard photometry only.
  \label{tab:nearbySN}}
\tabletypesize{\footnotesize}
\begin{tabular}{llll}
\hline
\hline
\footnotesize
\snia&Photometry source& \footnotesize{Magnitude system and filters}\\
\hline
SN 1981B&\scriptsize{\cite{ButaTurner83}} & Standard \emph{UBVR}\\
SN 1990N&\scriptsize{\cite{Lira98}} & Standard \emph{UBVRI} \\
SN 1994ae &\scriptsize{\cite{Riess05}} & AndyCam$^a$ \emph{BVRI}\\
SN 1998aq &\scriptsize{\cite{Riess05}} & 4Shooter/AndyCam \emph{UBVRI}\\
SN 1995al&\scriptsize{\cite{Riess09b}} & AndyCam \emph{UBVRI}\\
SN 2002fk &CfA3$^b$& 4Shooter2 \emph{UBVRI}\\
 & LOSS$^c$& KAIT3/NICKEL \emph{BVRI}\\
SN 2007af&CfA3 & Keplercam \emph{BVri} \\
& LOSS&KAIT3/KAIT4 \emph{BVRI}\\
SN 2007sr&CfA3 & Keplercam \emph{BVri}\\
&  LOSS&KAIT3/4 \emph{BVRI}\\
\hline
\end{tabular}
\end{center}
$^a$ A thin, back-illuminated CCD camera on the FLWO 1.2~m telescope \citep{Jha06}.\\
$^b$ \cite{Hicken09a}.\\
$^c$ \cite{ Ganeshalingam10}. Both CfA3 and LOSS photometry were available for the most recent three \sneia~, so we used combined photometry from both sources as described in Appendix~\ref{sec:consistency}.
\end{table}

\subsection{Supernova systematics}
\label{sec:SNsyst}
As a statistical sample, type Ia supernovae are high fidelity standard candles. However as astronomical objects, \sneia~are diverse and subject to systematics, with their measurable quantities (absolute brightness, observed colour and decline rate) dependent on factors which correlate with their progenitors and environments. Countless investigations into these correlations and their origins are partly motivated by the need to reduce residual scatter from these intrinsic \snia~variations. Observations of supernovae are also influenced by factors such as galactic extinction, misclassifications, and differing telescope magnitude systems. Most of these effects are not sufficiently well-understood or accurately modelled to correct for them entirely. It is therefore essential to quantify the size of systematics; even when efforts have been made to apply corrections we still wish to estimate the uncertainty in the correction.\\
\\
Our approach to accounting for \snia~uncertainties follows methods in JLA, which are largely based on those in the Supernova Legacy Survey \citep[hereafter SNLS;][]{Conley11}. These use individual covariance matrices for each systematic, tracking correlated uncertainties between different SN quantities (i.e.\ $m_B, X_1, C$), between different supernovae. Advantages of the covariance matrix method over the more traditional method of adding systematics in quadrature are discussed in \citet[][section~4]{Conley11}; these include the ability to fully capture correlations in uncertainties, and the ease of including or reproducing uncertainties in further analyses. Details are our computations are provided in Appendix~\ref{sec:systematics}. 
\subsection{Cuts on supernova sample} 
\label{sec:SNcuts}
We make quality cuts on our \snia~sample to eliminate potential biases from poorly constrained lightcurves and peculiar events, and to remain within the bounds of the SALT2 model. With the intent of replicating the sample in R11 as closely as possible, we broadly follow the cuts described in CfA3~\citep{Hicken09b} and LOSS~\citep{Ganeshalingam13}, also using cuts in SNLS and JLA -- described in~\citet[section~4.5]{Guy10}, \citet[section~2.1]{Conley11}, \citet[section~4.5]{Betoule14} -- as guidance or as alternate cuts. In summary our criteria are as follows:
\begin{itemize}
\item
low Milky Way extinction $E(B-V) < 0.2$
\item
exclude local \sneia~not in the Hubble flow $z > 0.01$ 
\item
goodness-of-fit from SALT2 $\chi^2/\rm{DoF} < 8$ 
\item
first detection by +5 days, relative to {\em{B}}-band maximum
\item
exclude stretch outliers $|X_1| < 3$
\item
exclude colour outliers  $|C| < 0.5$
\item
well-constrained stretch $\sigma_{X_1}< 0.8$
\item
well-constrained colour $ \sigma_C < 0.1$.
\end{itemize}
The above encompass cuts in CfA3 and LOSS, with stricter cuts on the date of first detection and lightcurve goodness-of-fit (originally at +10 days and $\chi^2/\rm{DoF}=15$ in CfA3 respectively), and with additional cuts on the uncertainties in $X_1$ and $C$ to further exclude supernovae which have large uncertainties in their stretch or colour. Our cuts are also informed by visual inspection of individual lightcurves and their SALT2 fits, particularly in placing boundaries for the lightcurve goodness-of-fit, uncertainties in stretch and colour, and date of first detection. In summary, we exclude supernovae at very low redshift (i.e.\ not yet in the Hubble flow), significantly extinguished by Milky Way dust, detected too late, with poorly constrained stretch and colour. We also exclude \sneia~with poor SALT2 fits, and SNe that are too blue or red or have very fast or slow decline to exclude peculiar objects and ensure our sample fit within the SALT2 model.\\
\\
Furthermore we test some alternate cuts, including some suggested in JLA and original CfA3/LOSS cuts which we have changed above. We repeat the SN-only fit with these cuts to test the effect on the SN fit parameters, carrying some through to the global fit. In particular, we follow R11 in raising the low-redshift cut to $z = 0.0233$,\footnote{This is to reduce possible bias from local coherent flows, or a possible local underdensity (a so-called `Hubble bubble'). The latter is discussed in R11 and \cite{Conley11}; however there is no conclusive evidence for its existence.} and test strengthening or relaxing the lightcurve goodness-of-fit threshold to $\chi^2/\rm{DoF} <5$ or $\chi^2/\rm{DoF} <15$, and relaxing the date of first detection to +10 days. Following JLA we examine the effects of imposing a stricter bound on the colour ($|C| < 0.3$), the uncertainty on the stretch ($\sigma_{X_1}< 0.5$), and Milky Way extinction. These tests are important as the influence of these alternate cuts on the fit results is not straightforward or obvious; moreover no particular cut is necessarily more valid than the others. We discuss these results and their significance in Section~\ref{sec:SNresults}. Histograms showing $X_1$ and $C$ distributions for several cuts are included in Appendix~\ref{sec:histcut}. 

\subsection{SN-only fit} 
\label{sec:snfit}
Analogous to the Cepheid-only fit in Section~\ref{sec:cepheidfit}, we perform a preliminary fit of only the low-$z$ \sneia~to Equation~\ref{eq:sn2} using the MCMC routine \texttt{emcee} (Section~\ref{sec:multinest}), to identify the dependence of the SN parameters on the different cuts in Section~\ref{sec:SNcuts}. To clearly separate the data and model in Equation~\ref{eq:sn2} we define the quantity $\mBcor$ for the apparent SN magnitude corrected for stretch and colour:
\begin{align}
  \mBcor &:= m_B +\alpha X_1 - \beta C,\notag\\
  \label{eq:sn3}
\text{with \quad } {\hat{m}^\dagger}_{B \rm{mod}} &= 5\log_{10}(czf(z)) + M_B^* - \mathcal{H}.
\end{align}
Explicitly the $\chi^2$ function for the low-$z$ SN fit is
\begin{align}
  \label{eq:chisqlowz}
\chi^2_{\rm{SN}} &=(\boldsymbol{\hat{m}_B^\dagger} - \boldsymbol{\mBcor}_{\rm{mod}})\boldsymbol{\cdot} \mathbf{C_{\mBcor}}^{-1}\boldsymbol{\cdot}(\boldsymbol{\hat{m}_B^\dagger} - \boldsymbol{\mBcor}_{\rm{mod}})^T
\end{align}
where the covariance matrix $\mathbf{C_{\mBcor}}$ is derived from covariances in all SN parameters $\{m_B, X_1, C\}$, as given in Equation~\ref{eq:C_mBcor} in Appendix~\ref{sec:systematics} along with detailed explanations of statistical and systematic contributions.\\
\\
It is evident from Equations~\ref{eq:sn2} and \ref{eq:sn3} that the SN-only fit is degenerate: we cannot constrain both $M_B$ and $\mathcal{H}$ simultaneously; the nearby SNe are necessary to constrain $M_B$. Instead we fit for the difference $\mathcal{M} := M_B - \mathcal{H}$, adopting the blinds for each $M_B$ and $\mathcal{H}$ noted in Section~\ref{sec:blind} i.e.\ with the transformations $m_B \mapsto m_B + \mathfrak{o}_{M_B}$ in Equations~\ref{eq:nearby2} and \ref{eq:sn2} and $\mathcal{H} \mapsto \mathcal{H} + \mathfrak{o}_{\mathcal{H}}$ in the likelihood incorporating Equation~\ref{eq:chisqlowz} (a function of both $M_B$ and $\mathcal{H}$ through Equation~\ref{eq:sn3}). The marginalised posterior distributions (mean and 1$\sigma$ width) for $\alpha$ and $\beta$ are presented in Table~\ref{tab:SNinit} and plotted in Fig.~\ref{fig:alphabetaSN}; these results are dependent on the choice of quality cuts on the SN sample described in Section~\ref{sec:SNcuts}. 

\subsection{Results of SN-only fit}
\label{sec:SNresults}
\begin{table}
  \begin{center}
\caption{Results of preliminary SN-only fits for various cuts.
  \label{tab:SNinit}}
  \footnotesize
  \begin{tabular}{lllll}
\hline
\hline
SN cut & $N_{\rm{SN}}$ & $\alpha$ & $\beta$&$\mathcal{M}$\\
\hline
\scriptsize{default} & 171 & 0.164 (0.013) & 3.07 (0.14) & -3.240 (0.036)\\
\scriptsize{higher $\chi^2$} & 175 & 0.167 (0.013) & 3.12 (0.13) & -3.244 (0.036)\\
\scriptsize{lower $\chi^2$} & 163 & 0.158 (0.013) & 3.04 (0.16) & -3.256 (0.038)\\
\scriptsize{$z> 0.0233$} & 96 & 0.163 (0.016) & 2.73 (0.17) & -3.252 (0.038)\\
\scriptsize{stricter $C$} & 160 & 0.158 (0.015) & 2.93 (0.18) & -3.238 (0.037)\\
\scriptsize{str. $\sigma_{X_1}, \sigma_C$} & 164 & 0.171 (0.014) & 3.10 (0.14) &  -3.232 (0.038)\\
\scriptsize{str. $\sigma_{X_1}$} & 165 & 0.171 (0.013) & 3.10 (0.15) &  -3.245 (0.037)\\
\scriptsize{str. $E(B-V)$} & 166 & 0.167 (0.013) & 3.06 (0.15) & -3.241 (0.036)\\
\scriptsize{$t_{1st} <+10$d} & 187 & 0.165 (0.013) & 3.11 (0.14) & -3.234 (0.035)\\
\hline
\end{tabular}
\end{center}
\end{table}

\begin{figure}
\includegraphics[width=0.5\textwidth]{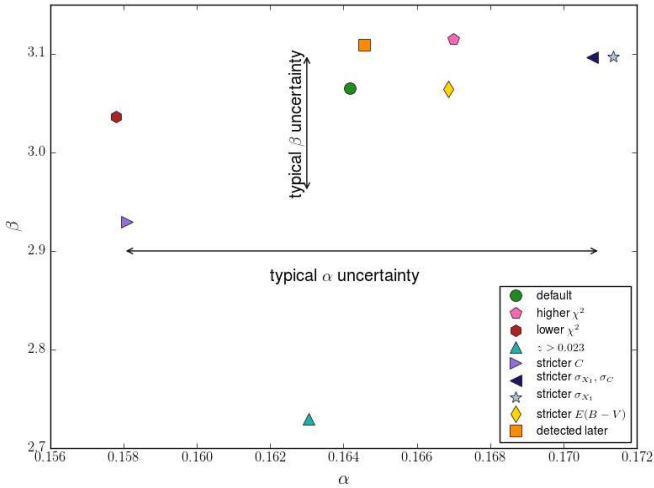}
\centering
\caption{The best-fitting values for $\alpha, \beta$ from all SN-only fits to Equation~\ref{eq:sn2},  assuming various cuts on the low-$z$ SNe. The different markers represent the cuts described in Section~\ref{sec:SNcuts}. The typical statistical uncertainties are indicated by the arrows. The variation in $\alpha$ is comparable to the statistical uncertainty, and the same is true for $\beta$ if we disregard the higher low-redshift cut.}
\label{fig:alphabetaSN}
\end{figure}
The results of the SN-only $\chi^2$-minimizing fit are presented in Table~\ref{tab:SNinit}, while Fig.~\ref{fig:alphabetaSN} shows the differences in fits with various SN cuts lie in the $\alpha,\beta$-plane (this is analogous to Fig.~\ref{fig:cepheidall}, which displays the numerous Cepheid fits in $b_W,Z_W$-space). We discuss the dependence of the fit results on the various cuts, and select cuts with results spanning the parameter space to carry forward to the global fit to assess the associated systematic uncertainty. 
\\
\\
The notable outlier is the higher low-redshift cut ($z>0.0233$), effecting a much lower value of $\beta$ than the other cuts. This cut, along with the stricter colour and stricter goodness-of-fit cuts, results in lower $\alpha$ also. The lowest and highest values of $\alpha$ correspond to lower~$\chi^2$, and stricter $\sigma_{X_1}$ respectively. Fig.~\ref{fig:zhigherhist} in Appendix~\ref{sec:histcut} shows normalised $X1$ and $C$ distributions for the $z>0.0233$ cut: there are marginally slower-declining SNe compared to the default, but overall the distributions appear similar. It does not appear that the discrepant fit results from this cut are the result of a change in the colour or stretch distribution of the sample; indeed our tests with jackknifed samples (described below) indicate this is likely the result of removing a large portion of the sample (over 40\% relative to the default). Disregarding the $z>0.0233$ cut, the variation in $\alpha$ and $\beta$ with the different cuts we test appears only slightly larger than the typical statistical uncertainties in these parameters (Fig.~\ref{fig:alphabetaSN}). 
\\\\
We use jackknife resampling to assess the statistical significance of the dependence of results on the cuts in Table~\ref{tab:SNinit}. For several cuts (the lower lightcurve $\chi^2$/\rm{DoF}, higher redshift cut and stricter cuts on colour or uncertainties in stretch and colour) we draw subsamples of size $N_{\rm{SN}}$ (Table~\ref{tab:SNinit}) of the 171 SNe selected by the default cut. For each cut we compare the systematic change in fit results (parameters $\alpha,\beta,\mathcal{M}$) from the new cut to results from repeated jackknifed subsamples of size $N_{\rm{SN}}$ and their scatter. These reveal a systematic variation of $1-3\sigma$ from the default for almost all combinations of parameters and cuts (where $\sigma$ is the scatter within the numerous jackknifed subsamples). Thus the differences between rows of Table~\ref{tab:SNinit} cannot be solely attributed to shot noise, and the variation due to different cuts must be propagated to the global fit (Section~\ref{sec:results}) and treated as a contribution to the total systematic uncertainty. However, we will find that the variation from the choice of SN cut is dwarfed by the analogous source of uncertainty from the choice of Cepheid fits.

\section{Global Fit Results}
\label{sec:results}
This section contains our final simultaneous fits to all Cepheid and \sneia~data. We set out parameters and equations for this fit, and present fit results for all parameters, including the dependence of results on choices within the individual Cepheid and supernova data sets. We summarise our uncertainties, and discuss their increase compared to other analyses of the same data. Finally, we break down the statistical and systematic contributions to the uncertainty budget.

\subsection{Global fit}
\label{sec:globalfit}

We fit all Cepheid and supernova data simultaneously to Equations~\ref{eq:pl2}, \ref{eq:nearby2}, \ref{eq:sn2} as described in Section~\ref{sec:globaleqns}. We minimize a global $\chi^2$ function (a function of $\Theta=\left\{\alpha, \beta,  M_B,\mathcal{H},b_W, Z_W, M_W, \mu_{4258},\{\Delta\mu_i\}\right\}$), which has contributions from the Cepheids and low-$z$ SNe remaining after cuts (given in Equations~\ref{eq:chisqcepheids} and \ref{eq:chisqlowz} respectively), and also an equivalent term to $\chi^2_{\rm{low-}z}$ for the eight nearby SNe:
\begin{align}
\label{eq:chisqglobal}
\chi^2_{\rm{global}} &= \chi^2_c + \chi^2_{\rm{SN}} + \chi^2_{\rm{nearby}}\\
\label{eq:chisqnearby}
\chi^2_{\rm{nearby}} &= (\boldsymbol{\hat{m}_B^\dagger} - \boldsymbol{\hat{m}_B^\dagger}_{\rm{mod}})\boldsymbol{\cdot} \mathbf{C_{\mBcor,n}}^{-1}\boldsymbol{\cdot}(\boldsymbol{\hat{m}_B^\dagger} -\boldsymbol{\hat{m}_B^\dagger}_{\rm{mod}})^T.
\end{align} 
The bolded quantities in Equation~\ref{eq:chisqnearby} are vectors, over the eight nearby \sneia. The terms contributing to the nearby covariance matrix $\mathbf{C_{\mBcor,n}}$ are covariances between SALT2 quantities $m_B, X_1$, and $C$, and the diagonal intrinsic scatter $\sigmaintSN$. 
\\
\begin{figure*}
   \centering
    \includegraphics[width=0.95\textwidth]{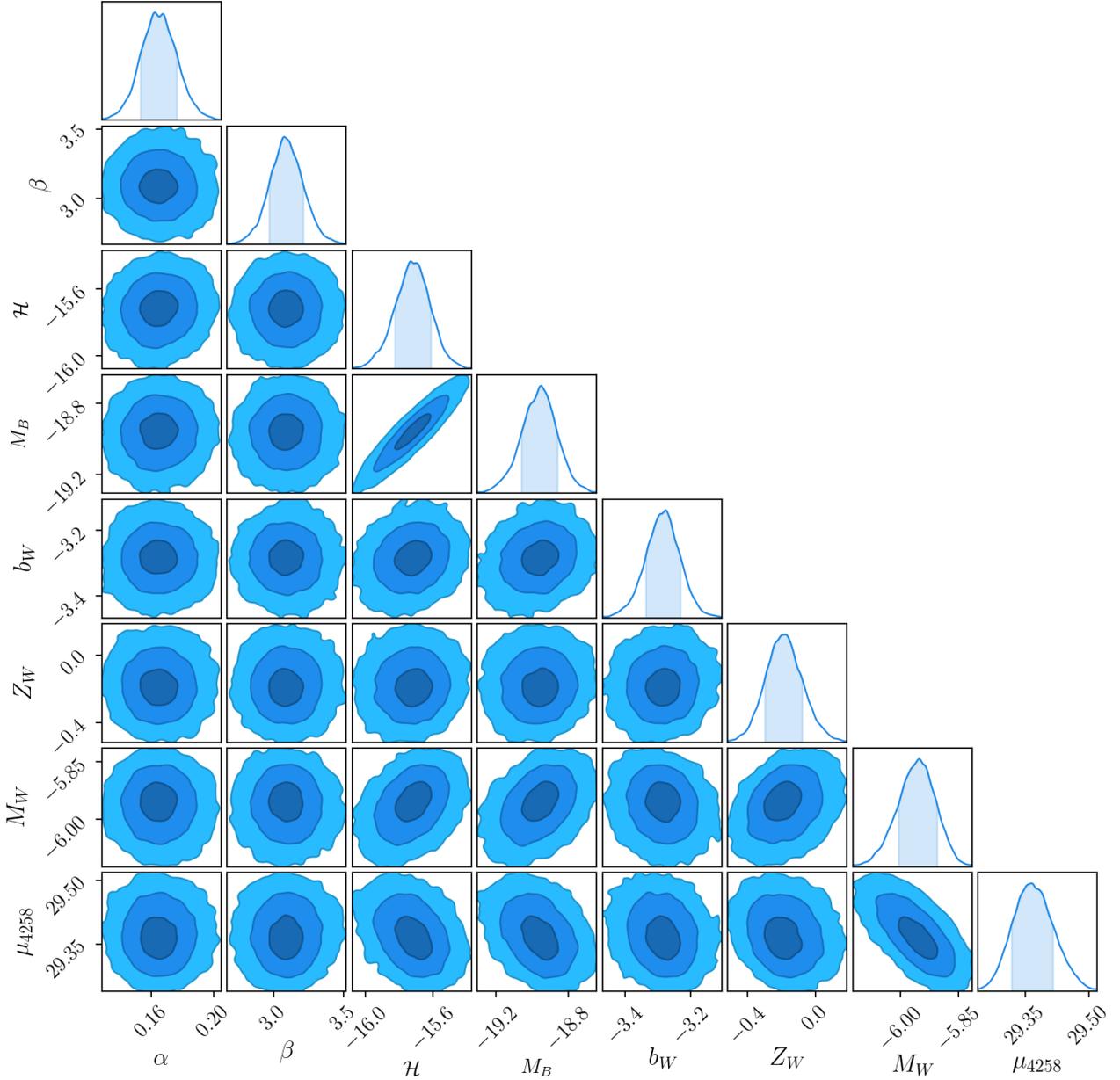}
    \caption{Constraints on parameters in $\Theta$ from an example global MultiNest fit (with $T=2.25$, NGC~4258+LMC+MW anchor, $P<60$~day cut Cepheid fit, default SN cuts) marginalised over $\{\Delta\mu_i\}$. The shaded regions in the PDFs represent 1$\sigma$ levels, and the 1$\sigma$, 2$\sigma$, and 3$\sigma$ regions are shown in the contours. Note the strong degeneracy between $\mathcal{H}$ and $M_B$, and slightly weaker degeneracies between $\mathcal{H}, M_B, \mu_{4258}$, and $M_W$. The other parameters appear uncorrelated.}
    \label{fig:8dim}
    \end{figure*}
\begin{figure}
  \centering
    \includegraphics[width=0.5\textwidth]{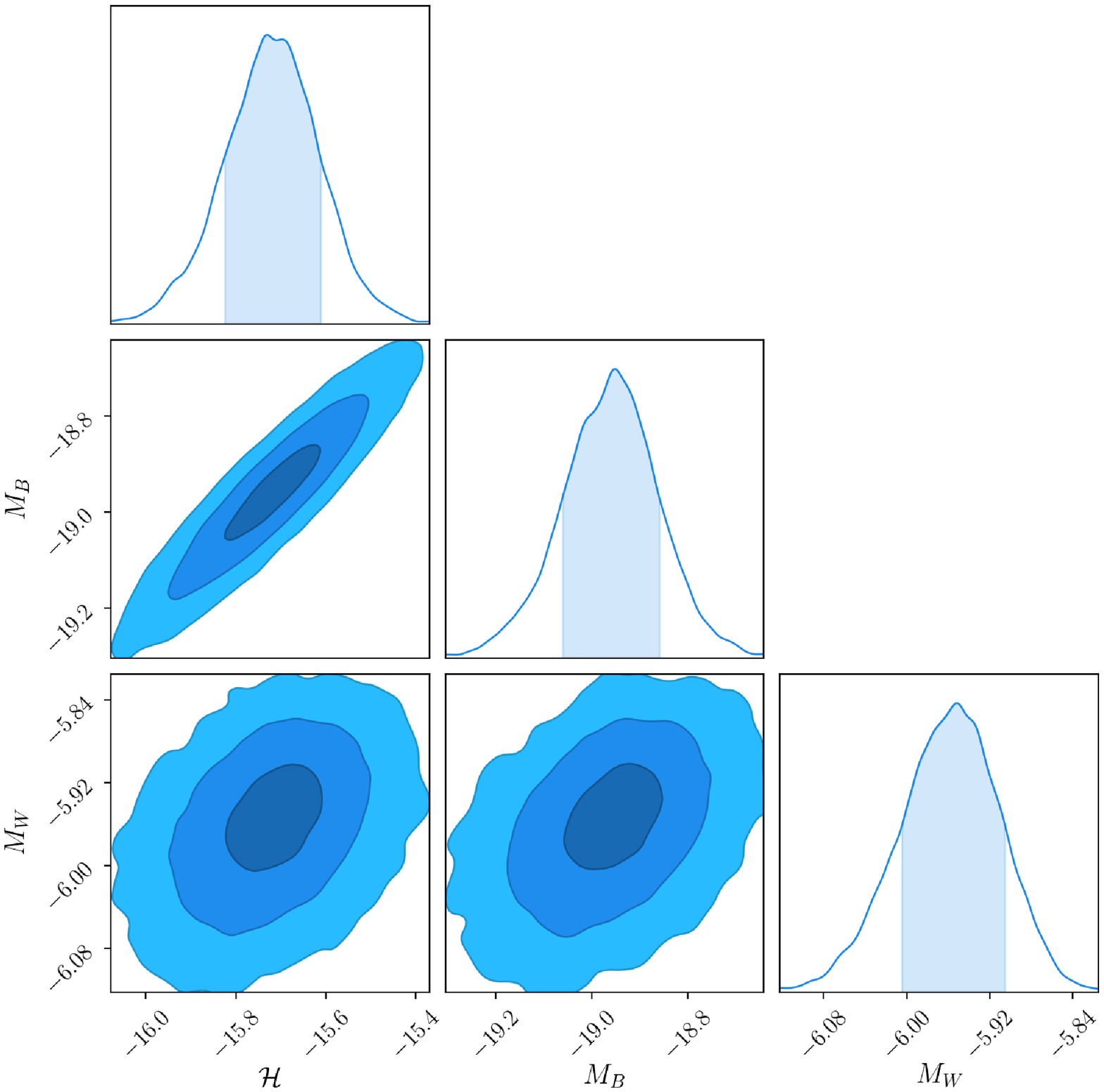}
    \caption{The same fit as Fig.~\ref{fig:8dim}, also marginalised over $\{\alpha,\beta,b_W,Z_W, \mu_{4258}\}$. This shows the three parameters that are the most highly correlated.}
    \label{fig:3dim} 
\end{figure}
\\
The global simultaneous fit is 16- or 17-dimensional (without and with the LMC included as a distance anchor respectively), and performed using MultiNest as described in Section~\ref{sec:multinest}. We are ultimately interested in $\mathcal{H}$, which contains the value of $H_0$. However to demonstrate degeneracies and correlations between parameters, we display in Figs~\ref{fig:8dim} and \ref{fig:3dim} marginalised contour plots of the posterior distribution of an example fit (with $T=2.25$, NGC~4258+LMC+MW anchor, $P<60$~day cut Cepheid fit and default SN cuts).\footnote{Figs~\ref{fig:8dim}, \ref{fig:3dim}, and 9 were created with the \texttt{ChainConsumer} package~\citep{Hinton16}.} The former posterior distribution is marginalised over the eight $\Delta \mu_i$, while the latter is also marginalised over $\mu_{4258}$ and the SN and Cepheid parameters which are strongly constrained by initial fits: $\{\alpha,\beta,b_W,Z_W\}$. Fig.~\ref{fig:8dim} shows a strong positive correlation between $M_B$ and $\mathcal{H}$ as expected from their degeneracy in the low-$z$ SN sample (Equation~\ref{eq:sn2}), and less apparent correlations between the `zero point-like' parameters $\{\mathcal{H}, M_B, M_W, \mu_{4258}\}$. In contrast, the five other parameters $\{\alpha,\beta,b_W,Z_W, \mu_{4258}\}$ each are largely independent of the other parameters (Fig.~\ref{fig:8dim}). 
\\\\
We repeat the global fit for each of 18 Cepheid fits in Table~\ref{tab:cepheidsymbols} and six supernova cuts determined in Section~\ref{sec:SNresults} from Fig.~\ref{fig:alphabetaSN}. Each Cepheid fit and SN cut corresponds to a subset of the total sample to use in the global fit, and associated values of best-fitting parameters, as well as $\sigmaintC$ for the Cepheids. In total there are 108 fits; the analysis of these results and the variation therein follows.

\begin{figure*}
\centering
\includegraphics[width=0.99\textwidth]{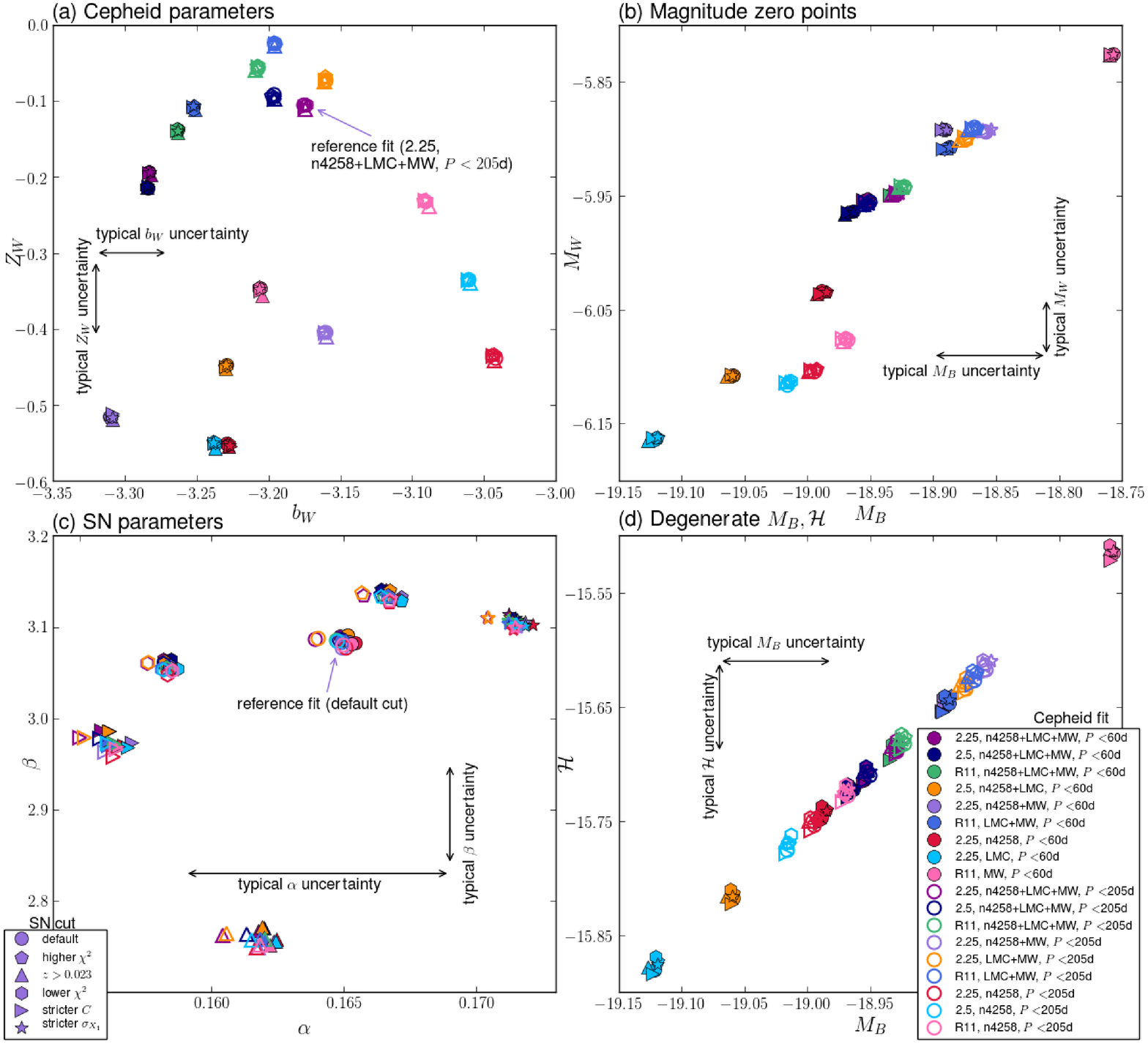}

\caption{Results of all global fits to Equations~\ref{eq:pl2}--\ref{eq:sn2} simultaneously, in the (a) $b_W,Z_W$- (b) $M_B,M_W$- (c) $\alpha,\beta$- and (d) $M_B,\mathcal{H}$-planes, when assuming various choices of SN cut and Cepheid fit. As shown in the legends the different combinations of colour and fill encapsulate information on the choice of Cepheid fit as described in Section~\ref{sec:selection}, while the different shapes represent difference cuts on the supernovae from Section~\ref{sec:SNresults}. The chosen reference fits (bolded in Table~\ref{tab:SNfinal} and \ref{tab:cepheidfinal}) are indicated by the violet arrows in (a) and (c). The overlap of points with the same colour and fill in (a) demonstrate that the Cepheid parameters $b_W, Z_W$ depend only on the Cepheid fit; similarly, the clusters of points with the same shape in (c) show that the \snia~parameters $\alpha,\beta$ depend mostly on the SN cut. Subplot (b) shows that $M_W$ and $M_B$ both depend predominantly on the choice of Cepheid fit, with the effect more strong in $M_W$. A strong degeneracy between $M_B$ and $\mathcal{H}$ is evident in (d), indicating that $\mathcal{H}$ depends primarily on the Cepheid fit, and secondarily on the SN cut. There is no systematic difference in $M_W, M_B$, and $\mathcal{H}$ between fits with and without an upper limit on Cepheid period.} 
  \label{fig:globalbig}
\end{figure*}

\subsection{Results of global fit}
\label{sec:globalresults}
The best-fitting values and uncertainties of parameters in $\Theta$ are given in Table~\ref{tab:globalfitfull} for each of 108 fits. Fig.~\ref{fig:globalbig} displays these fits in various subspaces of the 16- or 17-dimensional space spanned by $\Theta$, focussing on parameters $\{\alpha,\beta, b_W,Z_W, M_W, M_B, \mathcal{H}\}$. We discuss the dependences that this figure shows (which motivate the averaged tables and figures later), then present results for the nuisance parameters and the parameters of interest: the \snia~peak absolute magnitude $M_B$ and (proxy for the) Hubble constant $\mathcal{H}$, which are degenerate with each other.\\
\\
In the remainder of the section we depart from the distinction we make between statistical and systematic uncertainties in Appendix~\ref{sec:stat}: the uncertainties returned by MultiNest, reported in Table~\ref{tab:globalfitfull}, simply the 1$\sigma$ widths of the PDFs, do not distinguish between the statistical and systematic components of covariance matrices input into the fit in the likelihood. Henceforth, we refer to this uncertainty from the MultiNest fit as statistical, and the variation observed in e.g.\ Fig.~\ref{fig:globalbig} between global fits with differing supernova cuts or Cepheid fits as systematic.
\subsubsection{Dependence of parameters}
Fig.~\ref{fig:globalbig} highlights the following dependence of parameters on properties of the global fit: 
\begin{enumerate}
\item
  The Cepheid parameters $\{b_W,Z_W,M_W\}$ depend only on the choice of Cepheid fit (carried forward from Section~\ref{sec:selection}), reflecting the variation observed in Fig.~\ref{fig:cepheidall}. Thus there is negligible scatter in values for these parameters between fits with the same Cepheid data, regardless of the SN cut (Fig.~\ref{fig:globalbig}(a), (b)).
\item
 Similarly, the SN parameters $\{\alpha,\beta\}$ depend most strongly on cuts, and minimally on Cepheid fit, although there is more scatter than in $\{b_W, Z_W\}$. On average fits without an upper period cut on the Cepheids result in slightly lower $\alpha$ by ${\sim}0.01$, for each SN cut (Fig.~\ref{fig:globalbig}(c)). 
\item
  The Cepheid and SN zero points $M_W$ and $M_B$ both depend predominantly on the Cepheid fit (Fig.~\ref{fig:globalbig}(b)), reflecting the fact that the \sneia~are calibrated on the Cepheids. While $M_W$ depends directly on the Cepheid data (Equation~\ref{eq:pl2}), the influence on $M_B$ is through its interaction with $M_W$ via the distance modulus offsets $\{\Delta\mu_i\}$ (Equations~\ref{eq:pl2} and \ref{eq:nearby2}). We note that $M_W$ has negligible dependence on SN cut, whereas $M_B$ varies slightly with the choice of SN cut (with a spread of ${\sim}0.01$ within each choice of Cepheid fit). 
\item
  As mentioned in Section~\ref{sec:steps}, $\mathcal{H}$ is degenerate with $M_B$. Fig.~\ref{fig:globalbig}(d) shows this degeneracy between the parameters, and that the difference $\mathcal{M} = M_B - \mathcal{H}$ lies on a straight line. Within each choice of Cepheid fit there is slight systematic dependence only on the choice of SN cut. There is no systematic difference between these parameters from fits with and without a cut on Cepheid period.
\item
$\{\Delta\mu_i\}$: the values of the distance modulus offsets from the global fit depend significantly on the Cepheid fit, as shown in Figs~\ref{fig:globaloffsets} and \ref{fig:globaloffsetsrel}.
  
\end{enumerate}

In summary, it is expected that the SN cuts determine parameters $\{\alpha,\beta\}$, and the Cepheid fits determine parameters $\{b_W, Z_W, M_W\}$. However the interaction of the `zero point-like' parameters is more subtle, and emerges from the simultaneous fit of the three data samples, most obvious in Fig.~\ref{fig:globalbig}(b). Even though the parameter $M_B$ only appears in the SN apparent magnitudes (Equations~\ref{eq:nearby2}, \ref{eq:sn2}), it is most strongly influenced by the Cepheid data via $M_W$, as the two parameters are tied to their respective data sets through the distance modulus offsets $\{\Delta\mu_i\}$. Furthermore, $M_B$ and $\mathcal{H}$ are degenerate with their difference determined by the low-$z$ SNe. Thus the resultant value of $\mathcal{H}$, hence $H_0$, is sensitive both to the choice of SN cut (via the $M_B-\mathcal{H}$ degeneracy) and to the choice of Cepheid fit (via the influence of $M_W$ on $M_B$). Unsurprisingly, the most extreme values of $M_B$ and $\mathcal{H}$ (both driven by $M_W$, as seen in Fig.~\ref{fig:globalbig}(b)) arise from Cepheid fits anchored on only the LMC or MW (most and least negative, represented by dark purple and pink symbols, respectively). It is clear from Fig.~\ref{fig:globalbig}(d) that the variation with Cepheid fits (anchor and rejection) is at least an order of magnitude larger than the variation with SN cuts, even when the fits anchored on the LMC or MW only are excluded.

\subsubsection{Nuisance parameter results}
\label{sec:nuisance}

\begin{figure}
   \centering

    \includegraphics[width=0.48\textwidth]{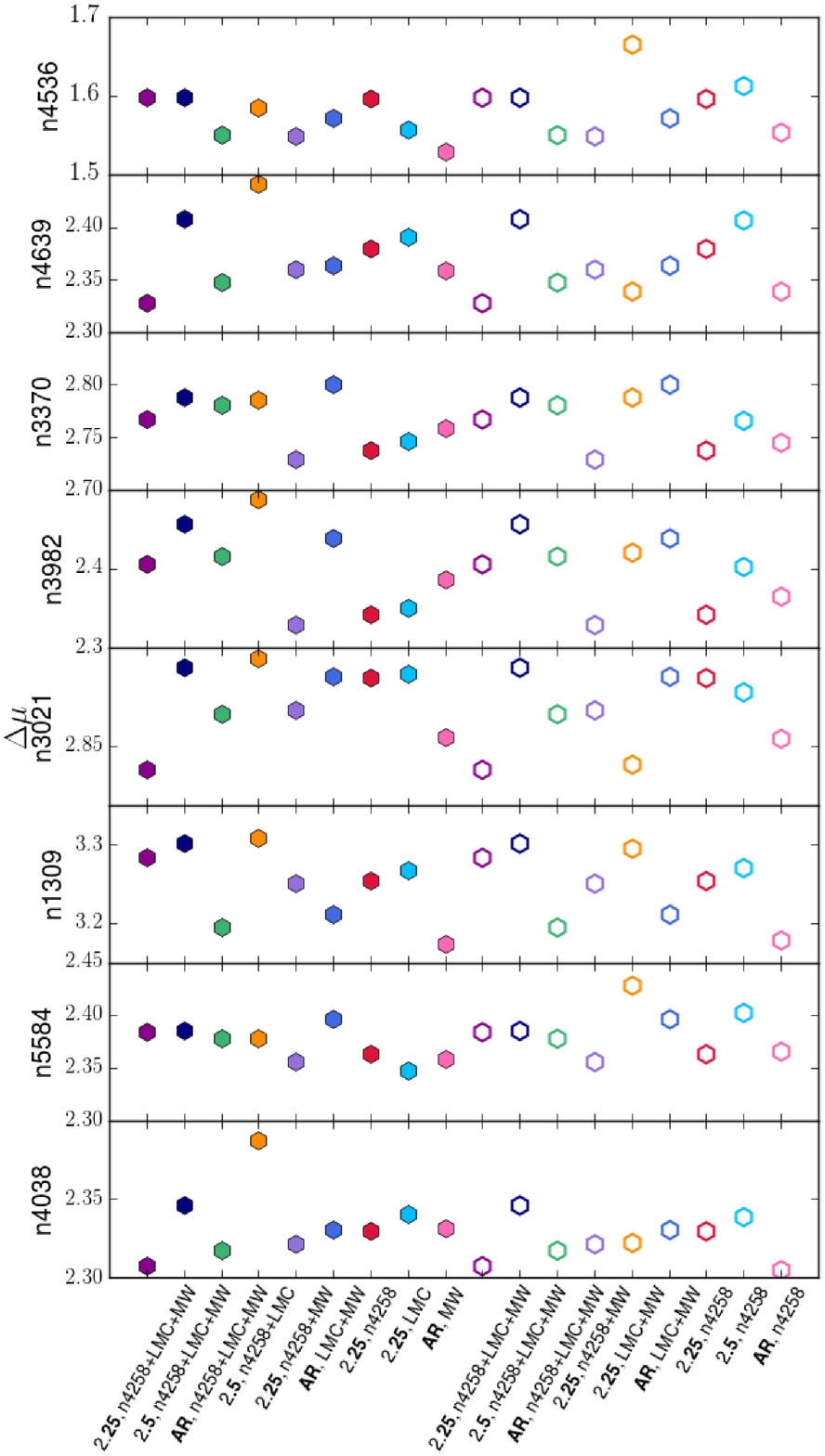}

    \caption{Visualisation of the best-fitting values for each $\Delta\mu_i$, which vary slightly with the different Cepheid fits in Section~\ref{sec:selection} (symbols shown in legends of Figs.~\ref{fig:globalbig} and~\ref{fig:scriptHhist}). Each horizontal subplot represents a different galaxy.} 
    \label{fig:globaloffsets}
    \includegraphics[width=0.46\textwidth]{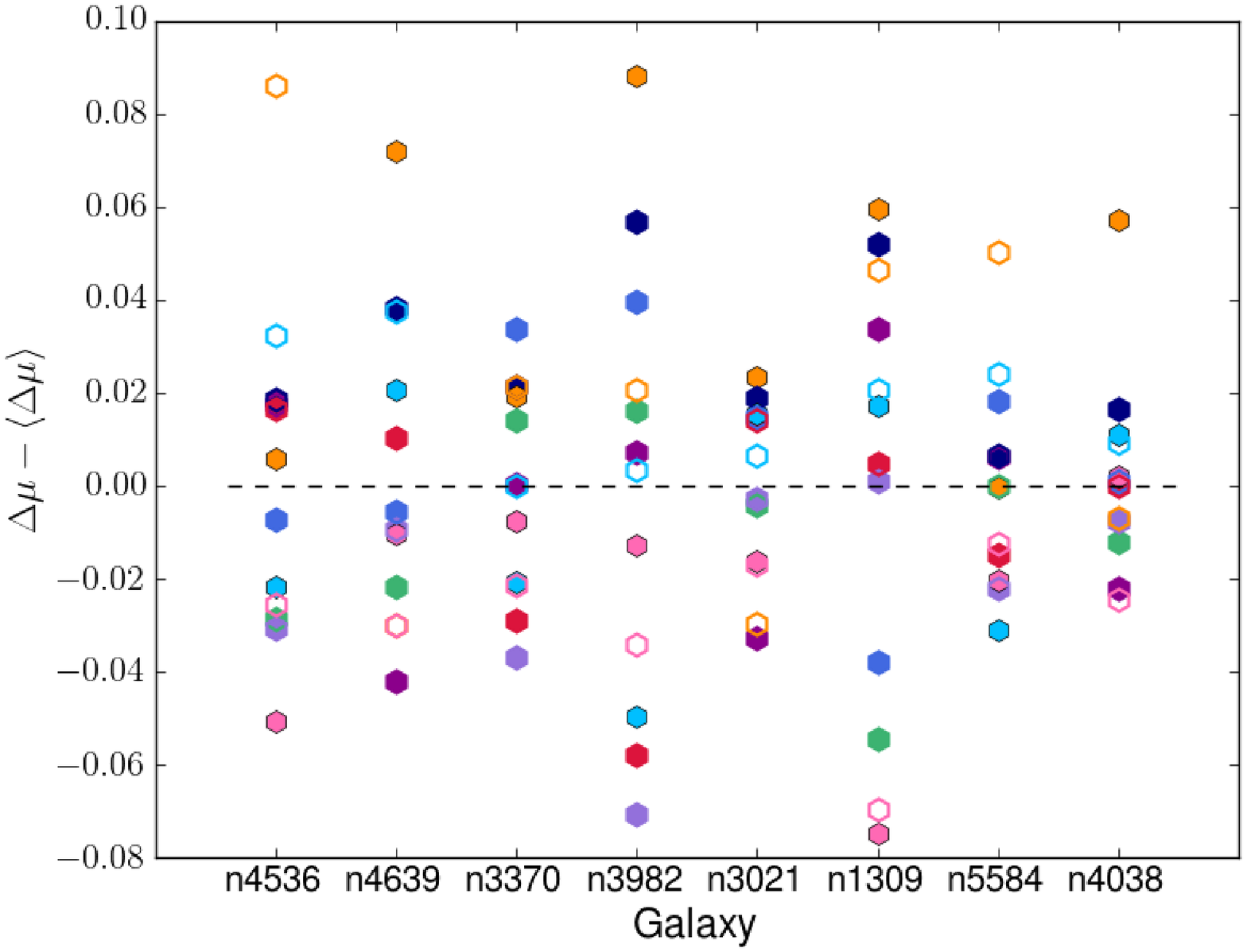}
    \caption{Visualisation of the relative value of the $\Delta\mu_i$ with respect to the mean over all Cepheid fits for each galaxy, marked by the black dashed line. The legends in Figs.~\ref{fig:globalbig} and~\ref{fig:scriptHhist} show the Cepheid fit (colour and fill).} 
    \label{fig:globaloffsetsrel} 

    \end{figure}
\begin{table}
  \begin{center}
  \caption{Global fit results for supernova parameters $\alpha,\beta$ for each SN cut, averaged over Cepheid fits. The default rejection (bolded) detailed in Section~\ref{sec:SNcuts} is chosen as our reference fit.
    \label{tab:SNfinal}}
  \begin{tabular}{lll}
\hline
\hline
SN cut & $\alpha$ & $\beta$\\
\hline
\textbf{default$\quad$} & \textbf{0.165} & \textbf{3.09} \\
higher $\chi^2$ & 0.167 & 3.134 \\
$z> 0.0233$ & 0.162 & 2.759 \\
lower $\chi^2$ & 0.158 & 3.057 \\
stricter $C$ & 0.156 & 2.974 \\
stricter $\sigma_{X_1}$ & 0.171 & 3.106 \\
\hline
\end{tabular}
\end{center}
\end{table}

\begin{table}
  \begin{center}
  \caption{Global fit results for Cepheid parameters $\{b_W, Z_W, M_W\}$ for each Cepheid fit, averaged over SN cuts. The bolded fit ($T=2.25$ rejection, all three anchors, and no upper cut on the period) is chosen as our reference fit.
\label{tab:cepheidfinal}}
  \begin{tabular}{llllll}
\hline
\hline
Rejection & Distance anchor & $P<60$d & $b_W$  & $Z_W$ & $M_W$ \\
\hline
2.25 &all$^a$ & Y & -3.28 & -0.19 & -5.95 \\
2.5 &all & Y & -3.28 & -0.21 & -5.96 \\
R11 &all & Y & -3.26 & -0.14 & -5.95 \\
2.5 & n4258+LMC & Y & -3.23 & -0.45 & -6.11 \\
2.25 & n4258+MW & Y & -3.31 & -0.52 & -5.89 \\
R11 & LMC+MW & Y & -3.25 & -0.11 & -5.91 \\
2.25 & n4258 & Y & -3.23 & -0.55 & -6.03 \\
2.25 & LMC & Y & -3.24 & -0.55 & -6.16 \\
R11 & MW & Y & -3.21 & -0.35 & -5.83 \\
\textbf{2.25} &  \textbf{all} & \textbf{N} & \textbf{-3.17} &  \textbf{-0.11} &  \textbf{-5.95} \\
2.5 &all & N & -3.20 & -0.10 & -5.96 \\
R11 &all & N & -3.21 & -0.06 & -5.94 \\
2.25 & n4258+MW & N & -3.16 & -0.41 & -5.89 \\
2.25 & LMC+MW & N & -3.16 & -0.07 & -5.90 \\
R11 & LMC+MW & N & -3.20 & -0.03 & -5.89 \\
2.25 & n4258 & N & -3.04 & -0.44 & -6.10 \\
2.5 & n4258 & N & -3.06 & -0.34 & -6.11 \\
R11 & n4258 & N & -3.09 & -0.23 & -6.08 \\
\hline
  \end{tabular}
\end{center}
$^a$ i.e. n4258+LMC+MW
\end{table}

Tables~\ref{tab:SNfinal} and \ref{tab:cepheidfinal} contain results for the supernova and Cepheid nuisance parameters, averaged over the Cepheid fits and SN cuts respectively. We choose to average over these aspects of the fit that have minimal effect on the parameters, as shown in Fig.~\ref{fig:globalbig}: the SN parameters in (c) predominantly depend on shape (SN cut) and not on colour (Cepheid fit), while the Cepheid parameters in (a) depend entirely on colour and not on shape. We omit statistical uncertainties of parameters in these tables as they can be obtained from the full set of results in Table~\ref{tab:globalfitfull}. For the nuisance parameters we select a single best fit (bolded in Table~\ref{tab:globalfitfull} and indicated in Fig.~\ref{fig:globalbig}). This is preferable to averaging over results in Tables~\ref{tab:SNfinal} and \ref{tab:cepheidfinal}, which are asymmetric, based on different premises (e.g.\ different distance anchors), and include more questionable fits (e.g.\ those SN cuts that reject a larger fraction of the total). Thus we use the maximal variation in these values to inform our systematic error budget, but not to influence the best fit.\\
\\
Final values for the nuisance parameters are taken from the bolded reference fits, which have the default SN cut and the Cepheid fit with all three anchors, $T=2.25$ rejection, and no cut on Cepheid period. We have chosen this fit because the results are representative and centred amongst the different choices. The Cepheid fit here also aligns with fits selected in E14 and R11. As in R11, we choose to not impose a cut on Cepheid period, and note the effects of including this cut on nuisance parameters described in Section~\ref{sec:cepheidresults}: both $b_W$ and $Z_W$ are more negative with the $P<60$~day cut, while there is no difference in $M_W$ when all three distance anchors are used. The global SN results differ slightly from the initial results in Table~\ref{tab:SNinit}, and we again note that the most deviant (lowest) values of $\alpha$ or $\beta$ are where a large number of SNe have been rejected; the remaining cuts are in agreement with values derived from the default cut. In summary, the fit parameters and uncertainties from Tables~\ref{tab:SNfinal} and \ref{tab:cepheidfinal} are: 
\begin{align}
  \alpha &= 0.165 \pm 0.010 (\rm{stat}) ^{+0.004}_{-0.005}(\rm{sys})\notag\\
  \beta &= 3.09 \pm 0.11 (\rm{stat}) ^{+0.04}_{-0.12} (\rm{sys})\notag\\
  b_W &=-3.17 \pm 0.04 (\rm{stat}) ^{+0.13}_{-0.11}(\rm{sys})\notag\\
  Z_W &= -0.11 \pm 0.09 (\rm{stat}) ^{+0.08}_{-0.10}(\rm{sys})\notag\\ 
  M_W &= -5.95 \pm 0.04 (\rm{stat}) ^{+0.06}_{-0.12}(\rm{sys}). \\\notag
\end{align}
These statistical uncertainties are found from Table~\ref{tab:globalfitfull}. We generally take the maximal variation measured from the reference fits in Tables~\ref{tab:SNfinal} and \ref{tab:cepheidfinal} as the systematic uncertainty, with the following exceptions. We disregard the higher low-redshift cut (associated with a large fraction of the SNe being discarded) in estimating the systematic uncertainty in $\beta$ -- see discussion in Section~\ref{sec:SNresults}. For the uncertainty in $Z_W$, we only consider the variation over fits which include both the LMC and MW in the distance anchor: the constraints on the metallicity dependence provided by different distance anchors are inconsistent with each other, so we only consider these fits for estimating the uncertainty for the nuisance parameter $Z_W$ alone (i.e.\ the other anchors are considered for estimating uncertainties on $M_B$ and $\mathcal{H}$, in Section~\ref{sec:MBscriptH}.) From Fig.~\ref{fig:globalbig}(c) it is clear that the statistical uncertainties in the SN parameters are around double the systematic uncertainty if we disregard the higher low-redshift SN cut. The opposite is true for the Cepheid parameters, where the statistical uncertainties are dwarfed by systematic variation with differing fits. If we restrict our analysis to only Cepheid fits anchored on all three galaxies, the statistical and systematic uncertainties are comparable in size. 
\\
\\
The systematic errors are asymmetric for most parameters, especially for $\beta$ (due to the outlying $z>0.0233$ cut) and $Z_W$. This can be observed in Fig.~\ref{fig:globalbig}(a)--(c), where it is evident our reference fits do not lie centrally within the parameter subspaces. Fig.~\ref{fig:globalbig}(b) shows that the MW as a distance anchor drives $M_W$ up, while the LMC (and to a lesser extent NGC~4258) drives $M_W$ down, an effect which propagates to $M_B$ and $\mathcal{H}$ (Fig.~\ref{fig:globalbig}(d)). Fits anchored on all three distance anchors lie centrally. Our Cepheid nuisance parameters remain consistent with R11 and E14 as initially found in Section~\ref{sec:cepheidcomparison}.\\
\\
We note that our best-fitting value for $\alpha$ is significantly higher than found in JLA (\cite{Betoule14}, table~10) and LOSS \citep{Ganeshalingam13}, by ${\sim}0.02$ (around double the total uncertainty in $\alpha$). This difference occurs consistently over a range of SN cuts. While the JLA analysis always determines $\alpha$ from the low-$z$ sample in conjunction with a higher-redshift sample, \cite{Ganeshalingam13} finds $\alpha = 0.146\pm 0.007$ from the LOSS sample, which overlaps with ours considerably and is over a similar redshift range. Our results for $\beta$ are consistent with the literature with the exception of the $z>0.0233$ SN cut, which results in a value ${\sim}1\sigma$ below the other cuts (the triangles in Fig.~\ref{fig:globalbig}(c)). The impact of this cut on $\mathcal{H}$ can be seen in Fig.~\ref{fig:globalbig}(d): the triangles (higher low-$z$ cut) have higher $\mathcal{H}$ than the other shapes (cuts) for each colour/fill (Cepheid fit). This effect is much smaller than the differences from varying the Cepheid fit. Nevertheless, it is in agreement with the increase of $H_0$ with increasing low-$z$ observed in R16, fig.~13.
\\
\\
The remaining nuisance parameters are the distance modulus offsets $\{\Delta\mu_i\}$, which, like all other nuisance parameters, are eventually marginalised over. Their values depend primarily on the Cepheid fits. The full table of fit values is left to Table~\ref{tab:offsetsfinal} in Appendix~D. The $\Delta\mu_i$ are visualised in Figs~\ref{fig:globaloffsets} and \ref{fig:globaloffsetsrel} with different colour/fill representing Cepheid fit. Fig.~\ref{fig:globaloffsets} gives some insight into the interplay and correlations between distance moduli of different galaxies, while Fig.~\ref{fig:globaloffsetsrel} shows the scatter and relative values of the $\Delta\mu_i$ from different fits. The statistical uncertainties in $\Delta\mu_i$ from individual fits range from 0.05 to 0.1, and is comparable to the scatter over different fits.

\subsubsection{Results for $M_B$ and $\mathcal{H}$}
\label{sec:MBscriptH}
\begin{figure}
\centering
\includegraphics[width=0.5\textwidth]{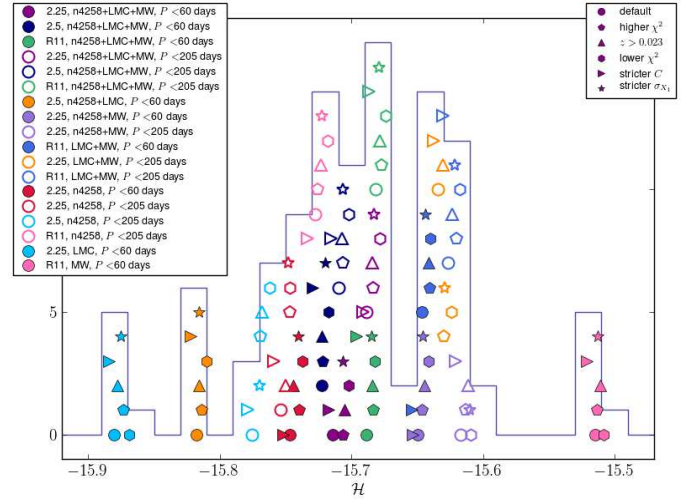}
\caption{Histogram of best-fitting values for $\mathcal{H}$ from all global fits: essentially a histogram of Fig.~\ref{fig:globalbig}(d) projected on to its $y-$axis. The blue line shows the binned histogram, while the individual points are plotted with their true $\mathcal{H}$ values and to reflect the distribution (i.e.\ they are stacked vertically for each bin). The frequency reflects the fits we chose to include in the global fit, i.e.\ we deliberately included more fits with all three anchors (and to a lesser extent, anchored on NGC~4258), rather than an inherent distribution. The legends are the same as in Fig.~\ref{fig:globalbig}, and reflect the Cepheid fit (colour and fill, with solidity of markers reflecting the inclusion of a Cepheid period cut) and SN cut (shape). The fits anchored on NGC~4258 only have a much broader spread in $\mathcal{H}$, and are responsible for the lowest values. The range in values in the NGC~4258-anchored fits is much greater, extending from left-filled pink markers to top-filled navy and dark purple markers and spanning ${\sim}0.11$~mag. In contrast the fits anchored on all three galaxies extend from the solid navy markers to the empty green markers, spanning ${\sim}0.04$~mag.}
  \label{fig:scriptHhist}
\end{figure}

We now consider the parameters $M_B$ and $\mathcal{H}$ which, together, directly reveal 
$H_0$. The degeneracy between them is apparent in Fig.~\ref{fig:globalbig}(d), which also shows that their primary dependence is on the Cepheid fits. Thus in Table~\ref{tab:MBscriptHfinal} we present the global fit results averaged over the SN cuts.\footnote{We report $M_B$ and $\mathcal{H}$ to 3 decimal places, unlike most other parameters which have been truncated to 2 decimal places (but not rounded in the analysis). These two quantities are of particular interest, and it is desirable to retain precision in both their values and uncertainties throughout this section.} Given that the fits in Table~\ref{tab:MBscriptHfinal} anchored on all three galaxies are spread out, we average these fits rather than choose a best fit, and take the maximal variation in these fits as the systematic uncertainty. There is a slight systematic difference between fits in Table~\ref{tab:MBscriptHfinal} with and without the upper period limit (on average, $\mathcal{H}$ is decreased by 0.015~mag where the $P<60$~day cut is applied). From a theoretical standpoint, we have no reason to preference one cut over the other. Thus our best estimates for $M_B$ and $\mathcal{H}$ are averaged over all fits anchored on all three galaxies (including fits both with and without the upper period limit), represented by solid and empty navy, green and dark purple markers in Figs.~\ref{fig:globalbig}--\ref{fig:scriptHhist}.
\\
\\
Our \textbf{best estimates} are
\begin{align}
  M_B&= -18.943 \pm 0.088 (\rm{stat}) \pm 0.024(\rm{sys})\notag\\
\mathcal{H} &= -15.698 \pm 0.093 (\rm{stat})\pm 0.023(\rm{sys}).\label{eq:final3anchor}\\\notag
\end{align}
Here the statistical uncertainties are found from relevant fits in Table~\ref{tab:globalfitfull}, in which a representative fit is bolded (with default SN cut, $T=2.25$ rejection, and no upper period limit). The above systematic uncertainties are given by the maximal variation in values with the combined NGC~4258+LMC+MW distance anchor. We impose this constraint on the anchor so that we can fairly assess the systematic uncertainty when all available distance information is used, and to allow better comparison with R11 and E14 who primarily report errors with all three anchors. In Section~\ref{sec:uncertainties} we investigate and discuss uncertainties in $\mathcal{H}$, including converting from an absolute error in the logarithmic quantity $\mathcal{H}$ to a relative error in $H_0$.\\
\\
We next consider fits anchored on NGC~4258 only, to estimate the uncertainty in this case, and for the sake of comparison with R11 and E14. These fits are represented by the empty turquoise and pink markers, and by all red markers in Figs.~\ref{fig:scriptHhist} and \ref{fig:globalbig}. We average these results from Table~\ref{tab:MBscriptHfinal} to find Equation~\ref{eq:finaln4258}, as with Equation~\ref{eq:final3anchor}. The systematic uncertainties are given by the maximal variation in values derived from these fits, and the statistical uncertainties are found from NGC~4258-anchored fits in Table~\ref{tab:globalfitfull}, with one representative fit bolded.
\begin{align}
 M_B&= -18.993 \pm 0.104 (\rm{stat}) \pm 0.023(\rm{sys})\notag\\
\mathcal{H} &= -15.748 \pm 0.107 (\rm{stat}) \pm 0.023(\rm{sys}).\label{eq:finaln4258}\\\notag
\end{align}
\begin{table}
  \begin{center}
   \caption{Global fit results for degenerate parameters $M_B$ and $\mathcal{H}$, averaged over SN cuts. The bolded fit ($T=2.25$ rejection, all three anchors, and no upper cut on the period) is chosen as our reference fit.
     \label{tab:MBscriptHfinal}}
   
  \begin{tabular}{lllll}
\hline
\hline
Rejection & Distance anchor & $P<60$d&  $M_B$ & $\mathcal{H}$\\
\hline
2.25 &all$^a$ & Y & -18.953 & -15.709 \\
2.5 &all & Y & -18.967 & -15.722 \\
R11 &all & Y & -18.932 & -15.687 \\
2.5 & n4258+LMC & Y & -19.061 & -15.816 \\
2.25 & n4258+MW & Y & -18.892 & -15.647 \\
R11 & LMC+MW & Y & -18.889 & -15.644 \\
2.25 & n4258 & Y & -18.988 & -15.743 \\
2.25 & LMC & Y & -19.122 & -15.877 \\
R11 & MW & Y & -18.759 & -15.514 \\
2.25 &all & N & -18.929 & -15.685 \\
2.5 &all & N & -18.953 & -15.708 \\
R11 &all & N & -18.924 & -15.679 \\
2.25 & n4258+MW & N & -18.859 & -15.614 \\
2.25 & LMC+MW & N & -18.875 & -15.631 \\
R11 & LMC+MW & N & -18.868 & -15.623 \\
2.25 & n4258 & N & -18.996 & -15.751 \\
2.5 & n4258 & N & -19.015 & -15.770 \\
R11 & n4258 & N & -18.970 & -15.725 \\
\hline
\end{tabular}
\end{center}
$^a$ i.e.\ n4258+LMC+MW\\
\end{table}

The resultant value of $\mathcal{H}$ in Equation~\ref{eq:finaln4258} is 0.05~mag lower (corresponding to a 2.3\% decrease in $H_0$) compared to where all three anchors are used (Equation~\ref{eq:final3anchor}). Moreover, $M_B$ (which is largely degenerate with $\mathcal{H}$) is also 0.05~mag lower (brighter). The systematic uncertainty (i.e.\ the spread in values between different fits) is the same, while the statistical uncertainties are larger, reflective of the fact that a distance scale is anchored on a smaller set of data. \\
\\
Our best estimate of the peak \snia~brightness $M_B$ (in Equation~\ref{eq:final3anchor}, from the three-galaxy anchor) appears mildly higher (dimmer) than values reported in JLA (assuming $H_0 = 70$~\kmsmpc), which are $M_B=-19.05\pm 0.02$ from all \snia~data, or $M_B=-19.02\pm 0.03$ from a lower-redshift subsample consisting of low-$z$ and SDSS supernovae (table~10 of \cite{Betoule14}). However, the supernova-only fit in JLA cannot constrain both $M_B$ and $H_0$, which are degenerate. As they have assumed a value for $H_0$ (while we have fitted separately using a distance ladder), our numerical values for $M_B$ are not directly comparable, but merely reflect the influence of different values of $H_0$.
\\\\
Returning to $H_0$, Equation~\ref{eq:final3anchor} corresponds to a value of $H_0 = 72.5 \pm 3.1 (\rm{stat}) \pm 0.77 (\rm{sys})$~\kmsmpc\ (total uncertainty of 4.4\%) from the combined NGC 4258+LMC+MW anchors. If we assume the older distance $\mu_{4258}=29.31$ in R11 (Footnote~\ref{foot:2}), our best estimate increases to $H_0 = 73.8 \pm 3.2 (\rm{stat}) \pm 0.78 (\rm{sys})$. These central values agree with R11 ($H_0 = 73.8 \pm 2.4$) and E14 ($H_0 = 72.5 \pm 2.5$), which respectively assume 
$\mu_{4258} = 29.31$ and 29.404. Using only NGC~4258 as a distance anchor (and the new \cite{Humphreys13} value of $\mu_{4258}=29.404$) gives $H_0 = 70.9 \pm 3.5 (\rm{stat}) \pm 0.75 (\rm{sys})$~\kmsmpc, which is $2.3\%$ lower than with the three anchors. The uncertainties in $\mathcal{H}$ are broken down in Section~\ref{sec:H0uncertainties} and summarised in Table~\ref{tab:errors}. We next discuss the uncertainties in $H_0$; their size informs the significance of the tension between values of the Hubble constant measured using different probes, so they are of equal interest to the values.

\subsection{Uncertainties} 
\label{sec:uncertainties}

We have emphasized the importance of quantifying and incorporating the scatter in parameters arising from varying aspects of the SN and Cepheid fits, and indeed we use this overall variation in results to gauge the systematic uncertainty in these parameters. However, we have also seen that the statistical uncertainty dominates for the supernova parameters $\alpha$ and $\beta$ (Fig.~\ref{fig:alphabetaSN} and \ref{fig:globalbig}(c)), as well as for $M_B$ and $\mathcal{H}$ when only considering the systematic variation between fits with all three anchors (Fig.~\ref{fig:globalbig}(d)). This dominance reflects the fact that the \snia~samples, especially the nearby sample (i.e.\ in Cepheid hosts), are relatively small with large errors when compared to the Cepheids. Hence the SNe are statistically limited while the Cepheids are not. 
\\\\
For clarity we divide the contributions to the total uncertainty in the parameters into three classes:
\begin{enumerate}
\item
  The statistical (in the usual sense) portion of the uncertainty reported by MultiNest, which is dominated by noise in the nearby and low-$z$ supernovae.
\item
  The systematic elements of $\mathbf{C_\eta}$, which make up remainder of the uncertainty reported by MultiNest, listed in Table~\ref{tab:relsize}. 
\item
  The systematic uncertainty associated with varying aspects of the fit between reasonable alternatives is dominated by the variation in the choice of Cepheid fit, as shown in Figs~\ref{fig:globalbig} and \ref{fig:scriptHhist}. For our final value and uncertainty of $H_0$ we focus on fits with all three anchors only (with some consideration of fits with only NGC~4258 as an anchor for the sake of comparison to R11 and E14). Then in effect we are only considering the variation with the rejection algorithm and the cut on Cepheid period.
\end{enumerate}

\subsubsection{Uncertainties in $H_0$}
\label{sec:H0uncertainties}
We now address the uncertainty in the Hubble constant $H_0$ explicitly, using results in Section~\ref{sec:MBscriptH} (Equations~\ref{eq:final3anchor} and \ref{eq:finaln4258}). As the quantity $\mathcal{H}$ is related to the logarithm of $H_0$, its absolute error informs the relative error in $H_0$, via
\begin{align}
  \label{eq:H0err}
  \frac{\sigma_{H_0}}{H_0}=\frac{\log(10)}{5}\sigma_{\mathcal{H}}.
\end{align}
\begin{table}
  \begin{center}
    \caption{Summary of uncertainties in $\mathcal{H}$ from Section~\ref{sec:MBscriptH} (Equation~\ref{eq:final3anchor} and \ref{eq:finaln4258}), converted to relative errors in $H_0$ using Equation~\ref{eq:H0err}, and added in quadrature in line with R11. The statistical error below are those reported by MultiNest (Table~\ref{tab:globalfitfull}) and include terms (i) and (ii) described at the start of Section~\ref{sec:uncertainties}. The systematic error is from the variation between fit results with different choices of Cepheid fits, and secondarily SN cuts, i.e. term (iii).
      \label{tab:errors}}
  \begin{tabular}{lll}
\hline
\hline
Anchor & all & NGC 4258 only\\
\hline
$\mathbf{\mathcal{H}}$&-15.698 & -15.748\\
\hline
$\mathbf{\sigma_{\mathcal{H}}}$ && \\
 Statistical & 0.093 & 0.107\\
 Systematic & 0.023&0.023\\
 \hline
\textbf{Relative $H_0$ error (\%)} && \\
 Statistical & 4.3 & 4.9\\
 Systematic & 1.1 & 1.1\\
 \hline
 \textbf{Total} &  \textbf{4.4}   & \textbf{5.0} \\
 \hline
  \end{tabular}
  \end{center}
\end{table}

Table~\ref{tab:errors} summarises our calculations of the final uncertainty in $H_0$ from Equations~\ref{eq:final3anchor} and \ref{eq:finaln4258}. We find using Equation~\ref{eq:H0err} relative errors in $H_0$ of 4.3\% statistical and 1.1\% systematic (corresponding to terms (i) and (ii) combined, and (iii) respectively as described at the start of Section~\ref{sec:uncertainties}) from all three distance anchors. From using only NGC~4258 as an anchor, these errors are 4.9\% statistical, 1.1\% systematic, 5.0\% total. The final uncertainty in $H_0$ (the bottom row of Table~\ref{tab:errors}) is the quadrature sum of the above statistical and systematic terms. Table~\ref{tab:errors} is comparable to the lower portion of table~5 of R11 (and subsequently table~7 of R16), which lists all systematic and statistical uncertainties contributing in quadrature to the uncertainty in $H_0$.

\subsubsection{Increase in error compared to R11 and E14}
\label{sec:bigerror}
Our final uncertainty in $H_0$ is 4.4\% total (4.3\% statistical and 1.1\% systematic, with the statistical term inclusive of contributions from \snia~covariance matrices) for the NGC~4258+LMC+MW distance anchor, which is significantly larger than previously found for the same data set (by 1\% absolutely, or a $\sim$20\% increase): R11 and E14\footnote{E14 adopts the error in R11 for the \snia-side of the analysis.} report total uncertainties of 3.3\%\footnote{The errors reported in table~5 of R11 are: 2.9\% statistical, 1.0\% systematic, 3.1\% total. However the final error given with all three distance anchors conservatively includes the larger statistical error associated with using two distance anchors instead of three, resulting in a total of 3.3\%.}
and 3.4\% respectively. If NGC~4258 alone is used as a distance anchor, the above errors increase to 5.0\% total (4.9\% statistical and 1.1\% systematic) for our fit, 4.1\% (4.0\% statistical, 1.0\% systematic) from R11, and 4.7\% total from E14. The difference between our errors and those found in E14 is smaller with the NGC~4258 anchor compared to when all three anchors are used -- however, both are significantly larger than found in R11. For the remainder of Section~\ref{sec:bigerror}, our discussion of errors pertains to fits with all three distance anchors. 
\\\\
Although it appears that the increase in our error lies in the statistical term (with the systematic term remaining the same), it is important to note the significant differences in how these terms are derived and defined in this work (given in points (i)--(iii) at the start of Section~\ref{sec:uncertainties}), compared to R11. Explicitly, the covariance matrices which quantify our \snia~systematic terms directly contribute to the statistical errors in our global fits (i.e.\ increase the widths of PDFs) via the likelihood, while our systematic term contains the variation in parameters resulting from changing features of the fits. In comparison, the errors in each part of the calibration chain from the distance calibrators to the \sneia~are separately given in R11, table~5. The total uncertainty is a quadrature sum of these individual terms, and the systematic variation described in R11, section~4. \\
\\
The two major differences in our analysis which potentially contribute to the increased error are the inclusion of the supernova covariance matrices, detailed in Appendix~\ref{sec:covmat}, and the simultaneous fit to all parameters, described in Section~\ref{sec:globaleqns}. As outlined above, it is not possible to make a direct comparison between contributions to our error and errors given in R11, with the aim of isolating the source of the discrepancy. However, we attempt to separate the influences of the covariance matrices and simultaneous fit, replicating the quadrature sum in R11 as closely as we can below.\\
\\
First, we isolate the effect of the supernova covariance matrices alone on the size of uncertainties, by considering the error in the intercept of the \snia~$m-z$ relation: this is $\mathcal{M}$ determined from our SN-only fit (Table~\ref{tab:SNinit} in Section~\ref{sec:snfit}), and is equivalent to $5a_V = 3.485\pm 0.010$ in R11. Our error in $\mathcal{M}$ is ${\sim}0.036$, over three times larger than in R11. This error is roughly halved to 0.019 if we only consider the strictly statistical covariance matrix, i.e.\ $\mathbf{C_{\rm{stat}}}$ in Equation~\ref{eq:covmats}.\footnote{Neglecting the uncertainty from the finiteness of the SALT2 training sample reduces the error slightly to 0.017, which reflects the statistical error in the observed \sneia~only (i.e.\ the tridiagonal matrix $\mathbf{C_{\rm{stat,diag}}}$).} For the same supernova data, our statistical-only error exceeds the total error in R11. Including the \snia~systematic covariance matrices doubles the error again. We infer that the increase in error in this analysis compared to R11 is attributable to both the covariance matrix method of accounting for correlated \snia~uncertainties, and to the individual systematic covariance matrices this method comprises.
\\\\
Next, we attempt to replicate the quadrature summation of terms in R11, table~5 (most of which unfortunately do not have equivalent terms in our analysis) using projected uncertainties from our global fit. It is important to note that this comparison is not directly equivalent, because we are marginalising simultaneously over all nuisance parameters. With this caveat, we break down the uncertainty in the overall value of $H_0$ into three components: the uncertainties in the distance anchor, in the calibration of the \snia~absolute magnitude $M_B$ via Cepheids, and in the measurement of the local expansion rate via \sneia~(given in the intercept $\mathcal{M}$).\footnote{This decomposition essentially follows equation~4 of R11.} These can be determined separately from three disjoint data sets, as follows. The anchor distance is constrained by external data: the megamaser distance to NGC~4258 has a 0.066~mag uncertainty, corresponding to 3.0\% in $H_0$. Only the low-$z$ \sneia~are used to constrain $\mathcal{M}$ (or $5a_V$), with a 0.036~mag or 1.7\% uncertainty. The calibration transfer from the Cepheids to the \sneia~occurs in the simultaneous fit of the nearby supernova and Cepheid data\footnote{For the uncertainty in $M_B$ to be independent of the error in $\mathcal{M}$, only these data can be included.} to Equations~\ref{eq:pl2} and \ref{eq:nearby2}. The resultant uncertainty in $M_B$ is 0.103~mag (with only the NGC~4258 anchor) and incorporates both the uncertainty in the \snia-Cepheid calibration and the uncertainty in the distance anchor; thus the former is 0.079~mag or 3.6\% in $H_0$.\footnote{The same calculation with all three anchors results in the same number. In the setup of R11, equation~4, this \snia-Cepheid calibration uncertainty is the error in $m^0_{v,4258}$, which is equivalent to $M_B + \mu_{4258}$.} In quadrature, these three terms sum to 5.0\% in $H_0$ using the NGC~4258 anchor, and 4.3\% with all three anchors. This decomposition, whilst approximate, indicates that a quadrature sum of uncertainties in independent components results in similar uncertainties to our simultaneous fit. Thus, the simultaneous fit does not by itself result in the increase in statistical error. 

\subsubsection{Relative size of \snia~uncertainties}
\label{sec:systresults}

\begin{figure}
  \centering
\includegraphics[width=0.5\textwidth]{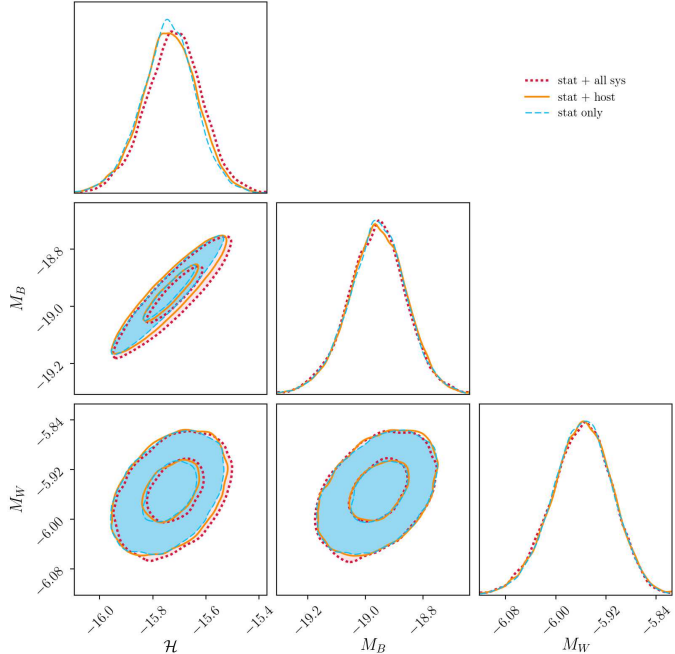}
\caption{Constraints on parameters $\mathcal{H}$, $M_B$, and $M_W$ from an example global MultiNest fit (with Cepheid fit $T=2.5$, NGC~4258 anchor, no priors, default SN cuts) with partial and full contributions to the full \snia~covariance matrix. Confidence contours are shown with the statistical contribution $\mathbf{C_{\rm{stat}}}$ only (turquoise filled), with one systematic term (the host mass correction) added i.e.\ $\mathbf{C_{\rm{stat}}+C_{\rm{host}}}$ (orange solid line), and with all \snia~systematics, i.e.\ $\mathbf{C_{\rm{stat}}+C_{\rm{sys}}}$ (red dashed). The inclusion of systematic terms only increases the widths of contours marginally relative to the $\mathbf{C_{\rm{stat}}}$-only (turquoise) contours, reflecting that the statistical contribution dominates the uncertainty in the parameters.}
  \label{fig:systcontours}
\end{figure}

We now examine the breakdown of uncertainties contributing to the statistical error, which include the multiple statistical and systematic uncertainties in \snia~parameters making up $\mathbf{C_\eta}$ as constructed in Appendix~\ref{sec:systematics}.
\\
\\
To visually assess the impact on confidence contours we compare results from MultiNest with different covariance matrix inputs. For an example global fit (with Cepheid fit $T=2.5$, NGC~4258 anchor, no priors, default SN cuts) we test each systematic, and compare their results from MultiNest. The full expression for the covariance matrix $\mathbf{C_\eta}$ for observed \snia~quantities is described in Appendix~\ref{sec:covmat}. As entries of $\mathbf{C_\eta}$ in MultiNest we try the following: only the statistical contribution $\mathbf{C_{\rm{stat}}}$ (described in Appendix~\ref{sec:stat}), each single systematic term added to $\mathbf{C_{\rm{stat}}}$, and all systematics added i.e.\ $\mathbf{C_{\rm{stat}}}+\mathbf{C_{\rm{sys}}}$ (the default for all global fits). The confidence contours with statistical uncertainties only and with all systematics are easily distinguishable in Fig.~\ref{fig:systcontours}, but the contours with individual systematics are not. Thus for clarity we only show in Fig.~\ref{fig:systcontours} the systematic term from the uncertainty in host mass correction corrections ($\mathbf{C_{\rm{host}}}$ in Equation~\ref{eq:covmats}, described in Appendix~\ref{sec:host}), in addition to $\mathbf{C_{\rm{stat}}}$ and $\mathbf{C_{\rm{stat}}}+\mathbf{C_{\rm{sys}}}$. The difference between the contours is slight, indicating that the uncertainties in the parameters only increase slightly when covariance matrices for different systematics are added to the statistical term $\mathbf{C_{\rm{stat}}}$.
\\

\begin{table}
  \begin{center}
    \caption{Relative contributions to the uncertainty in $H_0$ (i.e. the variance) from individual statistical and systematic sources uncertainties, calculated as described in \citet[][Section 6.2]{Betoule14}.
      \label{tab:relsize}}
  \begin{tabular}{lll}
\hline
\hline
Source of  & Relative contribution    & Described in  \\
uncertainty & to $\sigma^2(H_0)$ (\%) & Appendix \\
\hline
\textbf{Statistical:}&&\\
Lightcurves & 62.1 & \ref{sec:stat}\\
SALT2 training & 1.2&\ref{sec:stat}\\
 \textbf{Total statistical} & \textbf{63.3}\\
\hline
\textbf{Systematic:}&&\\
Malmquist bias & 13.7 & \ref{sec:malmquist},\ref{sec:systematics}\\
Host galaxy& 	13.0& \ref{sec:host}\\
Lightcurve model&	 6.8 &\ref{sec:systematics}\\
Calibration&	 3.1& \ref{sec:calibration}\\
Peculiar velocities&	 0.04& \ref{sec:pecvel}\\
Milky Way extinction & 0.03& \ref{sec:extinction}\\
\textbf{Total systematic} & \textbf{36.7} \\
\hline
  \end{tabular}
\end{center}
\end{table}

Following the method in JLA \cite[section~6.2]{Betoule14}, we quantify the relative contributions, replacing the parameters $\{\Omega_M, w,\alpha, \beta, M_B, \Delta M\}$ with our parameters $\{H_0, M_B, \alpha, \beta\}$ (the only parameters in $\Theta$ which can be influenced by the low-$z$ \snia~covariance matrices). The breakdown of relative contributions to the variance in $H_0$ from each term (the purely statistical term $\mathbf{C_{\rm{stat}}}$, and each systematic) are reported in Table~\ref{tab:relsize}. We emphasize that each of these numbers represents a proportion of the uncertainty (terms (i) and (ii) in Section~\ref{sec:uncertainties} combined) from the systematic or statistical term in question alone, rather than reflecting an uncertainty in $H_0$.
\\
\\
From Table~\ref{tab:relsize} and Fig.~\ref{fig:systcontours} it is clear that $\mathbf{C_{\rm{stat}}}$ is the largest component of $\mathbf{C_\eta}$. Even though the contributions to $\mathbf{C_\eta}$ from SN systematics are included in the statistical uncertainty, all of these systematics together are smaller than the SN statistical uncertainties: the relative contributions to the variance in $H_0$ are dominated by $\mathbf{C_{\rm{stat,diag}}}$ (Table~\ref{tab:relsize}), and the contours with and without systematic covariance matrices added to $\mathbf{C_{\rm{stat}}}$ in Figure~\ref{fig:systcontours} are similar to each other. Of the systematic terms (Appendices~\ref{sec:extinction} --\ref{sec:pecvel}), the most significant in their impact on $H_0$ are from uncertainties in the Malmquist bias correction (including the selection function) and the host mass correction (Appendices~\ref{sec:malmquist} and \ref{sec:host}), followed by the uncertainty in lightcurve model. While JLA had found the photometric calibration (Appendix~\ref{sec:calibration}), especially from low-$z$ SNe, to be the dominant uncertainty for $\Omega_M$ and $w$, its effect on $H_0$ is decidedly smaller. It is interesting to note that despite conservative estimates of both the uncertainties in Milky Way extinction and peculiar velocity correction (Appendices~\ref{sec:extinction} and \ref{sec:pecvel}), their effects on the error in $H_0$ are negligible.


\section{Conclusions and future work}
\label{sec:conclusion}

Our independent analysis of the \cite{Riess11} data complements the R11 and E14 analyses in understanding the local measurement of the Hubble constant from type Ia supernovae. This work occupies the unique position of combining the
precise Cepheid calibration of nearby \sneia~(R11) with the sophisticated, thorough treatment of supernova lightcurves and systematics within a SALT2 framework~\citep{Betoule14}. In the context of the present tension in $H_0$ we present the first blinded \snia-based determination of $H_0$, eliminating confirmation and other biases. This work is intentionally applied to the well understood historical work of R11 and E14, as a proof of concept. It is our intent to extend this analysis to the sample of \cite{Riess16}.
\\
\\
Our best estimate from R11 data is $H_0 = 72.5 \pm 3.1 (\rm{stat}) \pm 0.77 (\rm{sys})$~\kmsmpc~using a three-galaxy (NGC 4258+LMC+MW) anchor. The central value is in excellent agreement with both R11 (after correcting for the lower value of $\mu_{4258}$ adopted -- see Footnote~\ref{foot:2}) and the E14 reanalysis. Our above value and uncertainty imply tension with the Planck value at ${\sim}1.5\sigma$ significance, which is smaller than found in previous analyses of the R11 data, due to our increased uncertainties. However, our blinded affirmation of the central value in R11 signifies that the discrepancy between \snia- and CMB-derived values of the Hubble constant continues to exist. While this discrepancy is less significant in our analysis than in the original analysis of R11 data, it has potential to be magnified by the improved data set in R16 (which has smaller statistical uncertainties compared to R11), and hence remains of interest. It is thus necessary to apply the techniques in this paper to the R16 data, in order to make a contemporary assessment of the significance of the tension in the Hubble constant. 
\\
\\
Incidentally, we find a higher stretch coefficient $\alpha = 0.165\pm 0.010$ for our low-$z$ supernovae compared to LOSS (which find $\alpha=0.147\pm0.007$ over a similar redshift range) and JLA. This discrepancy at ${\sim}2\sigma$ is surprising, and prompts further investigation. While our \snia~zero point $M_B=-18.94\pm 0.09$ appears higher than in JLA, this difference arises because the parameters are degenerate using supernovae only: the JLA analysis assumes $H_0 = 70$~\kmsmpc, whereas we have used a distance ladder to constrain both $H_0$ and $M_B$.
\\\\
We find a larger relative uncertainty in $H_0$ (4.4\% total) compared to R11 and E14 analyses of the same data (3.3\% and 3.4\% total respectively), based on the NGC~4258+LMC+MW distance scale. The difference appears in the statistical error; our systematic term is similar to that in R11 (1.1\% and 1.0\% respectively), with the caveat that the separation of our total uncertainty into statistical and systematic components is not directly comparable to R11, as described in Section~\ref{sec:bigerror}. Our larger error primarily arises from our use of covariance matrices to estimate \snia~systematic uncertainties. Other significant differences in our analysis, which potentially contribute to the increased uncertainty, are our simultaneous fit of all three sets of data, allowing all parameters to interact, and our use of variation in results to quantify systematic error. These distinctions are in our view justified and desirable. Given the increase in uncertainty they produce compared to previous works, they are important to consider in future analyses.\\
\\
As found in R11, our results are limited by statistics in the supernova samples. Steps to reduce this statistical uncertainty have been implemented in R16, namely increasing the number of nearby galaxies to 18 and improving the \snia~photometry, to reduce the total uncertainty to 2.4\%. We envisage a similar relative increase in precision when the techniques in this work are applied to the same data set. R16 also includes important changes to data analysis of the Cepheids. Other contributions to our error budget are the systematic uncertainty, which is dominated by the variation in the different Cepheid fits, and the \snia~systematic terms in $\mathbf{C_\eta}$, the largest of which are $\mathbf{C_{\rm{bias}}}$ and $\mathbf{C_{\rm{host}}}$.
\\\\
Foremost, we find that both the use of covariance matrices and the simultaneous fit of data from different rungs of a distance ladder will be important in future analyses in order to wholly 
account for uncertainties. Furthermore, our findings recommend more sophisticated techniques for quantifying host galaxy dependence of \snia~magnitudes, and modelling of Malmquist bias -- both of which have the potential to diminish the systematic error in $H_0$. These techniques are continually improved in supernova analyses, particularly in the pursuit of more precise measurements of dark energy, for example in the Dark Energy Survey~\citep[DES; ][]{DES}. Meanwhile a uniform, non-targeted low-$z$ sample (e.g.\ the SkyMapper Transient Survey~\citep{Scalzo17}, or the Pan-STARRS1 Survey~\citep{Rest14}) will simplify photometric calibration and the selection function, reducing associated uncertainties, and will avoid peculiar velocity biases from coherent flows. Adopting these changes will benefit future \snia-based $H_0$ measurements.
\section*{Acknowledgements}

This research was supported by the Australian Research Council Centre of Excellence for All-sky Astrophysics (CAASTRO), through project number CE110001020, and ARC Laureate Fellowship Grant FL0992131. We also acknowledge support of a University of Southampton Diamond Jubilee International Visiting Fellowship. BZ thanks the Australian Astronomical Observatory and the John Shaw Foundation for their financial support.
\\
\\
We are grateful to the SNLS collaboration for providing us with the \texttt{snpca} code and the recalibrated SALT2 training surfaces, used in Sections~\ref{sec:stat} and \ref{sec:calibration} respectively. We thank Matthew Colless and Christina Magoulas for suggesting the 2M++ peculiar velocity model, Dan Scolnic for helpful discussions, and Anais M\"{o}ller for useful comments. This manuscript has greatly benefitted from thorough and constructive feedback from Adam Riess, to whom we are very grateful. Finally, we express our sincere thanks to the anonymous referee who has made particularly helpful suggestions and comments, which have substantially improved this work. Furthermore, we acknowledge their support and participation in the unique blinding process which contributed considerably both to the paper and their workload. 



\bibliographystyle{mnras}
\bibliography{mybibfile}


\appendix

\section{Dependence of Cepheid-only fit}
\subsection{Outlier rejection} 
\label{sec:rejection}
We perform fits in two ways: either assuming the outlier rejection in \cite{Riess11}, or following the rejection method in \cite{Efstathiou14}. The R11 algorithm rejects Cepheids from each galaxy (rather than from the global fit), based on their deviation from the best Leavitt law fit. This rejection does not take into account the size of the Cepheid uncertainties, so that points with small residuals but large uncertainties are selectively accepted (E14, section~3.1). Consequently a large fraction of the total number of Cepheids is rejected, including a set of subluminous low-metallicity Cepheids (later corrected in R16, as discussed in Appendix~\ref{sec:cepheidpriors}). Moreover, the intrinsic scatter is overestimated, resulting in a low reduced $\chi^2$.
\\
\\
These limitations in the R11 rejection are addressed in the upgraded algorithm in E14, which rejects a Cepheid from the global fit if its magnitude deviates from the global fit by more than the quantity $T\sqrt{m_{W,\mathrm{err}}^2+\sigmaintC^2}$ for a threshold $T$ (set to 2.25 or 2.5), where $m_{W,\mathrm{err}}$ and $\sigmaintC$ are the uncertainty in the individual Cepheid's measurement and the intrinsic scatter $\sigmaintC$ respectively. This process is iterative, with $\sigmaintC$ recalculated at each step (such that $\chi^2_c$ per degree of freedom ${\sim}1$) with increments of 0.1, where the sum in $\chi^2_c$ is always over only the Cepheids in NGC~4258 and SN hosts (i.e.\ not the LMC or MW). The rejection at each iteration is based on the best fit determined in the previous iteration, i.e.\ the mean and 1$\sigma$ uncertainty of the posterior distribution.\\
\\\\
Initially $\sigmaintC$ is set to 0.3. Then in each iteration we perform the following steps:
\begin{enumerate}
\item
perform a MultiNest fit to all remaining Cepheids, to find marginalised posterior distributions;
\item
find and remove outliers based on these parameters;
\item
compute the new $\sigmaintC$ for these parameters and the updated Cepheid sample.
\end{enumerate}
These steps are repeated until convergence, i.e.\ until no Cepheids are rejected in the second step.
\\\\
The variation in fit results from different outlier rejection is presented in Fig.~\ref{fig:cepheidall} and Section~\ref{sec:cepheidresults}. In general the R11 rejection results in less negative values of both $b_W$ and $Z_W$, attributable to the aforementioned subluminous and low-metallicity subsample that it rejects.
\\\\
The fit is forced to be good for all three rejection algorithms: $\sigmaintC$ is engineered to result in $\chi^2/\rm{DoF}{\sim}1$. Thus the algorithms cannot be compared statistically; the outlier rejection method has the drawback of not allowing the uncertainty on $\sigmaintC$ to be estimated, and the related consequence that we (by construction) cannot assess goodness-of-fit. Alternative statistical methods used in recent \snia~analysis can surpass these limitations, including Bayesian hierarchical models~\citep{March11,Shariff15}, the alternate Bayesian framework in \cite{Rubin15}, and Approximate Bayesian Computation~\cite{Jennings16}. Notably, these have been applied to determining $H_0$ from the R11 and R16 data sets in \cite{Cardona16}.

\subsection{Distance anchors}
\label{sec:anchors}
Our equations in Section~\ref{sec:data} assume NGC~4258 is the only distance calibrator. We can generalise these equations to allow for combinations of the three anchor galaxies in R11, adding Cepheids in the LMC and MW (data described in Section~\ref{sec:cepheidobs}). As these additional Cepheids do not have metallicity measurements, we adopt the mean values from E14 of $12 + \log_{10}[O/H]$ of 8.5 and 8.9 for LMC and MW Cepheids respectively. Here, we test the dependence of the Cepheid parameters on the distance anchor. For Cepheids in the LMC and MW the Leavitt law (Equation~\ref{eq:pl1}) takes the forms:
\begin{align}
m_{W,\mathrm{LMC}j} &= b_W (\log_{10} P_j - 1) - 0.4 Z_W + M_W +\mu_{\rm{LMC}}\label{eq:plLMC}\\
m_{W,\mathrm{MW}j} &= b_W (\log_{10} P_j - 1) + M_W \label{eq:plMW}.
\end{align}
We consider combinations of NGC~4258, LMC, and MW (seven in total) as distance calibrators. If NGC~4258 is not included, then no prior for $\mu_{4258}$ is imposed in MultiNest. However, the likelihood $\mathcal{L}$ in Equation~\ref{eq:likelihood} still depends on $\mu_{4258}$, which is indirectly constrained through the other anchors and $M_W$, and hence remains a fit parameter in $\Theta$. If the LMC is used as an anchor then it is necessary to include $\mu_{\rm{LMC}}$ as a parameter in $\Theta$; this always has a (Gaussian) prior set to reflect the \cite{Pietrzynski13} measurement from eclipsing binaries of $\mu_{\rm{LMC}} = 18.494 \pm 0.049$. The likelihood $\mathcal{L}$ is affected as a term $\chi^2_{\rm{LMC}}$ (Equation~\ref{eq:chisqcepheidsLMC}) is added to Equation~\ref{eq:chisqcepheids};  similarly if the MW is used as an anchor then $\chi^2_{\rm{MW}}$ (Equation~\ref{eq:chisqcepheidsMW}) is added. We assume $\sigmaintC = 0.113$ and $0.1$ for the LMC and MW respectively following E14. 
\begin{align}
\chi^2_{\rm{LMC}} & = \sum_{j}\frac{(\hat{m}_{W,\mathrm{LMC}j} -m_{W,\mathrm{LMC}j,\mathrm{mod}})^2}{\hat{m}_{\mathrm{LMC,err}j}^2 + \sigmaintC^2}\label{eq:chisqcepheidsLMC}\\
\chi^2_{\rm{MW}} & = \sum_{j}\frac{(\hat{m}_{W,\mathrm{MW}j} -m_{W,\mathrm{MW}j,\mathrm{mod}})^2}{\hat{m}_{\mathrm{MW,err}j}^2 + \sigmaintC^2}.\label{eq:chisqcepheidsMW}
\end{align}
A modification to the above is necessary if both the LMC and MW are used as distance anchors, to account for the calibration uncertainty between ground-based and HST photometry. We do this using the covariance matrix $\mathbf{C}_{\mathrm{LMC+MW}ij} = (\hat{m}_{W i}^2 + \sigmaintC^2)\delta_{ij} + \sigma_{\mathrm{cal}}^2$ with $\sigma_{\mathrm{cal}} = 0.04$ (R11). Instead of $\chi^2_{\rm{LMC}} + \chi^2_{\rm{MW}}$, we add the term

\begin{align}
 \chi^2_{\mathrm{LMC+MW}} = &(\boldsymbol{\hat{m}_{W,\mathrm{MW}}} -\boldsymbol{m_{W,\mathrm{MW,mod}}})\boldsymbol{\cdot}\mathbf{C_{LMC+MW}}^{-1}\boldsymbol{\cdot}\notag\\
 &(\boldsymbol{\hat{m}_{W,\mathrm{MW}}} -\boldsymbol{m_{W,\mathrm{MW,mod}}})^T\label{eq:chisqcepheidsLMCMW}
\end{align}
to the $\chi^2_c$ term going into $\mathcal{L}$. Here, bolded quantities represent vectors over all LMC and MW Cepheids.
\\
\\
The results of varying the distance anchor are discussed in Section~\ref{sec:cepheidresults}. Briefly, the inclusion of both the LMC and MW anchors constrains both $b_W$ and $Z_W$ more tightly.
\subsection{Longer-period cepheids}
\label{sec:longperiod}
The data include Cepheids with period ranging from ${\sim}10$~days, to over 200 days, and Cepheids of all periods are included in Leavitt law fits in R11 (except for those Cepheids marked `low~P' in R11, table~2). \cite{Bird09} examine longer-period ($P>80$~day) Cepheids and find that these Cepheids obey a flatter Leavitt law, with a less negative period dependence $b_W$. Accordingly, recent studies of the Leavitt law \citep[e.g.][]{Freedman11, Scowcroft11} have excluded Cepheids with period greater than 60~days. Similarly, E14 in their reanalysis of the R11 data have imposed the same upper limit on Cepheid period because of the observed change in slope. It is pragmatic to follow these examples in only using Cepheids over a period range where the slope remains constant; however, it is also useful the full range of periods to accommodate the change in slope and for the sake of comparison with R11. Rather than making an argument to include the $P<60$~day cut or not, we perform fits with and without an upper limit on the period.
\\
\\
E14 reasons that while including longer-period Cepheids decreases the magnitude of the Leavitt law slope $b_W$, there is little impact on $H_0$ \citep[][appendix~A]{Efstathiou14}, so they only include $P<60$~day Cepheids in their fits. Our priors on $b_W$ differ slightly from E14 (discussed in Appendix~\ref{sec:cepheidpriors}), and we are interested in the variation of Cepheid parameters with the choice of period cut (as with distance anchor and rejection algorithm in previous sections), so we test the dependence of fit results on the inclusion of an upper limit on period. Results of including longer-period Cepheids are lesser dependence on Cepheid slope and metallicity dependence (less negative $b_W$ and $Z_W$), as described in Section~\ref{sec:cepheidresults}.
\subsection{Slope and metallicity priors}
\label{sec:cepheidpriors}
We test and discuss the Gaussian priors on $b_W$ and $Z_W$ described in E14 (but not 
mentioned in R11), and explain our choices for our fits. E14 performs Cepheid fits with and without Gaussian priors centred at $b_W=-3.23$ and $Z_W=0$, motivated by expectations of what the slope and metallicity dependence should be. We test the same priors in our fits but ultimately decide to not include these different priors as one of the variables in our fit, for reasons which follow.\\
\\
Out of all the Cepheid data, the LMC Cepheids constrain the slope $b_W$ most tightly. Given the relative paucity of data on the Leavitt law, we always  include this information on the Leavitt law in all our fits, independent of whether the LMC is used as a distance anchor. For the fits where the LMC is not included as an anchor, we impose the same Gaussian prior on the slope as in E14: $\langle b_W\rangle = -3.23$, $\sigma_{b_W} = 0.10$. If the LMC is used as an anchor, there is already a contribution to the likelihood from these Cepheids, so it is inappropriate to reuse this information as a prior. Then, the inclusion of the prior on $b_W$ is prescribed 
by the distance anchor.
\\\\
The metallicity priors in E14 are motivated by the observed strong dependence of the Cepheids' period on metallicity, in contrast with expectations that $Z_W{\sim}0$, based on theoretical considerations and measurements in the LMC~\citep[][section~3.2, and references therein]{FreedmanMadore11, Efstathiou14}. However, the R11 sample of Cepheids demonstrates a strong metallicity dependence, with values of $Z_W$ around -0.3 or -0.5 for the R11 and E14 rejection algorithms respectively. The difference between values for $Z_W$ from the two approaches to outlier rejection (detailed in Appendix~\ref{sec:rejection}) can be traced to a sample of low metallicity Cepheids that are rejected by cuts in R11 but not E14. This systematic difference (discussed in E14, section~3.2) arose from the erroneous extrapolation of metallicity gradients to large radii, and was later corrected in R16. 
Including both the LMC and MW as distance anchors reduces the magnitude of the metallicity dependence $Z_W$. As we have observed that the R11 Cepheid data do not support the $Z_W{\sim}0$ priors (weak or strong) in E14, it is most appropriate to exclude these Gaussian priors in our analysis.

\section{Supernova lightcurves and data}

\subsection{SALT2 lightcurve fits}
\label{sec:SALT2}
For each supernova we use SALT2 to fit \snia~lightcurves (i.e.\ determine parameters $m_B, X_1, C$ in Equation~\ref{eq:snia}). The SALT2 model, based on its precursor 
SALT~\citep{Guy05}, is described in \cite{Guy07} along with details of its training. Two newer versions, SALT2.2 and SALT2.4, with additional training samples, have been released with the SNLS and JLA
analyses, respectively in \cite{Guy10} and \citet[][hereafter B14]{Betoule14}. Notably these include supernovae from the SDSS-II survey~\citep{Sako14} and high-$z$ SNe which have constrained the model better in the rest-frame ultraviolet region, eliminating the need to exclude {\em{U}}-band data as in R11.
\\
\\
Primary inputs for the SALT2 lightcurve-fitting routine \texttt{snfit} are photometry in each filter, heliocentric redshifts, and Milky Way extinction (obtained from \citet{SFD} dust maps) for each supernova. In addition, zero points and filter transmissions for each passband of each instrument are necessary. As CfA3 supernovae are included in the JLA analysis and the SALT2.4 release, we only need to create LOSS instruments KAIT1--4 and NICKEL. Adding these instruments involves including the filter transmissions provided by the Berkeley group\footnote{http://heracles.astro.berkeley.edu/sndb/info}, and determining zero points for BD+17$^\circ$4708\footnote{This is the fundamental SDSS standard star that the Vega-based magnitude system @VEGA2 in SALT2.4 is calibrated on.} using colour transformations in \citet[table~4]{Ganeshalingam10}. Our photometry and the instruments used are briefly mentioned in Sections~\ref{sec:SNobs} and \ref{sec:calibration}; for more details see the survey papers \cite{Hicken09a, Ganeshalingam10}.

\subsubsection{Photometry consistency checks}
\label{sec:consistency}
We are able to compare our SALT2 fit results to published values for a subsample of our \sneia, namely some of the CfA3 SNe which appear in SNLS~\citep[][hereafter C11]{Conley11} and JLA (B14)
. \cite{Hicken09b} and \cite{Ganeshalingam13} also report SALT2 fits of their samples, albeit with an older version of SALT2. We thoroughly check for consistency between these results and find agreement within quoted uncertainties, with no systematic differences.
\\\\
Furthermore we examine both the photometry and lightcurve fits for the 69 SNe in the CfA3-LOSS overlap, taking into account the different magnitude systems. For the directly comparable passbands (the Bessell-like {\em{BVRI}} filters) the photometry is consistent, while differences from the SDSS {\em{ri}} filters in Keplercam are in line with expectations. We also compare results of lightcurve fits using only CfA3 photometry, only LOSS photometry, and both combined in a single lightcurve. We find that SALT2 parameters $\{\alpha,\beta\}$, determined from each survey separately, generally average to the parameters determined from the combined fit (which lie well within reported uncertainties of from either CfA3 or LOSS). Occasionally one survey dominates in its influence over the SALT2 parameters; this occurs equally often with each survey and only when the lightcurve quality is discernibly superior in terms of number of points, sampling frequency, and size of uncertainties. Furthermore we test for systematic offsets in the three SALT2 fit parameters $m_B, X_1, C$, and find none. The comparison of combined lightcurves and their SALT2 fits supports the consistency of CfA3 and LOSS and favours neither over the other; thus we concatenate photometry from CfA3 and LOSS instruments to obtain the highest quality lightcurves available for these SNe.
\subsection{Malmquist bias correction}
\label{sec:malmquist}
In magnitude limited surveys, intrinsically brighter objects are preferentially detected, leading to Malmquist bias: a skewed estimate of the absolute magnitude distribution. The Malmquist bias can be estimated by modelling the selection efficiency (i.e.\ the rate of successful detection as a function of magnitude) to match observed distributions (in redshift, stretch, and colour), then simulating the survey with SNANA~\citep{SNANA} to obtain the bias $\delta \mu$ in distance modulus. 
This procedure, described in e.g.\ \citet{ScolnicKessler16}, is outside the scope of this work. Thus we adopt the estimate of the bias (for low-$z$ supernovae) in \citet[][section~5.3]{Betoule14}, which adopts a magnitude limited selection function, and uses the difference between the resultant bias and an unbiased regime as the uncertainty in the correction (the covariance matrix $\mathbf{C_{\rm{bias}}}$ in Appendix~\ref{sec:covmat}). The targeted discovery of supernovae in CfA3 and LOSS means they should not be magnitude limited; however, as observed in the JLA low-$z$ sample, the colour distribution grows more blue with redshift, suggesting that some selection effect is at play. Using the JLA approximation is justified as our supernova sample is similarly distributed to the low-$z$ sample in JLA, and the bias correction is inherently approximate and has a miniscule impact on $H_0$.\footnote{The difference between correcting for Malmquist bias and no correction is less than 0.01 in $\mathcal{H}$, or ${\sim}0.4\%$ in $H_0$.}

\subsection{Host galaxy dependence}
\label{sec:hostcorrection}

\begin{figure}
\centering
\includegraphics[width=0.5\textwidth]{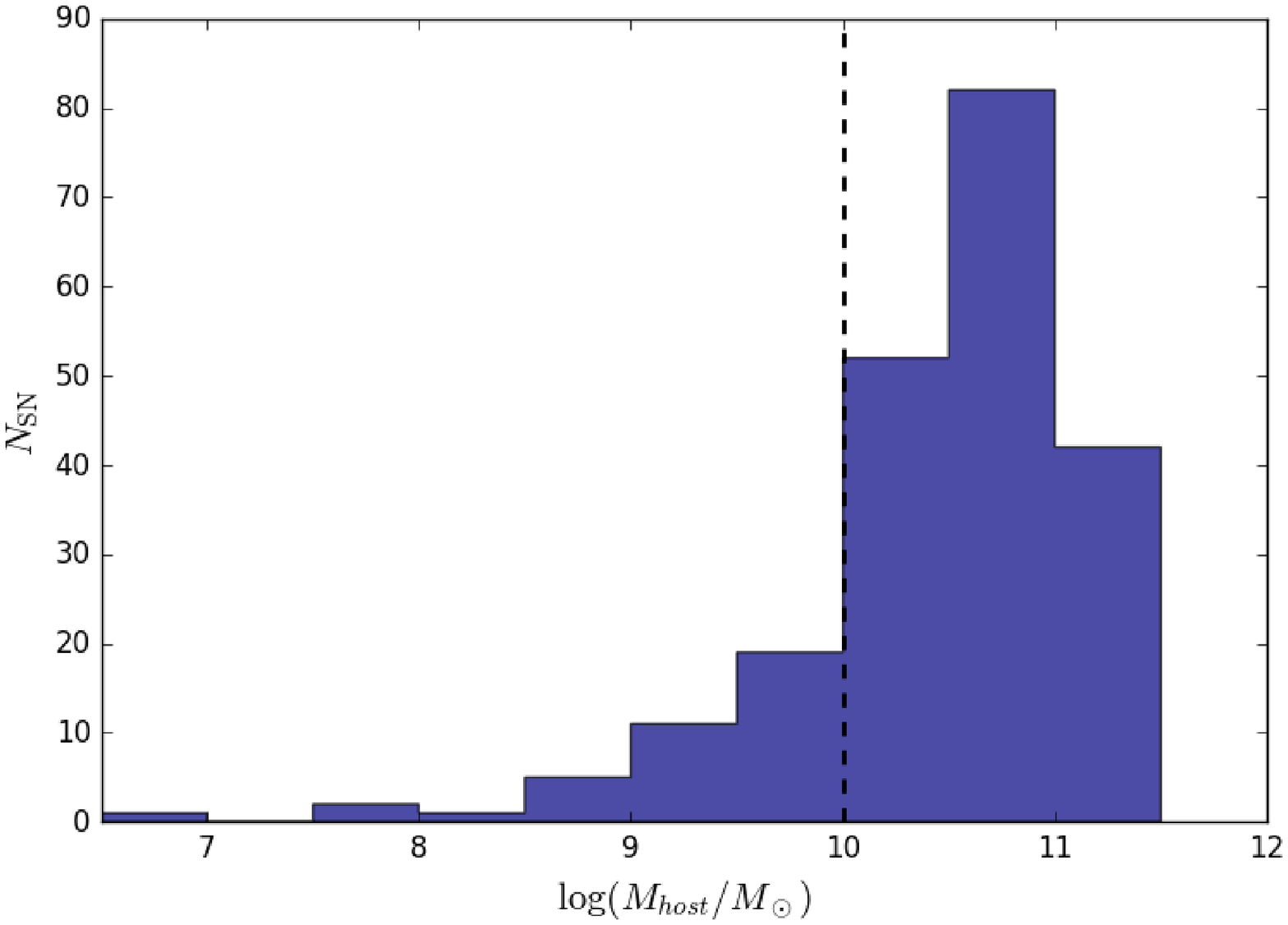} 
\caption{Histogram of host galaxy masses of low-$z$ \sneia~where available (total off 220 SNe). The dashed line indicates the boundary $10^{10} M_\odot$ which splits the absolute \snia~magnitude (Equation~\ref{eq:masssplit}).}
\label{fig:hosthist}
\end{figure}

The dependence of the intrinsic \snia~brightness on properties of their host galaxies is well established, with numerous studies finding that more massive galaxies (correlated with higher metallicity and lower specific star formation rates) host SNe which are on average ${\sim}0.08$ mag brighter \citep[e.g.][]{Sullivan10,Lampeitl10,Kelly10}. To mitigate the systematic error that this effect introduces to the cosmological analysis, we follow \cite{Sullivan11} and subsequent analyses (C11, B14) in adopting two discrete values for the \snia~absolute magnitude, using the variable
\begin{align}
\label{eq:masssplit}
M_B^* := 
\begin{cases}
M_B ,&\text{host galaxy mass} <10^{10} M_\odot \\
M_B + \Delta M_B,&\text{host galaxy mass} > 10^{10} M_\odot.
\end{cases}
\end{align}
We fit for the parameter $M_B$ as indicated in Section~\ref{sec:data}, and fix the offset $\Delta M_B$ based on analyses in C11; B14; \cite{Sullivan10, Childress13}, which determine $\Delta M_B = -0.08$ from \snia~samples greater in size and redshift range than ours. We consider fitting for the magnitude offset using our data, and find a larger absolute difference $\Delta M_B = -0.15 \pm 0.07$ (with some degeneracy with $\mathcal{M}$), which is still consistent with the established value in the literature. Given the large uncertainty on our value we adopt the more reliable reference value.
\\
\\
The host galaxy masses for our SNe are obtained from the literature where available, with 77 from JLA and 71 from a combination of \cite{Sako14,Childress13,Neill09,Kelly10}. The stellar masses of nearby galaxies are all given in \cite{Neill09}. We were able to derive mass estimates for 72 of the remaining galaxies using SDSS photometry, following standard methods in \cite{Sullivan06, Smith12}. We refer the reader to descriptions therein of the method, which relies on the \texttt{ZPEG} photometric redshift code~\citep{ZPEG02, ZPEG10} based on spectral energy densities from the P\'{E}GASE.2 spectral synthesis code. Where possible, we check for consistency between multiple sources and our estimates. The distribution of the host masses of the CfA3 and LOSS \sneia~(Fig.~\ref{fig:hosthist}) clearly shows that they predominantly exist in more massive galaxies, with 241 out of 280 SNe lying in the high mass bin. This is a consequence of the targeted nature of these surveys, in contrast to the magnitude-limited surveys SNLS and SDSS in JLA. Thus we assign the remaining 60 SNe with unknown masses to the high mass bin with a large associated uncertainty, even though unknown hosts in JLA are assigned to the low mass bin (B14, section 5.2). The propagation of uncertainties in this correction through to SN parameters is later described in Appendix~\ref{sec:host}.

\subsection{Peculiar velocity corrections}
\label{sec:velcorrection}
Peculiar velocities arise from motion other than from cosmological expansion, such as dipole or bulk flows, local galaxy infall, and higher order coherent flows. These perturb the observed redshifts via the Doppler effect,\footnote{A supernova's peculiar motion changes not only its redshift but also its observed luminosity \citep{HuiGreene06, Davis11} as it experiences relativistic beaming. This in turn induces a deviation in the supernova's peak magnitude; however this is approximately an order of magnitude smaller than the change in redshift.} and can impact cosmological analyses. \cite{HuiGreene06} show that neglecting correlations between peculiar velocity uncertainties at low-redshift results in a greatly underestimated zero point uncertainty, and degrades the precision of the dark energy equation of state parameter $w$. Moreover, correlated SN peculiar velocities can bias cosmological results: \cite{Davis11} show that neglecting coherent flows results in a shift of $\Delta w = 0.02$.
\\
\\
Thus, an effort to quantify the uncertainty induced by correlated motions is an essential part of any modern \snia~cosmological analysis. Approaches to this include the addition of large ($300-400$~\kms) uncertainties in redshifts to account for peculiar velocities \citep{Ganeshalingam13,Hicken09b}, and attempts to correct for peculiar velocities. The latter first appeared in SNLS (C11), which corrects redshifts on a supernova by supernova basis for the (line-of-sight) peculiar velocity at the location of the SN, as determined from a velocity field. 
\\
\\
C11 uses the velocity field in \cite{Hudson04}, derived from the galaxy density field from the IRAS PSCz redshift survey~\citep{Branchini99}. While B14 adopts the same correction, we apply the same method using updated density and velocity fields from the 2M++ redshift compilation.\footnote{Data available at http://cosmicflows.iap.fr/.} In each case the velocity field is derived from the respective density fields under a linear biasing approximation.\footnote{That is, the mass density and galaxy number density are proportional via the linear bias factor $b$, i.e.\ $\delta_g = b \delta$. In this regime, peculiar velocities are proportional to gravitational attraction:
\begin{align}
\boldsymbol{v} = \frac{\beta^*}{4\pi}\int_0^{R_{\rm{max}}} \delta_g(\boldsymbol{r}')\frac{(\boldsymbol{r}'-\boldsymbol{r})}{|\boldsymbol{r}'-\boldsymbol{r}|^3}\ d^3\boldsymbol{r}' + \boldsymbol{U},\label{eq:velfield}
\end{align}
where $\boldsymbol{U}$ represents a residual dipole (in 2M++ this is the dipole of the Local Group), with $\beta^* = \frac{f(\Omega_M)}{b} = 0.43$~\citep{Carrick15} where $f(\Omega_M) =\Omega_M^{0.55}$ for $\Lambda$CDM~\citep{WangSteinhardt98}.} 
\\\\
In correcting supernova redshifts for peculiar velocities the aim is to isolate the redshift due purely to expansion. The several redshifts of interest are: the heliocentric redshift $z_h$, the CMB frame redshift $z_{\rm{cmb}}$, and the cosmological redshift $\bar{z}$. The latter two differ in that $\bar{z}$ is corrected for peculiar velocities from coherent flows; it is intended to reflect a velocity derived only from the expansion of space and therefore this is the redshift that should be used in $v(z)$ in Hubble's law. The peculiar motions to consider in converting $z_h$ to $\bar{z}$ are the motions of the solar system ($\boldsymbol{v}_\odot^{\rm{pec}}$), and of the SN ($\boldsymbol{v}_{\rm{SN}}^{\rm{pec}}$), both relative to the CMB. Many SNe at low redshifts share some of the Local Group's motion; by converting to a heliocentric frame (i.e.\ correcting for the Sun's motion relative to the CMB) we are also overcorrecting for the motion of nearby SNe, necessitating the second correction. For a SN at position $\boldsymbol{n}$ from the Sun, these redshifts and velocity are related by\footnote{The minus sign in front of $\boldsymbol{v}_{\odot}^{\rm{pec}}$ arises because we have defined it is the motion of the Sun relative to the CMB, rather than the other way around.} \citep{Davis11}
\begin{align}
1 + z_h &= (1+\bar{z})(1+z_{\odot}^{\rm{pec}})(1+z_{\rm{SN}}^{\rm{pec}})\notag\\
&\approx (1+\bar{z})(1-\boldsymbol{v}_{\odot}^{\rm{pec}}\cdot \boldsymbol{n}/c+\boldsymbol{v}_{\rm{SN}}^{\rm{pec}}\cdot\boldsymbol{n}/c).\label{eq:velcorr}
\end{align}
For our analysis of the low-$z$ \sneia, we use $\bar{z}$ derived in this way as the CMB-frame redshift. Unless otherwise specified, this is the redshift meant by $z$.
\\
\\
The exact form of the luminosity distance introduced in Equation~\ref{eq:DL} actually requires both the heliocentric and cosmological redshifts:
\begin{align}
  \label{eq:DLprecise}
  D_L(z_h, \bar{z})= (1+z_h) D(\bar{z}).
\end{align}
This is because the factors affecting the $(1+z_h)$ prefactor (redshifting and beaming) depend on the total relative velocities, whereas the cosmological distance only depends on $\bar{z}$, the redshift due to expansion~\citep[][Appendix A]{Calcino16}. The difference resulting from using $\bar{z}$ for both is negligible so we do not differentiate in our analysis. \citet[][Section~4.2--4.3]{Calcino16} quantify the effect of possible redshift systematic errors on the derivation of $H_0$ and find that a systematic redshift error as small as ${\sim}2.6\times 10^{-4}$ can result in a ${\sim}0.3$\% bias in $H_0$. 
\\
\\
The peculiar velocity corrections we make here are reliant on predicted velocity fields, which are intrinsically approximate. We discuss and quantify these uncertainties in Appendix~\ref{sec:pecvel}, and propagate them to the SN fit parameters. Moreover, we ensure that our corrections do not bias our results: if the results of our SN-only fit (Section~\ref{sec:snfit}) varied significantly with the introduction of the correction, then this effect would need to be explored and quantified. In this scenario, the impact of performing a velocity correction would greater than the uncertainty contribution in $\mathbf{C_{\rm{pecvel}}}$ (Appendix~\ref{sec:pecvel}). However, we find a negligible effect on the zero point $\mathcal{M}$ (less than 10\% of the statistical uncertainty) in the SN-only fit when peculiar velocity corrections are omitted. Consequently the velocity correction cannot bias $H_0$ (as $\mathcal{H} = M_B - \mathcal{M}$).

\subsection{Histograms for SN cuts}
\label{sec:histcut}
\begin{figure}
    \includegraphics[width=0.46\textwidth]{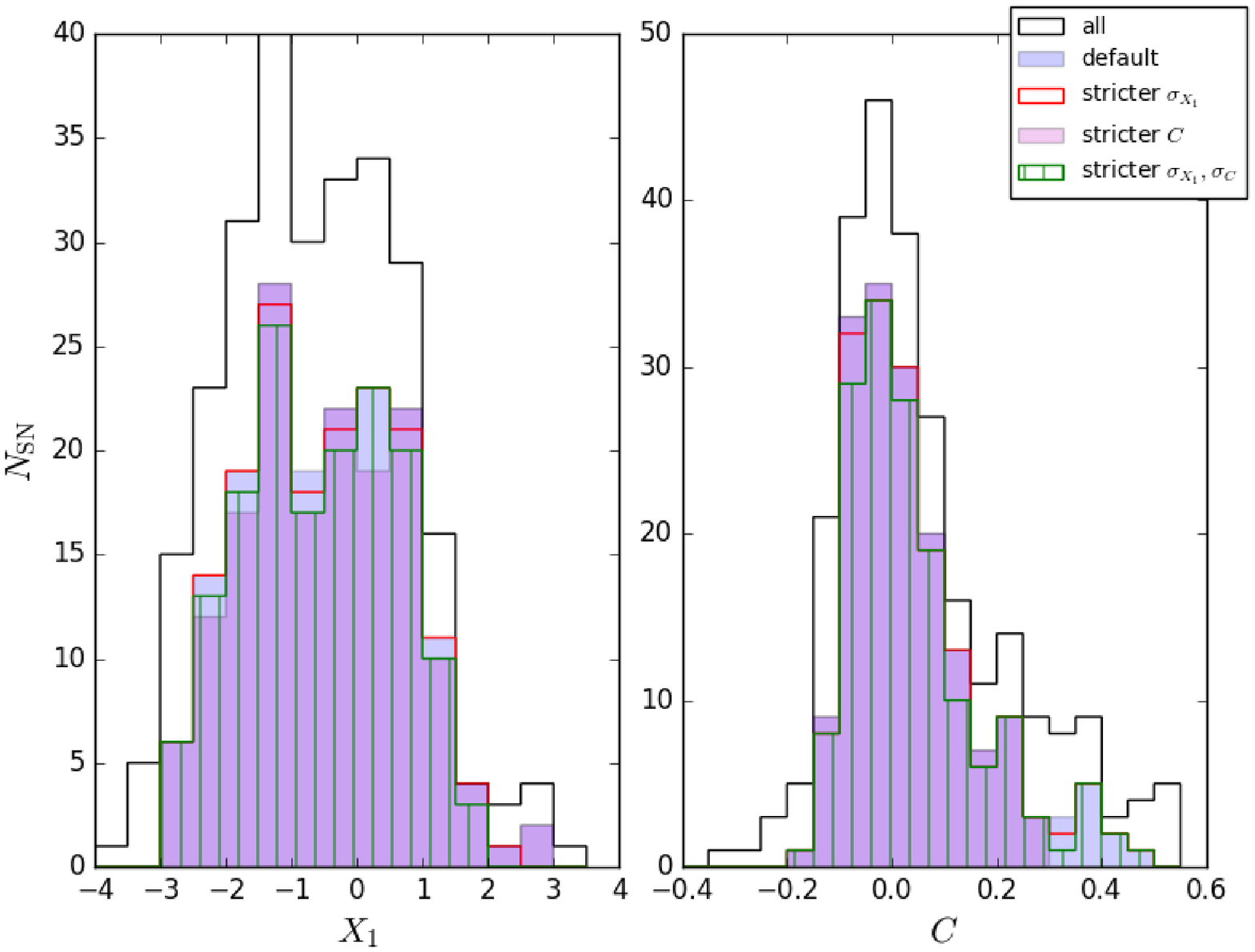}
    \caption{Histograms of the $X_1,C$ distributions with alternate cuts on their values and uncertainties (Section~\ref{sec:SNcuts}). These show that constraints on uncertainties in $X_1$ and $C$ remove the slowest-declining SNe, and that imposing a stricter cut on the colour affects the $C$ distribution asymmetrically. 
    }
    \label{fig:X1Chist}
    \includegraphics[width=0.46\textwidth]{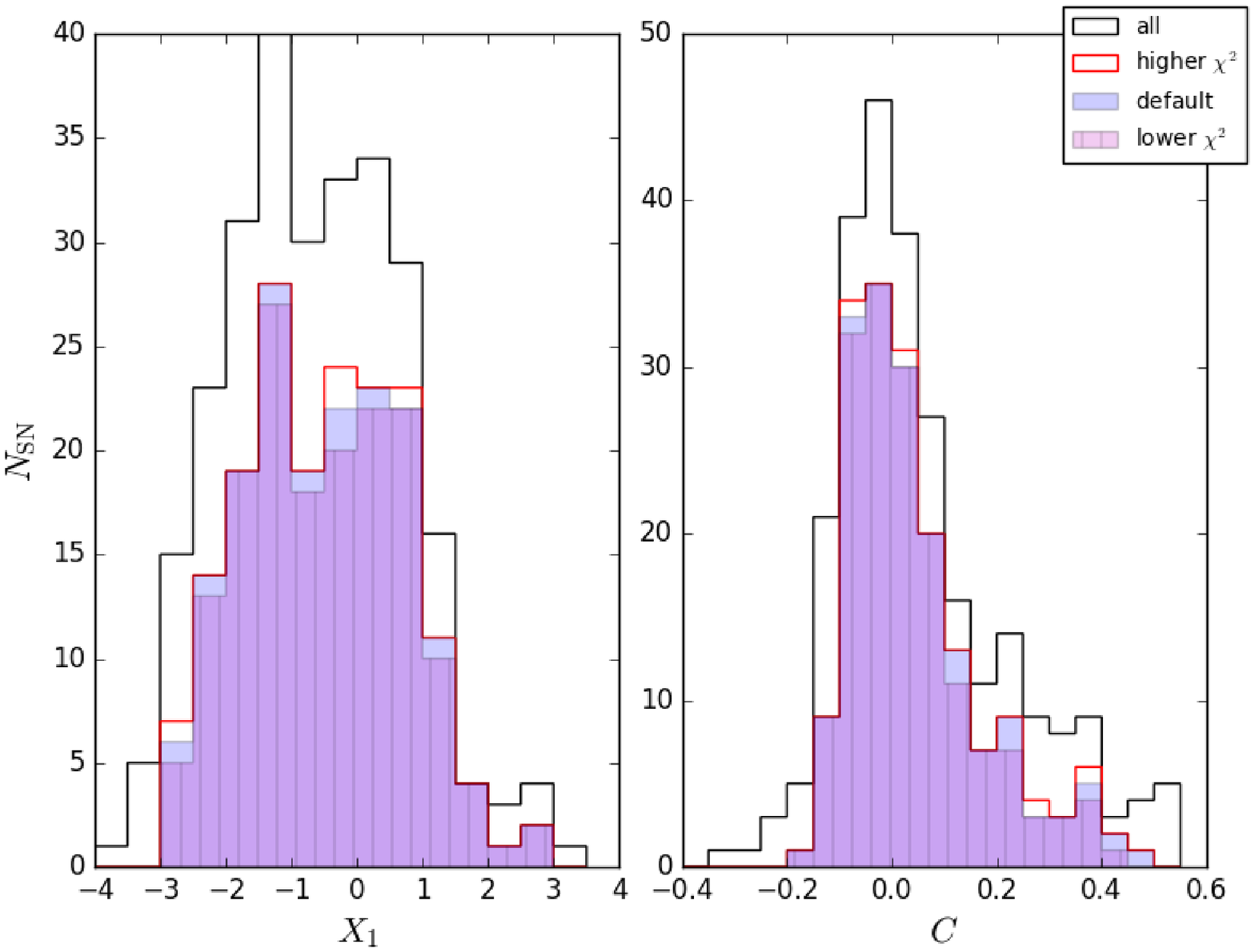}
    \caption{Histograms of the $X_1,C$ distributions with alternate cuts on the lightcurve goodness-of-fit $\chi^2/\rm{DoF}$ (Section~\ref{sec:SNcuts}). The \snia~distributions with these cuts and the default cut appear similar.}
    \label{fig:chisqhist}
   \includegraphics[width=0.46\textwidth]{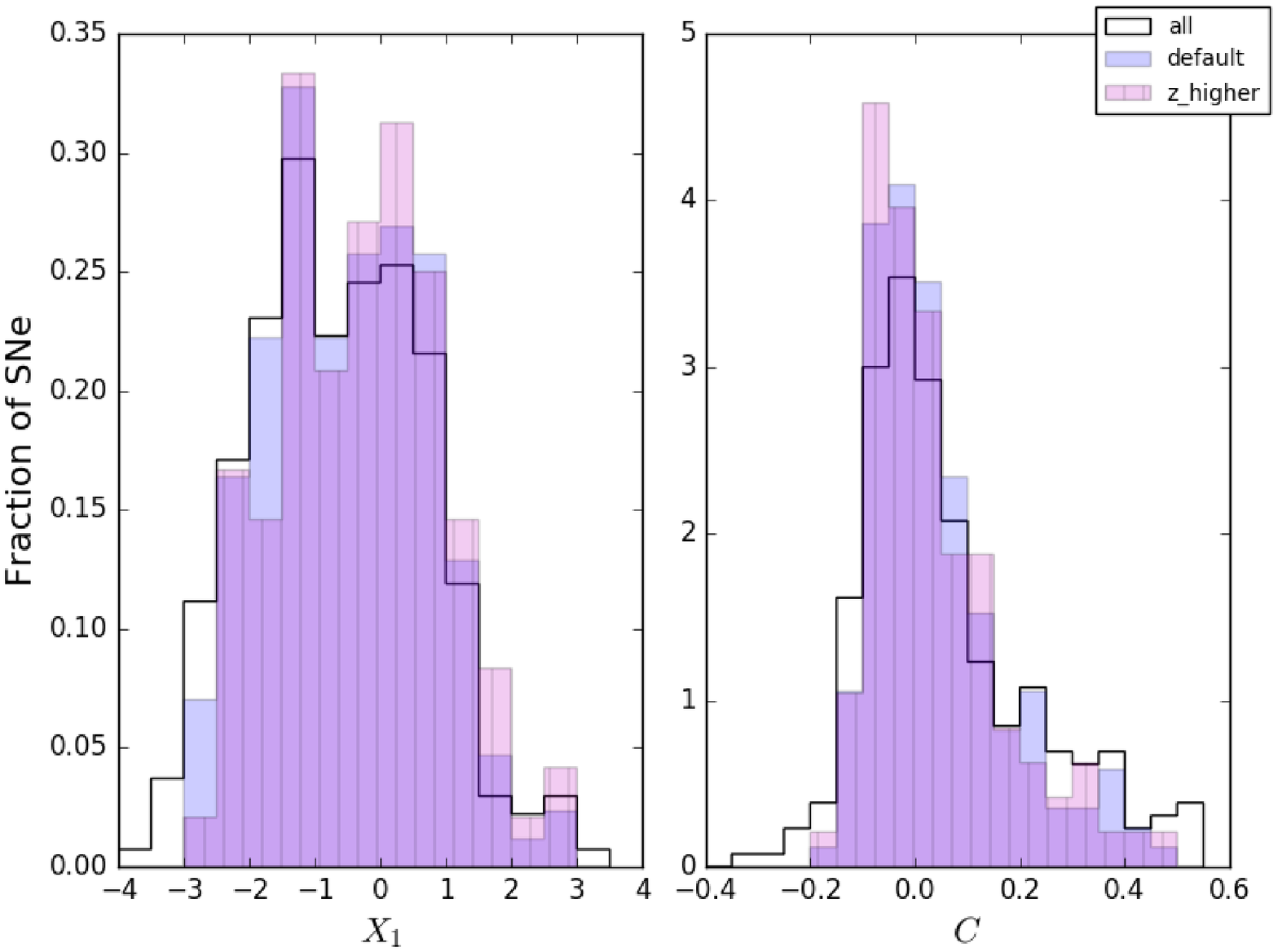}
    \caption{Normalised histograms of the $X_1,C$ distributions with a higher low-redshift cut of $z>0.0233$. The proportion of higher $X_1$ (slower-declining) supernovae is marginally higher with the $z>0.0233$ cut, but otherwise the relative distributions appear very similar.}
    \label{fig:zhigherhist}
    \centering
\end{figure}
We include in Figs~\ref{fig:X1Chist}, \ref{fig:chisqhist}, and \ref{fig:zhigherhist} the stretch and colour distributions of our low-$z$ \sneia~with various alternate cuts relative to the default, all described in Section~\ref{sec:SNcuts}. We observe that the $X_1$ distribution changes marginally with stricter cuts on $\sigma_{X_1}$ and a higher low-redshift cut. Naturally, the $C$ distribution is affected by a stricter cut on $C$. Otherwise, there is no significant impact on the stretch and colour distributions from alternate cuts, and in particular no evidence that the variation observed in Fig.~\ref{fig:alphabetaSN} in Section~\ref{sec:SNresults} arises from such biases.

\newpage

\section{Computing supernova systematics}

\label{sec:systematics}
This section is devoted to the construction of the \snia~covariance matrices. We break down systematic and statistical terms that contribute to the error budget, and describe their propagation to the supernova magnitudes.
\subsection{Overview of covariance matrices}
\label{sec:covmat}

Each covariance matrix tracks uncertainties in the vector $\eta = \{m_{Bi}, X_{1i}, C_i\}_{1 \leq i \leq N}$, which contain the SALT2 quantities for all $N=280$ low-$z$ SNe. These matrices sum to the $3N\times 3N$ matrix $\mathbf{C_\eta}$ which encompasses all covariances in $\eta$, and are independent of $\alpha$ and $\beta$. For fitting the low-$z$ SN data we require covariances in supernova magnitudes in the form of $\mathbf{C_{\mBcor}}$ (Equation~\ref{eq:chisqlowz}), which is derived from $\mathbf{C_\eta}$ by conjugation with the $N\times 3N$ matrix $\mathbf{A}$ (with $\mathbf{A}_{ij} = \delta_{3i,j} + \alpha\delta_{3i+1,j} + \beta\delta_{3i+2,j}$):
\begin{align}
\label{eq:C_mBcor}
\mathbf{C}_{\mBcor} &= \mathbf{A\cdot C_\eta\cdot A}^T + \diag\left(\frac{5\sigma_z}{z\log10}\right)^2\\ &+ \diag(\sigma_{\rm{lens}}^2) + \diag(\sigmaintSN^2).
\end{align}
The remaining terms are diagonal uncertainties, which affect each SN individually, ascribed to uncorrelated uncertainties in redshift due to peculiar velocity uncertainties (distinct from the uncertainty in their corrections, described in Appendix~\ref{sec:pecvel}), and perturbances in \snia~magnitudes caused by gravitational lensing and intrinsic scatter. We adopt the values for these used in C11 and B14, of $c\sigma_z =$~150~\kms, $\sigma_{\rm{lens}} = 0.055z$, and $\sigmaintSN = 0.12$.\\
\\
To understand $\mathbf{C_\eta}$, we first separate it into statistical and systematic components, and later explain the distinction in Appendix~\ref{sec:stat}. The contributions to $\mathbf{C_{\rm{sys}}}$ we consider are from uncertainties in the following sources: peculiar velocity corrections, Milky Way extinction, host galaxy mass dependence, photometric calibration, Malmquist bias correction, and lightcurve model.\footnote{This is equation~11 of B14 without the contamination term $\mathbf{C_{\rm{nonIa}}}$, which only concerns higher redshift SNe.} For the nearby SNe (i.e.\ those in galaxies containing Cepheids), we only include 
$\mathbf{C_{\rm{diag}}}$ and $\mathbf{C_{\rm{host}}}$. The host mass correction has the potential to shift the magnitude scale by up to ${\sim}0.08$~mag, and is important in the context of the dependence of \snia~magnitude on host galaxy properties (Appendix~\ref{sec:host}). The other correction terms, for Malmquist bias and peculiar velocities, are redshift-dependent effects hence irrelevant for this sample. The remaining covariance matrices are not tied to the corrections in Appendices~\ref{sec:malmquist}--\ref{sec:velcorrection}, and are more precise than warranted, given the inhomogeneity and larger uncertainties in these data, so we neglect them. 
\begin{align}
\mathbf{C_\eta} &= \mathbf{C_{\rm{stat}} + C_{\rm{sys}}};\notag\\
\mathbf{C_{\rm{sys}}} &=  \mathbf{C_{\rm{bias}}  +  C_{\rm{cal}} + C_{\rm{dust}} + C_{\rm{host}}+ C_{\rm{model}}+ C_{\rm{pecvel}}}.\label{eq:covmats}
\end{align}
We follow standard techniques to compute each covariance matrix, which is to enfold partial derivatives of SN parameters with respect to each systematic, with the typical size of systematics:
\begin{align}
\label{eq:partial}
\mathbf{C_{\rm{sys}}}_{ij} = \sum_{k} \left(\frac{\partial \eta_i}{\partial k}\right)\left(\frac{\partial \eta_j}{\partial k}\right)(\Delta k)^2.
\end{align}
Here the sum is over all systematics $k$, each of size $\Delta k$. Equation~\ref{eq:partial} is applied directly to compute $\mathbf{C_{\rm{dust}}}$ and $\mathbf{C_{\rm{cal}}}$. These calculations are intrinsically approximate, yet even as estimates they are invaluable for gauging the contribution of each systematic term affecting \sneia, and affirming that we sufficiently account for each effect. Section~\ref{sec:systresults} presents our assessment of these uncertainties. 
\\\\
We first digress to discuss the statistical term $\mathbf{C_{\rm{stat}}}$, then address the construction of each systematic in turn. We describe our calculations of the first four systematics from first principles. Computations of the bias and model uncertainties, as well as the non-diagonal part of $\mathbf{C_{\rm{stat}}}$ require estimates of the sample's selection function (as discussed in Section~\ref{sec:malmquist}), thorough end-to-end simulations with several lightcurve models, and in-depth deconstruction and analysis of the SALT2 model. These have been performed in section~5.3 of B14, \cite{Mosher14}, and \citet[Appendix~A3][]{Guy10} respectively. We obtain our best estimate of these matrices for our low-$z$ \snia~sample, and refer the reader to the aforementioned references for details.

\subsection{Statistical uncertainties}
\label{sec:stat}

The distinction between statistical and systematic errors blurs, as many uncertainties have sources for which both descriptors are appropriate. We adopt the separation used in C11, which defines statistical uncertainties as those that can be reduced by increasing the size of some data set. In this case the data sets are the measured low-$z$ supernovae, and the training set used to define the SALT2 parameters (\cite{Guy10}, updated in B14). We separate these two terms into matrices $\mathbf{C_{\rm{stat,diag}}}$ and $\mathbf{C_{\rm{stat,model}}}$ respectively.\footnote{In B14 these two terms are combined as $\mathbf{C_{\rm{stat}}}$, while C11 sums $\mathbf{C_{\rm{stat,diag}}}$ and the three diagonal terms in our Equation~\ref{eq:C_mBcor} to their $\mathbf{D_{\rm{stat}}}$.} 
\\
\\
The matrix $\mathbf{C_{\rm{stat,diag}}}$ arises from uncertainties in the measurement of lightcurves, constructed directly from correlated uncertainties in $m_B, X_1, C$ (a $3\times 3$ covariance matrix for each SN) reported in SALT2 outputs. These terms are uncorrelated between different supernovae, so $\mathbf{C_{\rm{stat,diag}}}$ is tridiagonal (i.e.\ only a diagonal strip of width 3 is nonzero).
\\
\\
The latter matrix $\mathbf{C_{\rm{stat,model}}}$ contains the uncertainty in the SALT2 model from the finiteness of the training sample, which could theoretically be decreased by training SALT2 on more supernovae. Its calculation requires propagating statistical uncertainties in the lightcurve model through to supernova fit parameters $\eta$, as described in~\citet[Appendix~A3][]{Guy10} and implemented in the \texttt{snpca} package.\footnote{Private SNLS communication.} We use the code \texttt{salt2\_stat} from this package to directly compute $\mathbf{C_{\rm{stat,model}}}$.

\subsection{Milky Way extinction}
\label{sec:extinction}
The calculation of our first systematic covariance matrix directly follows Equation~\ref{eq:partial}. This contains the uncertainty in $\eta$ due to the uncertainty in the Milky Way extinction. The sole systematic $k$ is the uncertainty in the $E(B-V)$ value from dust maps~\citep{SFD}; we follow the conservative estimate in B14 of a 20\% relative uncertainty. Perturbing the value of the extinction (encoded in the dust keyword @MWEBV in SALT2 inputs) and refitting lightcurves give the partial derivatives in Equation~\ref{eq:partialdust}:
\begin{align}
\label{eq:partialdust}
\mathbf{C_{\rm{dust}}}_{ij} = \left(\frac{\partial \eta_i}{\partial E(B-V)}\right)\left(\frac{\partial \eta_j}{\partial E(B-V)}\right)(0.2\times E(B-V))^2.
\end{align}
We verify that the partial derivatives of SN parameters $\eta$ with respect to Milky Way extinction are independent of the size of perturbation over a satisfactory range, and that our resultant $\mathbf{C_{\rm{dust}}}$ is identical to the same matrix reported in B14 for the 60 \sneia~in common.

\subsection{Calibration}
\label{sec:calibration}
B14 and C11 emphasize the significant contribution of uncertainties in the calibration of individual surveys to the total error budget. We follow the methods therein and in \cite{Betoule13} to reproduce the calibration covariance matrix relevant to our SN sample and the instruments used to observe them. Computing the calibration matrix $\mathbf{C_{\rm{cal}}}$ relies on the same principle as in Appendix~\ref{sec:extinction}, but over multiple systematics. Calibration uncertainties are grouped into two types of systematics: uncertainties in the magnitude zero point (shifting the overall flux scale) and in the effective wavelength (shifting the transmission function in wavelength space), for each filter. These are contained in the vector $\kappa$ and enumerated in Table~\ref{tab:kappainst}.\\
\begin{table}
  \begin{center}
  \caption{Systematics in $\kappa$
    \label{tab:kappainst}}
  \begin{tabular}{llll}
\hline
\hline
Instrument& Filters& ZP index& $\lambda_{\rm{eff}}$ index\\
\hline
MegaCam & {\em{griz}} &0-3 & 50-53\\
Standard & {\em{UBVRI}} & 4-8 & 54-58\\
KeplerCam & {\em{UsBVri}}$^a$ & 9-13 & 59-63\\ 
4Shooter2 & {\em{UsBVRI}} & 14-18 & 64-68 \\ 
Swope & {\em{ugriBV}} & 19-24 & 69-74\\
SDSS & {\em{ugriz}} & 25-29 & 75-79\\
KAIT1-4 & {\em{BVRI}}  & 30-45 & 80-95\\
NICKEL & {\em{BVRI}} & 46-49 & 96-99\\
\hline
  \end{tabular}
  \end{center}
  $^a$ {\em{Us}} indicates the standard Landolt {\em{U}} passband, derived from \cite{Bessell90} (see Section~\ref{sec:SNobs}).\\
\end{table}

\\
The instruments and passbands to consider in $\kappa$ are those used for observing the low-$z$ SNe: 4Shooter2 and Keplercam for CfA3, and KAIT1--4 and Nickel for LOSS, and those involved in the training of SALT2 (i.e.\ used to observe the SNe in the training sample). The latter, and sizes of systematics in these passbands, are given in B14,~table~5. It is necessary to include these training instruments and passbands as they influence measured magnitudes of training SNe hence the SALT2 model.
\\\\
We directly consider the covariance matrix of calibration systematics $\mathbf{C_{\kappa ij}} = \langle \sigma_{\kappa_i}\sigma_{\kappa_j}\rangle$, which captures the correlations between systematics in different instruments and passbands. Then Equation~\ref{eq:partial} is equivalent to $\mathbf{C_{\rm{cal}}} = \mathbf{J \cdot C_\kappa\cdot J}^T$ where $\mathbf{J}_{ij} = \frac{\partial \eta_i}{\partial \kappa_j}$ is the Jacobian matrix, encoding partial derivatives of SN parameters. Then finding $\mathbf{C_{\rm{cal}}}$ amounts to constructing $\mathbf{C_\kappa}$, and calculating $\mathbf{J}$ from first principles. For the LOSS instruments we achieve the latter by either perturbing an element of $\kappa$ (i.e.\ shift the zero point of effective wavelength). For the other instruments, which were involved in training SALT2, we change to a different SALT2 model altogether i.e.\ one that was trained with the systematic shift in question applied. Each SN lightcurve is fitted again to find the difference, and the resultant Jacobian is smoothed in accordance with footnote~9 of B14.
\\\\
To find $\mathbf{C_\kappa}$ we start with the same matrix from JLA and reindex it according to Table~\ref{tab:kappainst}, appending the LOSS instruments and removing HST instruments NICMOS and ACSWF (which do not contribute to the SALT2 training sample). We approximate the elements of $\mathbf{C_\kappa}$ for LOSS instruments as diagonal: this is exact for the $\lambda_{\rm{eff}}$ elements, and a good approximation for the zero point. Using \cite{Ganeshalingam10} as a guide, we take the zero point and $\lambda_{\rm{eff}}$ uncertainties to be 0.03~mag and 10~\AA~respectively. As LOSS observations (with KAIT1--4 and Nickel) of \sneia~were not used for SALT2 training, only SNe in the sample with LOSS measurements have nonzero partial derivatives with respect to these instruments.

\subsection{Host galaxy mass}
\label{sec:host}
The \snia~magnitude zero point $M_B$ is corrected for the magnitude offset $\Delta_M$ between high and low host galaxy (stellar) mass bins, as described in Appendix~\ref{sec:hostcorrection}. The uncertainty in this correction is propagated to SN parameters in $\mathbf{C_{\rm{host}}}$. As in B14, we treat the systematic associated with this correction as having two components: from potentially having attributed an individual supernova to the wrong host mass bin, and from the arbitrariness of the $10^{10} M_\odot$ cut. Both effects are discrete, so the computation of $\mathbf{C_{\rm{host}}}$ differs from those of $\mathbf{C_{\rm{dust}}}$ and $\mathbf{C_{\rm{cal}}}$ which take partial derivatives with respect to continuous quantities.
\\
\\
Our calculation follows identically the method in B14. As discussed in Appendix~\ref{sec:hostcorrection}, our data do not justify fitting for $\Delta_M$, and instead we adopt a fixed value from literature. Only supernova magnitudes are affected: the only indices of nonzero components of $\mathbf{C_{\rm{host}}}$ correspond to the $m_B$ components of $\eta$, but for compatibility for the other matrices we make $\mathbf{C_{\rm{host}}}$ the same size ($3N\times 3N$). The vectors $\boldsymbol{H_h}$ and $\boldsymbol{H_l}$ are indicator functions of (the magnitudes of) the SNe with masses within an order of magnitude higher and lower respectively than the mass boundary $10^{10} M_\odot$, while $\boldsymbol{B}$ is and indicator function for the SNe whose host mass estimates and uncertainties in combination imply that they could be be assigned to either bin. Then the covariance matrix for the host mass correction is: 
\begin{align}
\label{eq:host}
\mathbf{C_{\rm{host}}} = \Delta_M^2(\boldsymbol{H_hH_h^T} + \boldsymbol{H_lH_l^T} + \diag(\boldsymbol{B})).
\end{align}
For the overlap of our SNe with B14, results are very similar, with differences arising only from SNe with masses newly obtained or updated (Appendix~\ref{sec:hostcorrection}).

\subsection{Peculiar velocities}
\label{sec:pecvel}
As described in Appendix~\ref{sec:velcorrection} we have corrected individual SN redshifts for peculiar motion, using the 2M++ velocity field corrections. However, there is intrinsic uncertainty in these models, with variation between velocity fields generated from different galaxy density fields, and in some cases limited agreement between predicted and measured velocities~\citep{Springob14, Scrimgeour16}. \\
\\
Thus the significant contribution in the correction model itself must be taken into account. Below, we adopt the approach in C11 and B14, which is to use the uncertainty in the velocity field to inform $\mathbf{C_{\rm{pecvel}}}$, the contribution to $\mathbf{C_\eta}$ from peculiar velocities. We emphasize that this is distinct from the diagonal term $\sigma_z$ in Equation~\ref{eq:C_mBcor}.
\\
\\
For a given density field $\delta_g$, the velocity field derived through Equation~\ref{eq:velfield} can be parametrized by $\beta^*$, the ratio of the growth rate of density perturbations to the linear bias factor. In C11, $\beta^*$ is the systematic which encompasses the uncertainty in the peculiar velocity model; that is, $\mathbf{C_{\rm{pecvel}}}$ is derived through Equation~\ref{eq:partial} with $k = \beta^*$. As this treatment of uncertainty lies within only one density field and model (that is, it doesn't account fully for velocities derived from different realisations/measurements of galaxy densities) we are conservative in using it; like C11 we perturb $\beta^*$ by five times its uncertainty.\footnote{\cite{PikeHudson05} find $\beta^* = 0.49\pm 0.04$ so C11 vary $\beta^*$ between 0.3 and 0.7.} Likewise we adopt $\beta^* = 0.43 \pm 0.02$~\citep{Carrick15} in the correction. To compute $\mathbf{C_{\rm{pecvel}}}$, we measure the shift in $z_{\rm{cmb}}$ when $\beta^*$ is set to $0.33$ or $0.53$ instead. The resultant difference in $z_{\rm{cmb}}$ is propagated to $m_B$ using the derivative of Equation~\ref{eq:sn1}:
\begin{align}
\sigma_{m_B} &= \frac{5}{\log 10} \left(\frac{1}{z} + \frac{f'(z)}{f(z)}\right)\sigma_{z_{\rm{cmb}}}.
\end{align} 
This has no impact on the stretch and colour of SNe, so only the $m_B$ elements of $\mathbf{C_\eta}$ have non-zero entries from $\mathbf{C_{\rm{pecvel}}}$.

\pagebreak 

\onecolumn 

\section{Full tables of fit results} 
This appendix supplements Sections~\ref{sec:cepheids} and \ref{sec:results} with full tables of the Cepheid and global fits, and the averaged results for $\{\Delta\mu_i\}$ from the global fit.
\subsection{Results of Cepheid-only fit}
\label{app:cepheid}

\begin{deluxetable}{llllllll}
\tablecaption{Results of the Cepheid-only fits described in Section~\ref{sec:cepheidfit} from each combination of distance anchor, rejection and period cut for each Cepheid fit. The best fit Cepheid parameters $\{b_W, Z_W, M_W\}$ are given, as well as the number of Cepheids remaining after rejection and intrinsic scatter. 
\label{tab:cepheidall}}
\tablewidth{0pt}
\tabletypesize{\small}
\tablehead{
  \colhead{Rejection}&
    \colhead{Distance anchor}&
  \colhead{$P<60$d}&
\colhead{$N_{\rm{cepheids}}$}&
  \colhead{$\sigmaintC$} &
\colhead{$b_W$}&
\colhead{$Z_W$}& 
\colhead{$M_W$}
}
\startdata
 $T=2.25$ &n4258 & Y & 439 & 0.17 & -3.23 (0.07) & -0.54 (0.13) & -6.03 (0.07)\\
 $T=2.5$ & n4258 &Y & 463 & 0.27 & -3.22 (0.08) & -0.49 (0.14) & -6.06 (0.07)\\
R11 & n4258 &  Y & 379 & 0.21 & -3.18 (0.07) & -0.32 (0.14) & -6.05 (0.08)\\
 $T=2.25$ &LMC & Y & 439 & 0.17 & -3.24 (0.05) & -0.54 (0.13) & -6.16 (0.07)\\
 $T=2.5$ & LMC &Y & 464 & 0.27 & -3.24 (0.05) & -0.50 (0.14) & -6.14 (0.08)\\
R11 & LMC &  Y & 379 & 0.21 & -3.22 (0.05) & -0.32 (0.14) & -6.07 (0.07)\\
 $T=2.25$ &MW & Y & 439 & 0.17 & -3.24 (0.07) & -0.54 (0.13) & -5.83 (0.05)\\
 $T=2.5$ & MW &Y & 463 & 0.27 & -3.24 (0.07) & -0.49 (0.15) & -5.83 (0.05)\\
R11 & MW &  Y & 379 & 0.21 & -3.20 (0.07) & -0.32 (0.15) & -5.82 (0.05)\\
 $T=2.25$ &n4258+LMC & Y & 439 & 0.18 & -3.23 (0.05) & -0.46 (0.11) & -6.10 (0.05)\\
 $T=2.5$ & n4258+LMC &Y & 466 & 0.28 & -3.23 (0.05) & -0.42 (0.12) & -6.10 (0.05)\\
R11 & n4258+LMC &  Y & 379 & 0.21 & -3.22 (0.05) & -0.29 (0.12) & -6.05 (0.05)\\
 $T=2.25$ &n4258+MW & Y & 437 & 0.17 & -3.31 (0.06) & -0.50 (0.12) & -5.89 (0.04)\\
 $T=2.5$ & n4258+MW &Y & 464 & 0.27 & -3.30 (0.07) & -0.46 (0.14) & -5.90 (0.04)\\
R11 & n4258+MW &  Y & 379 & 0.21 & -3.26 (0.06) & -0.30 (0.14) & -5.89 (0.04)\\
 $T=2.25$ &LMC+MW & Y & 435 & 0.16 & -3.27 (0.05) & -0.12 (0.10) & -5.91 (0.05)\\
 $T=2.5$ & LMC+MW &Y & 464 & 0.28 & -3.27 (0.05) & -0.15 (0.11) & -5.92 (0.06)\\
R11 & LMC+MW &  Y & 379 & 0.21 & -3.25 (0.05) & -0.08 (0.11) & -5.90 (0.06)\\
 $T=2.25$ & n4258+LMC+MW & Y & 434 & 0.16 & -3.28 (0.05) & -0.17 (0.10) & -5.95 (0.04)\\
 $T=2.5$ & n4258+LMC+MW &Y & 463 & 0.27 & -3.28 (0.05) & -0.19 (0.10) & -5.96 (0.04)\\
R11 & n4258+LMC+MW &  Y & 379 & 0.21 & -3.26 (0.05) & -0.12 (0.11) & -5.94 (0.04)\\
 $T=2.25$ & n4258 & N & 521 & 0.2 & -3.04 (0.05) & -0.42 (0.12) & -6.10 (0.07)\\
 $T=2.5$ & n4258 &N & 540 & 0.26 & -3.06 (0.06) & -0.32 (0.13) & -6.11 (0.07)\\
R11 & n4258 &  N & 444 & 0.21 & -3.09 (0.06) & -0.21 (0.13) & -6.08 (0.07)\\
 $T=2.25$ & LMC & N & 523 & 0.21 & -3.11 (0.04) & -0.39 (0.12) & -6.12 (0.07)\\
 $T=2.5$ & LMC &N & 544 & 0.28 & -3.12 (0.04) & -0.26 (0.13) & -6.06 (0.07)\\
R11 & LMC &  N & 444 & 0.21 & -3.13 (0.04) & -0.20 (0.13) & -6.04 (0.07)\\
 $T=2.25$ & MW & N & 521 & 0.2 & -3.07 (0.05) & -0.42 (0.12) & -5.80 (0.05)\\
 $T=2.5$ & MW &N & 539 & 0.26 & -3.09 (0.05) & -0.30 (0.13) & -5.81 (0.05)\\
R11 & MW &  N & 444 & 0.21 & -3.12 (0.06) & -0.20 (0.13) & -5.81 (0.05)\\
 $T=2.25$ & n4258+LMC & N & 523 & 0.21 & -3.11 (0.04) & -0.37 (0.11) & -6.09 (0.05)\\
 $T=2.5$ & n4258+LMC &N & 539 & 0.26 & -3.12 (0.04) & -0.30 (0.11) & -6.08 (0.05)\\
R11 & n4258+LMC &  N & 444 & 0.21 & -3.13 (0.04) & -0.21 (0.12) & -6.05 (0.05)\\
 $T=2.25$ & n4258+MW & N & 520 & 0.2 & -3.16 (0.05) & -0.40 (0.12) & -5.89 (0.04)\\
 $T=2.5$ & n4258+MW &N & 538 & 0.26 & -3.16 (0.05) & -0.29 (0.13) & -5.90 (0.04)\\
R11 & n4258+MW &  N & 444 & 0.21 & -3.17 (0.06) & -0.17 (0.13) & -5.89 (0.04)\\
 $T=2.25$ & LMC+MW & N & 519 & 0.2 & -3.14 (0.04) & -0.05 (0.10) & -5.89 (0.05)\\
 $T=2.5$ & LMC+MW &N & 546 & 0.29 & -3.14 (0.04) & -0.01 (0.10) & -5.89 (0.06)\\
R11 & LMC+MW &  N & 444 & 0.21 & -3.16 (0.04) & -0.01 (0.10) & -5.88 (0.05)\\
 $T=2.25$ & n4258+LMC+MW & N & 517 & 0.19 & -3.15 (0.04) & -0.09 (0.09) & -5.95 (0.04)\\
 $T=2.5$ & n4258+LMC+MW &N & 544 & 0.28 & -3.16 (0.04) & -0.08 (0.10) & -5.96 (0.04)\\
R11 & n4258+LMC+MW &  N & 444 & 0.21 & -3.17 (0.04) & -0.04 (0.10) & -5.94 (0.04)\\
\enddata
\end{deluxetable}

\subsection{Full results for $\{\Delta\mu_i\}$}
\label{app:offsets}

\begin{deluxetable}{lllllllllllll}
\tablecaption{Global fit results for distance modulus offsets $\{\Delta\mu_i\}$ for the eight galaxies.
\label{tab:offsetsfinal}}
\tablewidth{0pt}
\tabletypesize{\small}
\rotate
\tablehead{
  \colhead{Cepheid rejection} & \colhead{Distance anchor} &
    \colhead{$P<60$d}&
\colhead{ $\Delta\mu_{4536}$} & 
\colhead{  $\Delta\mu_{4639}$} & 
\colhead{ $\Delta\mu_{3370}$} & 
\colhead{ $\Delta\mu_{3982}$} & 
\colhead{ $\Delta\mu_{3021}$} & 
\colhead{ $\Delta\mu_{1309}$} & 
\colhead{ $\Delta\mu_{5584}$} & 
\colhead{ $\Delta\mu_{4038}$}
}
\startdata
2.25 & n4258+LMC+MW & Y & 1.56 (0.05) & 2.34 (0.07) & 2.77 (0.06) & 2.43 (0.07) & 2.86 (0.09) & 3.30 (0.07) & 2.37 (0.05) & 2.31 (0.09)\\
2.5 & n4258+LMC+MW & Y & 1.58 (0.06) & 2.41 (0.08) & 2.79 (0.06) & 2.47 (0.09) & 2.89 (0.09) & 3.30 (0.08) & 2.36 (0.06) & 2.35 (0.10)\\
R11 & n4258+LMC+MW & Y & 1.54 (0.06) & 2.35 (0.08) & 2.78 (0.06) & 2.43 (0.08) & 2.86 (0.09) & 3.18 (0.08) & 2.37 (0.06) & 2.32 (0.10)\\
2.5 & n4258+LMC & Y & 1.59 (0.06) & 2.44 (0.08) & 2.79 (0.07) & 2.49 (0.09) & 2.89 (0.09) & 3.31 (0.08) & 2.38 (0.06) & 2.39 (0.10)\\
2.25 & n4258+MW & Y & 1.52 (0.05) & 2.37 (0.07) & 2.73 (0.06) & 2.34 (0.08) & 2.87 (0.09) & 3.26 (0.07) & 2.33 (0.06) & 2.32 (0.09)\\
R11 & LMC+MW & Y & 1.56 (0.06) & 2.37 (0.08) & 2.80 (0.06) & 2.45 (0.08) & 2.88 (0.09) & 3.20 (0.08) & 2.38 (0.06) & 2.33 (0.10)\\
2.25 & n4258 & Y & 1.56 (0.06) & 2.39 (0.07) & 2.74 (0.06) & 2.35 (0.08) & 2.88 (0.09) & 3.26 (0.07) & 2.35 (0.06) & 2.34 (0.09)\\
2.25 & LMC & Y & 1.56 (0.06) & 2.39 (0.07) & 2.75 (0.06) & 2.35 (0.08) & 2.89 (0.09) & 3.27 (0.07) & 2.35 (0.06) & 2.34 (0.09)\\
R11 & MW & Y & 1.53 (0.06) & 2.36 (0.08) & 2.76 (0.07) & 2.39 (0.09) & 2.85 (0.10) & 3.17 (0.08) & 2.36 (0.06) & 2.33 (0.10)\\
2.25 & n4258+LMC+MW & N & 1.64 (0.05) & 2.31 (0.07) & 2.76 (0.05) & 2.39 (0.07) & 2.81 (0.08) & 3.27 (0.06) & 2.40 (0.05) & 2.30 (0.07)\\
2.5 & n4258+LMC+MW & N & 1.62 (0.06) & 2.41 (0.08) & 2.79 (0.06) & 2.44 (0.08) & 2.89 (0.09) & 3.30 (0.07) & 2.41 (0.05) & 2.35 (0.08)\\
R11 & n4258+LMC+MW & N & 1.57 (0.06) & 2.34 (0.08) & 2.78 (0.06) & 2.41 (0.08) & 2.87 (0.09) & 3.21 (0.07) & 2.39 (0.05) & 2.32 (0.08)\\
2.25 & n4258+MW & N & 1.58 (0.05) & 2.35 (0.07) & 2.72 (0.05) & 2.32 (0.08) & 2.87 (0.08) & 3.24 (0.07) & 2.38 (0.05) & 2.32 (0.07)\\
2.25 & LMC+MW & N & 1.67 (0.05) & 2.34 (0.07) & 2.79 (0.05) & 2.42 (0.07) & 2.84 (0.08) & 3.30 (0.07) & 2.43 (0.05) & 2.32 (0.07)\\
R11 & LMC+MW & N & 1.59 (0.06) & 2.36 (0.08) & 2.80 (0.06) & 2.43 (0.08) & 2.89 (0.09) & 3.22 (0.07) & 2.41 (0.05) & 2.33 (0.08)\\
2.25 & n4258 & N & 1.64 (0.05) & 2.37 (0.07) & 2.73 (0.05) & 2.34 (0.08) & 2.89 (0.08) & 3.25 (0.07) & 2.38 (0.05) & 2.32 (0.07)\\
2.5 & n4258 & N & 1.61 (0.06) & 2.41 (0.08) & 2.77 (0.06) & 2.40 (0.08) & 2.88 (0.09) & 3.27 (0.07) & 2.40 (0.06) & 2.34 (0.08)\\
R11 & n4258 & N & 1.55 (0.06) & 2.34 (0.08) & 2.75 (0.06) & 2.37 (0.08) & 2.85 (0.09) & 3.18 (0.07) & 2.37 (0.06) & 2.31 (0.08)\\
\enddata
\end{deluxetable}

\subsection{Full results of global fit} 
\label{app:global}

\begin{deluxetable}{llllllllllll}
  \tablecaption{Results of all global fits described in Section~\ref{sec:globalfit}, from each combination of Cepheid fit and SN cut. The SN parameters $\{\alpha,\beta\}$, the Cepheid parameters $\{b_W, Z_W, M_W\}$, and the zero points $\{\mathcal{H}, M_B$, and $\mu_{4258}\}$ are displayed. The indicative fits from which statistical uncertainties for parameters are retrieved are bolded.
\label{tab:globalfitfull}}
\tablewidth{0pt}
\tabletypesize{\small}
\rotate
\tablehead{
  \colhead{\scriptsize{Cepheid}}&  \colhead{\scriptsize{Distance}} &  \colhead{\scriptsize{$P<60$d}} & \colhead{SN cut}&\colhead{$\alpha$} & \colhead{$\beta$} & \colhead{$\mathcal{H}$} &\colhead{$M_B$} &\colhead{$b_W$} &\colhead{$Z_W$} &\colhead{$M_W$} & \colhead{$\mu_{4258}$}\\
  \colhead{\scriptsize{rejection}}&  \colhead{\scriptsize{anchor}} &  \colhead{} & \colhead{}&\colhead{}&\colhead{}&\colhead{}&\colhead{}&\colhead{}&\colhead{}&\colhead{}&\colhead{}
}
\startdata
2.25 & all & Y & default & 0.165 (0.010) & 3.09 (0.11) & -15.714 (0.094) & -18.955 (0.089) & -3.28 (0.05) & -0.19 (0.10) & -5.95 (0.04) & 29.36 (0.04)\\
2.25 & all & Y & higher $\chi^2$ & 0.167 (0.010) & 3.14 (0.11) & -15.706 (0.092) & -18.952 (0.087) & -3.28 (0.05) & -0.19 (0.10) & -5.95 (0.04) & 29.36 (0.04)\\
2.25 & all & Y & $z> 0.0233$ & 0.162 (0.012) & 2.77 (0.13) & -15.705 (0.093) & -18.954 (0.087) & -3.28 (0.05) & -0.20 (0.10) & -5.95 (0.04) & 29.36 (0.04)\\
2.25 & all & Y & lower $\chi^2$ & 0.158 (0.010) & 3.06 (0.12) & -15.702 (0.092) & -18.953 (0.087) & -3.28 (0.05) & -0.19 (0.09) & -5.95 (0.04) & 29.36 (0.04)\\
2.25 & all & Y & stricter $C$ & 0.156 (0.011) & 2.99 (0.14) & -15.717 (0.092) & -18.956 (0.087) & -3.28 (0.05) & -0.20 (0.10) & -5.95 (0.04) & 29.36 (0.05)\\
2.25 & all & Y & stricter $\sigma_{X_1}$ & 0.171 (0.010) & 3.11 (0.11) & -15.706 (0.093) & -18.950 (0.088) & -3.28 (0.05) & -0.19 (0.09) & -5.95 (0.04) & 29.36 (0.04)\\
2.5 & all & Y & default & 0.165 (0.010) & 3.09 (0.10) & -15.722 (0.095) & -18.964 (0.090) & -3.28 (0.05) & -0.22 (0.10) & -5.96 (0.04) & 29.35 (0.05)\\
2.5 & all & Y & higher $\chi^2$ & 0.166 (0.010) & 3.14 (0.11) & -15.722 (0.094) & -18.967 (0.089) & -3.28 (0.05) & -0.21 (0.10) & -5.96 (0.04) & 29.35 (0.05)\\
2.5 & all & Y & $z> 0.0233$ & 0.162 (0.012) & 2.77 (0.13) & -15.722 (0.096) & -18.971 (0.091) & -3.28 (0.05) & -0.21 (0.10) & -5.97 (0.04) & 29.35 (0.05)\\
2.5 & all & Y & lower $\chi^2$ & 0.158 (0.010) & 3.06 (0.12) & -15.717 (0.094) & -18.968 (0.089) & -3.29 (0.05) & -0.21 (0.10) & -5.96 (0.04) & 29.35 (0.04)\\
2.5 & all & Y & stricter $C$ & 0.156 (0.011) & 2.99 (0.14) & -15.729 (0.095) & -18.968 (0.090) & -3.28 (0.05) & -0.21 (0.10) & -5.96 (0.04) & 29.35 (0.05)\\
2.5 & all & Y & stricter $\sigma_{X_1}$ & 0.171 (0.010) & 3.11 (0.11) & -15.720 (0.094) & -18.964 (0.090) & -3.28 (0.05) & -0.21 (0.10) & -5.96 (0.04) & 29.35 (0.04)\\
R11 & all & Y & default & 0.165 (0.010) & 3.08 (0.11) & -15.688 (0.096) & -18.930 (0.091) & -3.26 (0.05) & -0.14 (0.10) & -5.95 (0.05) & 29.35 (0.05)\\
R11 & all & Y & higher $\chi^2$ & 0.167 (0.010) & 3.13 (0.10) & -15.684 (0.095) & -18.929 (0.090) & -3.26 (0.05) & -0.14 (0.11) & -5.95 (0.04) & 29.35 (0.05)\\
R11 & all & Y & $z> 0.0233$ & 0.162 (0.013) & 2.76 (0.13) & -15.684 (0.094) & -18.933 (0.089) & -3.26 (0.05) & -0.14 (0.11) & -5.95 (0.05) & 29.35 (0.04)\\
R11 & all & Y & lower $\chi^2$ & 0.158 (0.010) & 3.06 (0.12) & -15.682 (0.093) & -18.934 (0.088) & -3.26 (0.05) & -0.14 (0.10) & -5.95 (0.05) & 29.35 (0.05)\\
R11 & all & Y & stricter $C$ & 0.156 (0.011) & 2.97 (0.14) & -15.696 (0.094) & -18.935 (0.089) & -3.26 (0.05) & -0.14 (0.10) & -5.95 (0.04) & 29.35 (0.04)\\
R11 & all & Y & stricter $\sigma_{X_1}$ & 0.172 (0.011) & 3.11 (0.11) & -15.685 (0.094) & -18.929 (0.090) & -3.26 (0.05) & -0.14 (0.11) & -5.95 (0.04) & 29.35 (0.04)\\
\textbf{2.25} & \textbf{all} & \textbf{N} & \textbf{default} & \textbf{0.164 (0.010)} & \textbf{3.09 (0.11)} & \textbf{-15.689 (0.093)} & \textbf{-18.929 (0.088)} & \textbf{-3.18 (0.04)} & \textbf{-0.11 (0.09)} & \textbf{-5.95 (0.04)} & \textbf{29.34 (0.04)}\\
2.25 & all & N & higher $\chi^2$ & 0.166 (0.010) & 3.13 (0.10) & -15.683 (0.092) & -18.928 (0.088) & -3.17 (0.04) & -0.11 (0.09) & -5.95 (0.04) & 29.34 (0.04)\\
2.25 & all & N & $z> 0.0233$ & 0.160 (0.012) & 2.76 (0.13) & -15.684 (0.091) & -18.933 (0.086) & -3.17 (0.04) & -0.11 (0.09) & -5.95 (0.04) & 29.34 (0.04)\\
2.25 & all & N & lower $\chi^2$ & 0.158 (0.010) & 3.06 (0.12) & -15.678 (0.092) & -18.929 (0.088) & -3.17 (0.04) & -0.11 (0.09) & -5.95 (0.04) & 29.34 (0.04)\\
2.25 & all & N & stricter $C$ & 0.155 (0.011) & 2.98 (0.15) & -15.691 (0.093) & -18.929 (0.088) & -3.17 (0.04) & -0.11 (0.09) & -5.95 (0.04) & 29.34 (0.04)\\
2.25 & all & N & stricter $\sigma_{X_1}$ & 0.170 (0.010) & 3.11 (0.11) & -15.683 (0.093) & -18.926 (0.088) & -3.17 (0.04) & -0.11 (0.09) & -5.95 (0.04) & 29.34 (0.04)\\
2.5 & all & N & default & 0.165 (0.010) & 3.09 (0.11) & -15.710 (0.093) & -18.951 (0.089) & -3.20 (0.04) & -0.09 (0.10) & -5.96 (0.04) & 29.33 (0.04)\\
2.5 & all & N & higher $\chi^2$ & 0.166 (0.010) & 3.14 (0.11) & -15.707 (0.093) & -18.952 (0.089) & -3.20 (0.04) & -0.09 (0.10) & -5.96 (0.04) & 29.33 (0.04)\\
2.5 & all & N & $z> 0.0233$ & 0.161 (0.012) & 2.76 (0.13) & -15.707 (0.094) & -18.955 (0.089) & -3.20 (0.04) & -0.10 (0.10) & -5.96 (0.04) & 29.33 (0.04)\\
2.5 & all & N & lower $\chi^2$ & 0.158 (0.010) & 3.06 (0.12) & -15.702 (0.093) & -18.953 (0.089) & -3.20 (0.04) & -0.09 (0.10) & -5.96 (0.04) & 29.33 (0.04)\\
2.5 & all & N & stricter $C$ & 0.156 (0.011) & 2.98 (0.14) & -15.715 (0.095) & -18.953 (0.090) & -3.20 (0.04) & -0.10 (0.10) & -5.96 (0.04) & 29.33 (0.04)\\
2.5 & all & N & stricter $\sigma_{X_1}$ & 0.171 (0.010) & 3.11 (0.11) & -15.706 (0.092) & -18.950 (0.088) & -3.20 (0.04) & -0.10 (0.10) & -5.96 (0.04) & 29.33 (0.04)\\
R11 & all & N & default & 0.165 (0.010) & 3.09 (0.11) & -15.682 (0.095) & -18.923 (0.090) & -3.21 (0.04) & -0.06 (0.10) & -5.94 (0.04) & 29.34 (0.04)\\
R11 & all & N & higher $\chi^2$ & 0.167 (0.010) & 3.13 (0.11) & -15.677 (0.095) & -18.923 (0.090) & -3.21 (0.04) & -0.05 (0.10) & -5.94 (0.04) & 29.34 (0.04)\\
R11 & all & N & $z> 0.0233$ & 0.162 (0.013) & 2.76 (0.13) & -15.678 (0.094) & -18.927 (0.089) & -3.21 (0.04) & -0.06 (0.10) & -5.94 (0.04) & 29.34 (0.05)\\
R11 & all & N & lower $\chi^2$ & 0.158 (0.010) & 3.06 (0.12) & -15.674 (0.092) & -18.926 (0.087) & -3.21 (0.04) & -0.05 (0.10) & -5.94 (0.04) & 29.34 (0.04)\\
R11 & all & N & stricter $C$ & 0.156 (0.011) & 2.97 (0.15) & -15.687 (0.094) & -18.925 (0.089) & -3.21 (0.04) & -0.06 (0.10) & -5.94 (0.04) & 29.34 (0.04)\\
R11 & all & N & stricter $\sigma_{X_1}$ & 0.171 (0.011) & 3.11 (0.11) & -15.679 (0.095) & -18.923 (0.090) & -3.21 (0.04) & -0.06 (0.10) & -5.94 (0.04) & 29.34 (0.04)\\
2.5 & n4258+LMC & Y & default & 0.165 (0.010) & 3.09 (0.11) & -15.818 (0.097) & -19.059 (0.093) & -3.23 (0.05) & -0.45 (0.12) & -6.11 (0.06) & 29.43 (0.05)\\
2.5 & n4258+LMC & Y & higher $\chi^2$ & 0.167 (0.010) & 3.14 (0.11) & -15.814 (0.097) & -19.059 (0.092) & -3.23 (0.05) & -0.45 (0.12) & -6.11 (0.06) & 29.43 (0.05)\\
2.5 & n4258+LMC & Y & $z> 0.0233$ & 0.162 (0.013) & 2.77 (0.13) & -15.816 (0.096) & -19.065 (0.091) & -3.23 (0.05) & -0.45 (0.12) & -6.11 (0.05) & 29.43 (0.05)\\
2.5 & n4258+LMC & Y & lower $\chi^2$ & 0.158 (0.010) & 3.06 (0.12) & -15.810 (0.096) & -19.061 (0.091) & -3.23 (0.05) & -0.45 (0.12) & -6.11 (0.05) & 29.43 (0.05)\\
2.5 & n4258+LMC & Y & stricter $C$ & 0.156 (0.011) & 2.99 (0.15) & -15.822 (0.098) & -19.060 (0.092) & -3.23 (0.05) & -0.45 (0.12) & -6.11 (0.05) & 29.43 (0.05)\\
2.5 & n4258+LMC & Y & stricter $\sigma_{X_1}$ & 0.171 (0.011) & 3.11 (0.11) & -15.816 (0.098) & -19.060 (0.093) & -3.23 (0.05) & -0.45 (0.12) & -6.11 (0.05) & 29.43 (0.05)\\
2.25 & n4258+MW & Y & default & 0.165 (0.010) & 3.08 (0.11) & -15.650 (0.097) & -18.891 (0.092) & -3.31 (0.06) & -0.52 (0.12) & -5.89 (0.04) & 29.32 (0.04)\\
2.25 & n4258+MW & Y & higher $\chi^2$ & 0.167 (0.010) & 3.13 (0.10) & -15.646 (0.097) & -18.892 (0.092) & -3.31 (0.06) & -0.52 (0.12) & -5.89 (0.04) & 29.32 (0.04)\\
2.25 & n4258+MW & Y & $z> 0.0233$ & 0.162 (0.013) & 2.76 (0.13) & -15.645 (0.098) & -18.893 (0.093) & -3.31 (0.06) & -0.52 (0.12) & -5.89 (0.04) & 29.32 (0.04)\\
2.25 & n4258+MW & Y & lower $\chi^2$ & 0.159 (0.010) & 3.06 (0.12) & -15.642 (0.095) & -18.893 (0.091) & -3.31 (0.06) & -0.52 (0.12) & -5.89 (0.04) & 29.32 (0.04)\\
2.25 & n4258+MW & Y & stricter $C$ & 0.157 (0.011) & 2.97 (0.15) & -15.654 (0.098) & -18.893 (0.094) & -3.31 (0.06) & -0.51 (0.12) & -5.89 (0.04) & 29.32 (0.04)\\
2.25 & n4258+MW & Y & stricter $\sigma_{X_1}$ & 0.172 (0.010) & 3.11 (0.11) & -15.646 (0.097) & -18.890 (0.092) & -3.31 (0.06) & -0.52 (0.12) & -5.89 (0.04) & 29.32 (0.04)\\
2.25 & n4258+MW & N & default & 0.165 (0.010) & 3.08 (0.11) & -15.617 (0.093) & -18.858 (0.089) & -3.16 (0.05) & -0.40 (0.11) & -5.89 (0.04) & 29.29 (0.04)\\
2.25 & n4258+MW & N & higher $\chi^2$ & 0.167 (0.010) & 3.13 (0.10) & -15.613 (0.093) & -18.859 (0.089) & -3.16 (0.05) & -0.41 (0.12) & -5.89 (0.04) & 29.29 (0.04)\\
2.25 & n4258+MW & N & $z> 0.0233$ & 0.162 (0.012) & 2.75 (0.13) & -15.612 (0.094) & -18.861 (0.089) & -3.16 (0.05) & -0.41 (0.11) & -5.89 (0.04) & 29.29 (0.04)\\
2.25 & n4258+MW & N & lower $\chi^2$ & 0.158 (0.010) & 3.05 (0.12) & -15.609 (0.095) & -18.861 (0.090) & -3.16 (0.05) & -0.40 (0.11) & -5.89 (0.04) & 29.29 (0.04)\\
2.25 & n4258+MW & N & stricter $C$ & 0.156 (0.011) & 2.96 (0.14) & -15.621 (0.095) & -18.859 (0.090) & -3.16 (0.05) & -0.41 (0.11) & -5.89 (0.04) & 29.28 (0.04)\\
2.25 & n4258+MW & N & stricter $\sigma_{X_1}$ & 0.172 (0.010) & 3.10 (0.11) & -15.610 (0.096) & -18.854 (0.091) & -3.16 (0.05) & -0.41 (0.11) & -5.89 (0.04) & 29.28 (0.04)\\
R11 & LMC+MW & Y & default & 0.165 (0.010) & 3.09 (0.11) & -15.646 (0.102) & -18.887 (0.098) & -3.25 (0.05) & -0.11 (0.11) & -5.91 (0.06) & 29.29 (0.07)\\
R11 & LMC+MW & Y & higher $\chi^2$ & 0.167 (0.010) & 3.14 (0.11) & -15.640 (0.101) & -18.886 (0.097) & -3.25 (0.05) & -0.11 (0.11) & -5.91 (0.06) & 29.29 (0.07)\\
R11 & LMC+MW & Y & $z> 0.0233$ & 0.162 (0.012) & 2.76 (0.13) & -15.641 (0.102) & -18.890 (0.097) & -3.25 (0.05) & -0.11 (0.11) & -5.91 (0.06) & 29.29 (0.07)\\
R11 & LMC+MW & Y & lower $\chi^2$ & 0.158 (0.010) & 3.06 (0.12) & -15.640 (0.102) & -18.891 (0.098) & -3.25 (0.05) & -0.11 (0.11) & -5.91 (0.06) & 29.29 (0.07)\\
R11 & LMC+MW & Y & stricter $C$ & 0.156 (0.011) & 2.97 (0.14) & -15.654 (0.101) & -18.893 (0.097) & -3.25 (0.05) & -0.11 (0.11) & -5.91 (0.06) & 29.29 (0.07)\\
R11 & LMC+MW & Y & stricter $\sigma_{X_1}$ & 0.171 (0.011) & 3.11 (0.11) & -15.644 (0.102) & -18.888 (0.098) & -3.25 (0.05) & -0.11 (0.11) & -5.91 (0.06) & 29.29 (0.07)\\
2.25 & LMC+MW & N & default & 0.164 (0.010) & 3.09 (0.10) & -15.634 (0.098) & -18.875 (0.094) & -3.16 (0.04) & -0.07 (0.09) & -5.90 (0.05) & 29.26 (0.06)\\
2.25 & LMC+MW & N & higher $\chi^2$ & 0.166 (0.010) & 3.14 (0.10) & -15.630 (0.099) & -18.874 (0.095) & -3.16 (0.04) & -0.07 (0.10) & -5.90 (0.05) & 29.26 (0.06)\\
2.25 & LMC+MW & N & $z> 0.0233$ & 0.161 (0.012) & 2.76 (0.13) & -15.631 (0.098) & -18.879 (0.093) & -3.16 (0.04) & -0.07 (0.09) & -5.90 (0.05) & 29.26 (0.06)\\
2.25 & LMC+MW & N & lower $\chi^2$ & 0.158 (0.010) & 3.06 (0.12) & -15.624 (0.101) & -18.875 (0.097) & -3.16 (0.04) & -0.07 (0.10) & -5.90 (0.06) & 29.26 (0.07)\\
2.25 & LMC+MW & N & stricter $C$ & 0.155 (0.011) & 2.98 (0.14) & -15.637 (0.099) & -18.875 (0.095) & -3.16 (0.04) & -0.07 (0.10) & -5.90 (0.05) & 29.26 (0.06)\\
2.25 & LMC+MW & N & stricter $\sigma_{X_1}$ & 0.170 (0.010) & 3.11 (0.11) & -15.629 (0.100) & -18.873 (0.096) & -3.16 (0.04) & -0.07 (0.10) & -5.90 (0.06) & 29.26 (0.07)\\
R11 & LMC+MW & N & default & 0.165 (0.010) & 3.08 (0.11) & -15.626 (0.102) & -18.867 (0.097) & -3.20 (0.04) & -0.02 (0.10) & -5.89 (0.06) & 29.26 (0.07)\\
R11 & LMC+MW & N & higher $\chi^2$ & 0.167 (0.010) & 3.13 (0.10) & -15.620 (0.101) & -18.866 (0.097) & -3.20 (0.04) & -0.03 (0.10) & -5.89 (0.06) & 29.26 (0.07)\\
R11 & LMC+MW & N & $z> 0.0233$ & 0.162 (0.012) & 2.76 (0.13) & -15.623 (0.102) & -18.872 (0.098) & -3.20 (0.04) & -0.03 (0.10) & -5.89 (0.06) & 29.26 (0.07)\\
R11 & LMC+MW & N & lower $\chi^2$ & 0.158 (0.010) & 3.05 (0.12) & -15.618 (0.100) & -18.869 (0.095) & -3.20 (0.05) & -0.02 (0.10) & -5.89 (0.06) & 29.26 (0.07)\\
R11 & LMC+MW & N & stricter $C$ & 0.156 (0.011) & 2.97 (0.15) & -15.630 (0.103) & -18.869 (0.098) & -3.20 (0.05) & -0.03 (0.10) & -5.89 (0.06) & 29.26 (0.07)\\
R11 & LMC+MW & N & stricter $\sigma_{X_1}$ & 0.171 (0.010) & 3.10 (0.11) & -15.621 (0.103) & -18.865 (0.098) & -3.20 (0.04) & -0.02 (0.10) & -5.89 (0.06) & 29.26 (0.07)\\
2.25 & n4258 & Y & default & 0.165 (0.010) & 3.08 (0.11) & -15.747 (0.109) & -18.989 (0.104) & -3.23 (0.07) & -0.55 (0.12) & -6.03 (0.07) & 29.40 (0.06)\\
2.25 & n4258 & Y & higher $\chi^2$ & 0.167 (0.010) & 3.13 (0.11) & -15.740 (0.108) & -18.985 (0.104) & -3.23 (0.07) & -0.55 (0.12) & -6.03 (0.07) & 29.40 (0.06)\\
2.25 & n4258 & Y & $z> 0.0233$ & 0.162 (0.013) & 2.76 (0.13) & -15.744 (0.107) & -18.993 (0.102) & -3.23 (0.07) & -0.55 (0.12) & -6.04 (0.07) & 29.41 (0.06)\\
2.25 & n4258 & Y & lower $\chi^2$ & 0.158 (0.010) & 3.05 (0.12) & -15.737 (0.110) & -18.989 (0.105) & -3.23 (0.07) & -0.55 (0.12) & -6.03 (0.07) & 29.40 (0.06)\\
2.25 & n4258 & Y & stricter $C$ & 0.157 (0.011) & 2.97 (0.15) & -15.752 (0.109) & -18.990 (0.105) & -3.23 (0.07) & -0.55 (0.12) & -6.04 (0.07) & 29.41 (0.06)\\
2.25 & n4258 & Y & stricter $\sigma_{X_1}$ & 0.172 (0.011) & 3.10 (0.11) & -15.741 (0.108) & -18.985 (0.104) & -3.23 (0.07) & -0.55 (0.12) & -6.03 (0.07) & 29.40 (0.06)\\
2.25 & n4258 & N & default & 0.165 (0.010) & 3.08 (0.11) & -15.754 (0.107) & -18.995 (0.104) & -3.04 (0.05) & -0.44 (0.11) & -6.10 (0.07) & 29.40 (0.06)\\
2.25 & n4258 & N & higher $\chi^2$ & 0.167 (0.010) & 3.13 (0.11) & -15.747 (0.107) & -18.992 (0.103) & -3.04 (0.05) & -0.43 (0.11) & -6.10 (0.07) & 29.40 (0.06)\\
2.25 & n4258 & N & $z> 0.0233$ & 0.162 (0.013) & 2.75 (0.13) & -15.750 (0.106) & -18.999 (0.102) & -3.04 (0.05) & -0.44 (0.11) & -6.10 (0.07) & 29.40 (0.06)\\
2.25 & n4258 & N & lower $\chi^2$ & 0.158 (0.010) & 3.05 (0.12) & -15.747 (0.105) & -18.998 (0.102) & -3.04 (0.05) & -0.43 (0.11) & -6.10 (0.07) & 29.41 (0.06)\\
2.25 & n4258 & N & stricter $C$ & 0.156 (0.011) & 2.96 (0.15) & -15.759 (0.107) & -18.998 (0.103) & -3.04 (0.05) & -0.44 (0.11) & -6.10 (0.07) & 29.40 (0.06)\\
2.25 & n4258 & N & stricter $\sigma_{X_1}$ & 0.171 (0.010) & 3.10 (0.11) & -15.748 (0.106) & -18.992 (0.102) & -3.04 (0.05) & -0.44 (0.11) & -6.10 (0.07) & 29.40 (0.06)\\
2.5 & n4258 & N & default & 0.165 (0.010) & 3.08 (0.11) & -15.775 (0.107) & -19.017 (0.103) & -3.06 (0.06) & -0.33 (0.13) & -6.12 (0.07) & 29.41 (0.06)\\
2.5 & n4258 & N & higher $\chi^2$ & 0.166 (0.010) & 3.13 (0.10) & -15.769 (0.108) & -19.015 (0.104) & -3.06 (0.06) & -0.34 (0.13) & -6.11 (0.07) & 29.40 (0.06)\\
2.5 & n4258 & N & $z> 0.0233$ & 0.162 (0.012) & 2.76 (0.13) & -15.768 (0.108) & -19.017 (0.104) & -3.06 (0.06) & -0.34 (0.13) & -6.11 (0.07) & 29.40 (0.06)\\
2.5 & n4258 & N & lower $\chi^2$ & 0.158 (0.010) & 3.05 (0.12) & -15.762 (0.108) & -19.013 (0.104) & -3.06 (0.06) & -0.33 (0.13) & -6.11 (0.07) & 29.40 (0.06)\\
2.5 & n4258 & N & stricter $C$ & 0.156 (0.011) & 2.97 (0.15) & -15.779 (0.110) & -19.017 (0.106) & -3.06 (0.06) & -0.34 (0.13) & -6.11 (0.07) & 29.40 (0.06)\\
2.5 & n4258 & N & stricter $\sigma_{X_1}$ & 0.172 (0.010) & 3.10 (0.11) & -15.770 (0.107) & -19.014 (0.103) & -3.06 (0.06) & -0.34 (0.13) & -6.11 (0.07) & 29.40 (0.06)\\
R11 & n4258 & N & default & 0.165 (0.010) & 3.08 (0.11) & -15.728 (0.109) & -18.969 (0.105) & -3.09 (0.06) & -0.23 (0.13) & -6.08 (0.07) & 29.40 (0.06)\\
R11 & n4258 & N & higher $\chi^2$ & 0.167 (0.010) & 3.13 (0.10) & -15.725 (0.107) & -18.971 (0.103) & -3.09 (0.06) & -0.23 (0.13) & -6.08 (0.07) & 29.40 (0.06)\\
R11 & n4258 & N & $z> 0.0233$ & 0.162 (0.013) & 2.75 (0.13) & -15.723 (0.109) & -18.971 (0.104) & -3.09 (0.06) & -0.24 (0.13) & -6.08 (0.07) & 29.40 (0.06)\\
R11 & n4258 & N & lower $\chi^2$ & 0.159 (0.010) & 3.05 (0.12) & -15.718 (0.110) & -18.970 (0.106) & -3.09 (0.06) & -0.23 (0.13) & -6.07 (0.07) & 29.40 (0.06)\\
R11 & n4258 & N & stricter $C$ & 0.156 (0.011) & 2.97 (0.14) & -15.733 (0.110) & -18.972 (0.106) & -3.09 (0.06) & -0.23 (0.13) & -6.08 (0.07) & 29.40 (0.06)\\
R11 & n4258 & N & stricter $\sigma_{X_1}$ & 0.171 (0.011) & 3.10 (0.11) & -15.723 (0.109) & -18.967 (0.105) & -3.09 (0.06) & -0.23 (0.13) & -6.08 (0.07) & 29.41 (0.06)\\
2.25 & LMC & Y & default & 0.165 (0.010) & 3.08 (0.11) & -15.880 (0.104) & -19.122 (0.100) & -3.24 (0.05) & -0.55 (0.12) & -6.16 (0.07) & 29.54 (0.07)\\
2.25 & LMC & Y & higher $\chi^2$ & 0.167 (0.010) & 3.13 (0.11) & -15.873 (0.102) & -19.119 (0.098) & -3.24 (0.05) & -0.55 (0.12) & -6.16 (0.07) & 29.54 (0.07)\\
2.25 & LMC & Y & $z> 0.0233$ & 0.162 (0.013) & 2.76 (0.13) & -15.878 (0.104) & -19.127 (0.098) & -3.24 (0.05) & -0.56 (0.12) & -6.16 (0.07) & 29.54 (0.07)\\
2.25 & LMC & Y & lower $\chi^2$ & 0.159 (0.010) & 3.05 (0.12) & -15.869 (0.105) & -19.121 (0.101) & -3.24 (0.05) & -0.55 (0.12) & -6.16 (0.07) & 29.54 (0.07)\\
2.25 & LMC & Y & stricter $C$ & 0.157 (0.011) & 2.97 (0.15) & -15.884 (0.103) & -19.123 (0.099) & -3.24 (0.05) & -0.55 (0.12) & -6.16 (0.07) & 29.53 (0.07)\\
2.25 & LMC & Y & stricter $\sigma_{X_1}$ & 0.172 (0.011) & 3.10 (0.11) & -15.876 (0.103) & -19.119 (0.099) & -3.24 (0.05) & -0.55 (0.12) & -6.16 (0.07) & 29.53 (0.07)\\
R11 & MW & Y & default & 0.165 (0.010) & 3.08 (0.11) & -15.515 (0.111) & -18.757 (0.106) & -3.21 (0.07) & -0.35 (0.14) & -5.83 (0.05) & 29.19 (0.07)\\
R11 & MW & Y & higher $\chi^2$ & 0.167 (0.010) & 3.13 (0.11) & -15.513 (0.112) & -18.758 (0.108) & -3.21 (0.07) & -0.35 (0.14) & -5.83 (0.05) & 29.19 (0.07)\\
R11 & MW & Y & $z> 0.0233$ & 0.162 (0.013) & 2.75 (0.14) & -15.511 (0.111) & -18.760 (0.106) & -3.20 (0.07) & -0.36 (0.14) & -5.83 (0.05) & 29.19 (0.07)\\
R11 & MW & Y & lower $\chi^2$ & 0.158 (0.010) & 3.06 (0.12) & -15.508 (0.111) & -18.760 (0.107) & -3.21 (0.07) & -0.35 (0.14) & -5.83 (0.05) & 29.19 (0.07)\\
R11 & MW & Y & stricter $C$ & 0.156 (0.011) & 2.97 (0.15) & -15.521 (0.112) & -18.759 (0.107) & -3.21 (0.07) & -0.35 (0.14) & -5.83 (0.05) & 29.19 (0.07)\\
R11 & MW & Y & stricter $\sigma_{X_1}$ & 0.171 (0.011) & 3.10 (0.11) & -15.513 (0.110) & -18.757 (0.106) & -3.21 (0.07) & -0.35 (0.14) & -5.83 (0.05) & 29.19 (0.07)\\
\enddata
\end{deluxetable}

\bsp	
\label{lastpage}
\end{document}